\documentclass[twocolumn,english]{IEEEtran}
\usepackage[T1]{fontenc}
\usepackage[latin9]{inputenc}
\usepackage{color}
\usepackage{float}
\usepackage{amsmath}
\usepackage{amsthm}
\usepackage{amssymb}
\usepackage{graphicx}
\usepackage[numbers]{natbib}

\makeatletter

\floatstyle{ruled}
\newfloat{algorithm}{tbp}{loa}
\providecommand{\algorithmname}{Algorithm}
\floatname{algorithm}{\protect\algorithmname}

\theoremstyle{plain}
\newtheorem{thm}{\protect\theoremname}
\theoremstyle{definition}
\newtheorem{defn}[thm]{\protect\definitionname}
\theoremstyle{plain}
\newtheorem{prop}[thm]{\protect\propositionname}
\theoremstyle{plain}
\newtheorem{lem}[thm]{\protect\lemmaname}

\usepackage{subfigure}

\floatstyle{ruled}
\newfloat{algorithm}{tbp}{loa}
\providecommand{\algorithmname}{Algorithm}
\floatname{algorithm}{\protect\algorithmname}

\newcounter{algo}

\ifCLASSINFOpdf
\else
\fi
\usepackage[footnotesize]{caption}
\usepackage{flushend}

\makeatother

\usepackage{babel}
\providecommand{\definitionname}{Definition}
\providecommand{\lemmaname}{Lemma}
\providecommand{\propositionname}{Proposition}
\providecommand{\theoremname}{Theorem}

\begin{document}

\title{\vspace{-0.55cm}
 \hspace{0.9cm} Parallel and Distributed Methods for Nonconvex Optimization$-$Part
II: Applications}

\author{Gesualdo Scutari, Francisco Facchinei, Lorenzo Lampariello, Peiran
Song, and Stefania Sardellitti\vspace{-.8cm}\thanks{Scutari is with the School of Industrial Engineering and the Cyber Center (Discovery Park), Purdue University, West-Lafayette, IN, USA; email: \texttt{gscutari@purdue.edu}. Facchinei and Lampariello are with the Dept. of Computer, Control, and Management Engineering, University of Rome ``La Sapienza'',  Rome, Italy; emails: \texttt{<facchinei, lampariello>@diag.uniroma1.it}. Song is with  the School of Information and Communication Engineering, Beijing Information Science and Technology University, Beijing,  China; email: \texttt{peiransong@bistu.edu.cn}.  Sardellitti is with the Dept. of   Information, Electronics and Telecommunications, University of Rome ``La Sapienza'',  Rome, Italy; email: \texttt{sardellitti@diag.uniroma1.it}.\newline The work of  Scutari was supported by the USA National Science Foundation under Grants CMS 1218717, CCF 1564044, and CAREER Award 1254739. The work of Facchinei was partially supported by the MIUR project PLATINO, Grant  n. PON01$\_$01007.  Part of this work appeared in \cite{ScuFaccLampSonICASSP14,ScuFaccLampSonICASSP16}.}}
\maketitle
\begin{abstract}
In Part I of this paper, we proposed and analyzed a novel algorithmic
framework for the minimization of a nonconvex (smooth) objective function,
subject to nonconvex constraints, based on inner convex approximations. This Part II is devoted to the application of the framework
to some resource allocation problems in communication networks. In
particular, we consider two non-trivial case-study applications, namely: (generalizations
of) i) the rate profile maximization in MIMO interference broadcast
networks; and the ii) the max-min fair multicast multigroup beamforming
problem in a multi-cell environment. We develop a new class of algorithms
enjoying the following distinctive features: i) they are \emph{distributed}
across the base stations (with limited signaling) and lead to subproblems
whose solutions are computable in closed form; and ii) differently
from current relaxation-based schemes (e.g., semidefinite relaxation),
they are proved to always converge to d-stationary solutions of the
aforementioned class of nonconvex problems. Numerical results show
that the proposed (distributed) schemes achieve larger worst-case
rates (resp. signal-to-noise interference ratios) than state-of-the-art
centralized ones while having comparable computational complexity.\vspace{-.3cm}
\end{abstract}

\section{Introduction}

In Part I of this paper \citep{ScuFacLamPartI}, we proposed a novel
general algorithmic framework for the minimization of a nonconvex
(smooth) objective function $U:\,\mathcal{K}\to\mathbb{R}$ subject
to convex constraints $\mathcal{K}$ and nonconvex smooth ones $g_{j}(\mathbf{x})\leq0$,
with $g_{j}:\,\mathcal{K}\to\mathbb{R}$,\vspace{-.2cm} 
\begin{equation}
\begin{array}{cl}
\underset{\mathbf{x}}{\textnormal{min}} & \,\,U(\mathbf{x})\smallskip\\
\mbox{s.t.} & \vspace{-0.4cm}\\
 & \hspace{-0.1cm}\left.\begin{array}{l}
g_{j}(\mathbf{x})\le0,\;j=\,1,\ldots,m\\[5pt]
\mathbf{x}\in\mathcal{K}
\end{array}\right\} \triangleq\mathcal{X}.
\end{array}\label{eq:nonnonU}
\end{equation}
Building on the idea of inner convex approximation, our approach consists in solving
a sequence of \emph{strongly convex inner} approximations of (\ref{eq:nonnonU})
in the form: given $\mathbf{x}^{\nu}\in\mathcal{X}$, \vspace{-.3cm}

\begin{equation}
\begin{array}{cl}
\hat{{\mathbf{x}}}(\mathbf{x}^{\nu})=\underset{\mathbf{x}}{\textnormal{argmin}} & \,\,\tilde{U}(\mathbf{x};\mathbf{x}^{\nu})\medskip\\
\qquad\qquad\mbox{s.t.} & \vspace{-0.4cm}\\
 & \hspace{-0.1cm}\left.\begin{array}{l}
\tilde{g}_{j}(\mathbf{x};\mathbf{x}^{\nu})\le0,\;j=\,1,\ldots,m\\[5pt]
\mathbf{x}\in\mathcal{K}
\end{array}\right\} \triangleq\mathcal{X}(\mathbf{x}^{\nu}),
\end{array}\label{eq:convexified-subproblem}
\end{equation}
where $\tilde{U}(\mathbf{x};\mathbf{x}^{\nu})$ is a strongly convex
surrogate function of $U(\mathbf{x})$ and $\tilde{g}_{j}(\mathbf{x};\mathbf{x}^{\nu})$
is an upper convex approximant of $g_{j}(\mathbf{x})$, both depending on
the current iterate $\mathbf{x}^{\nu}$; and $\mathcal{X}(\mathbf{x}^{\nu})$
is the feasible set of \eqref{eq:convexified-subproblem}. Denoting
by $\hat{{\mathbf{x}}}(\mathbf{x}^{\nu})$ the unique solution of
(\ref{eq:convexified-subproblem}), the main iterate of the algorithm reads \citep{ScuFacLamPartI}: given $\mathbf{x}^{\nu}\in\mathcal{X}$,\vspace{-.1cm}
\begin{equation}
\mathbf{x}^{\nu+1}=\mathbf{x}^{\nu}+\gamma^{\nu}\,\left(\hat{{\mathbf{x}}}(\mathbf{x}^{\nu})-\mathbf{x}^{\nu}\right),\qquad\nu\geq0,\label{eq:main-iterate}\vspace{-.1cm}
\end{equation}
were $\{\gamma^{\nu}\}$ is a step-size sequence. The proposed
scheme represents a gamut of new algorithms, each of them corresponding
to a specific choice of the surrogate functions $\tilde{{U}}$ and
$\tilde{{g}_{j}}$, and the step-size sequence $\{\gamma^{\nu}\}$.
Several choices offering great flexibility to control iteration complexity,
communication overhead and convergence speed, while all guaranteeing
convergence are discussed in Part I of the paper, see \citep[Th. 1]{ScuFacLamPartI}.
Quite interestingly, the scheme leads to new efficient \emph{distributed}
algorithms even when customized to solve well-researched problems.
Some examples include power control problems in cellular systems \citep{DC-BranchAndBound-1,DC-BranchAndBound-5,DC-Linearization-1,Chiang-WeiTan-PalomarOneil-Julian_ITWC-GP},
MIMO relay optimization \citep{DC-Polynomial}, dynamic spectrum management
in DSL systems \citep{DC-BranchAndBound-2,CendrillonHuangChiangMoonen_TSP06},
sum-rate maximization, proportional-fairness and max-min optimization
of SISO/MISO/MIMO ad-hoc networks \citep{SchmidtShiBerryHonigUtschick-SPMag,KimGiannakisIT11,ScutFacchSonPalPang13,ZhangCui_RateProfileTSP10,MochaourabCaoJorswieck_RateProfile_arxiv13,QiuZhangLuoCui_maxminTSP11,LiuZhangChuai_RateProfileTWC12,RazaviyayniHongLuo_maxminSP13},
robust optimization of CR networks \citep{KimGiannakisIT11,DallAnese2012,YangScutariPalomar_JSAC13,Wang-Krunz-Cui_JSTSP08},
transmit beamforming design for multiple co-channel multicast groups
\citep{KariSidiLuo08,ChriChatOtte14,HsuWangSuLin14,XianTaoWang13},
and cross-layer design of wireless networks \citep{ConvexSumSeparable-2,Chiang_Hande_Lan_Tan_book_PC,Palomar-Chiang_ACTran07-Num}.
\\\indent Among the problems mentioned above, in this Part II, we
focus as case-study on two important resource allocation designs (and
their generalizations) that are representative of some key challenges
posed by the next-generation communication networks, namely: 1) the
rate profile maximization in MIMO interference broadcast networks;
and 2) the max-min fair multicast multigroup beamforming problem in
multi-cell systems. The interference management problem as in 1) has
become a compelling task in 5G densely deployed multi-cell networks,
where interference seriously limits the achievable rate performance
if not properly managed. Multicast beamforming as in 2) is a part
of the Evolved Multimedia Broadcast Multicast Service in the Long-Term
Evolution standard for efficient audio and video streaming and has
received increasing attention by the research community, due to the
proliferation of multimedia services in next-generation wireless networks.
Building on the framework developed in Part I, we propose a new class
of convex approximation-based algorithms for the aforementioned two problems enjoying
several desirable features. First, they provide better performance
guarantees than ad-hoc state-of-the art schemes (e.g., \citep{RazaviyayniHongLuo_maxminSP13,KariSidiLuo08,SidiDaviLuo06}),
both theoretically and numerically. Specifically, our algorithms achieve
(d-)stationary solutions of the problems under considerations, whereas
current relaxation-based algorithms (e.g., semidefinite relaxation)
either converge just to feasible points or to stationary points of
a related problem, which are not proved to be stationary for the original
problems. Second, our schemes represent the first class of \emph{distributed}
algorithms in the literature for such problems: at each iteration,
a convex problem is solved, which naturally decomposes across the
Base-Stations (BSs), and thus is solvable in a distribute way (with
limited signaling among the cells). Moreover, the solution of the
subproblems is computable in closed form by each BS. We remark that
the proposed parallel and distributed decomposition across the cells
naturally matches modern multi-tiers network architectures wherein
high-speeds wired links are dedicated to coordination and data exchange
among BSs. Third, our algorithms are quite flexible and applicable
also to generalizations of the original formulations 1) and 2). For
instance, i) one can readily add additional constraints, including
interference constraints, null constraints, per-antenna peak and average
power constraints, and quality of service constraints; and/or ii) one can consider several other objective functions (rather than the max-min
fairness), such as weighted users' sum-rate or rates' weighted geometric
mean; all of this without affecting the convergence of the resulting
algorithms. This is a major improvement on current solution methods,
which are instead rigid ad-hoc schemes that are not applicable to
other (even slightly different) formulations. 

The rest of the paper is organized as follows. Sec.$\,$\ref{sec:Interference-Broadcast-Networks}
focuses on the rate profile maximization in MIMO interference broadcast
networks and its generalizations: after reviewing the state of the
art (cf. Sec.$\,$\ref{sub:Related-works_IBC}), we propose centralized
and distributed schemes along with their convergence properties in
Sec. \ref{sub:Algorithmic-design-maxmin-centr} and Sec. \ref{sub:Distributed-implementation},
respectively, while some experiments are reported in Sec.$\,$\ref{sub:Numerical-results_IBC}.
Sec.$\,$\ref{sec:Multigroup-Multicast-Beamforming} studies the max-min
fair multicast multigroup beamforming problem in multi-cell systems:
the state of the art is summarized in Sec.$\,$\ref{sub:MFB_Related-works};
centralized algorithms based on alternative convexifications are introduced
in Sec.$\,$\ref{sec:Centralized-MMF}; whereas distributed schemes
are presented in Sec.$\,$\ref{sub:Distributed-implementation_MMF};
finally, Sec.$\,$\ref{sec:Simulations_MMFB} presents some numerical
results. Conclusions are drawn in Sec.$\,$\ref{sec:Conclusions}.\vspace{-.3cm}

\section{Interference Broadcast Networks\label{sec:Interference-Broadcast-Networks}\vspace{-.1cm}}

\subsection{System model}

\label{sec:System models} 
\begin{figure}[t]
\vspace{-1.6cm}\center \includegraphics[scale=0.25]{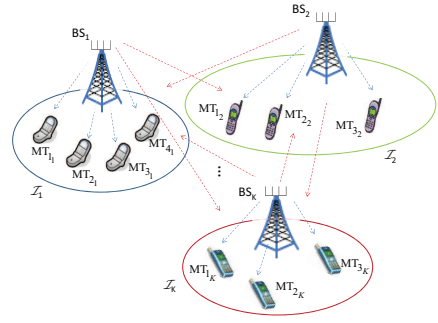}\vspace{-.2cm}
\caption{Interference Broadcast network.}

\label{fig:IBC} \vspace{-.5cm}
\end{figure}

Consider the Interference Broadcast Channel (IBC),
modeling a cellular system composed of $K$ cells; each cell $k\in\mathcal{K}_{\texttt{BS}}\triangleq\{1,\ldots,K\}$,
contains one Base Station (BS) equipped with $T_{k}$ transmit antennas
and serving $I_{k}$ Mobile Terminals (MTs); see Fig. \ref{fig:IBC}.
We denote by $i_{k}$ the $i$-th user in cell $k$, equipped with
$M_{i_{k}}$ antennas; the set of users in cell $k$ and the set of
all the users are denoted by $\mathcal{I}_{k}\triangleq\left\{ i_{k}:\,1\leq i\leq I_{k}\right\} $
and $\mathcal{I}\triangleq\{i_{k}\,:\,i_{k}\in\mathcal{I}_{k}\mbox{ and }k\in\mathcal{K}_{\texttt{BS}}\}$,
respectively. We denote by $\mathbf{Q}_{k}\triangleq(\mathbf{Q}_{i_{k}})_{i=1}^{I_{k}}$
the tuple of covariance matrices of the signals transmitted by each
BS $k$ to the $I_{k}$ users in the cell, with each $\mathbf{Q}_{i_{k}}\in\mathbb{C}^{T_{k}\times T_{k}}$
being the covariance matrix of the information symbols of user $i_{k}$. 

\noindent \indent Each BS $k$ is subject to power constraints in
the form 
\begin{equation}
\!\!\!\!\mathcal{Q}_{k}\triangleq\left\{ \begin{array}{ll}
\mathbf{Q}_{k}\in\mathcal{W}_{k}: & \!\!\!\!\mathbf{Q}_{i_{k}}\succeq\mathbf{0},\,\forall i_{k}\in\mathcal{I}_{k},\\
 & \!\!\!\!\sum_{i=1}^{I_{k}}\textrm{tr}(\mathbf{Q}_{i_{k}})\leq P_{k}
\end{array}\right\} ,\label{eq:feasible-set-BS_k}
\end{equation}
where the set $\mathcal{W}_{k}$, assumed to be closed and convex
(with nonempty relative interior) captures possibly additional constraints,
such as: i) per-antenna limits $\sum_{i=1}^{I_{k}}[\mathbf{Q}_{i_{k}}]_{tt}\leq\alpha_{t},$
with $t=1,\ldots,T_{k}$; ii) null constraints $\mathbf{U}_{k}^{H}\sum_{i=1}^{I_{k}}\mathbf{Q}_{i_{k}}=\mathbf{0}$,
where $\mathbf{U}_{k}$ is a given matrix whose columns contain the
directions (angle, time-slot, or frequency bands) along with BS $k$
is not allowed to transmit; and iii) soft-shaping $\sum_{i=1}^{I_{k}}\text{{tr}}(\mathbf{G}_{k}^{H}\mathbf{Q}_{i_{k}}\mathbf{G}_{k})\leq\beta_{k}$
and peak-power constraints $\sum_{i=1}^{I_{k}}\lambda_{\max}(\mathbf{G}_{k}^{H}\mathbf{Q}_{i_{k}}\mathbf{G}_{k})\leq\bar{{\beta}}_{k}$,
which limit respectively the total average and peak average power
radiated along the range space of matrix $\mathbf{G}_{k}$, where
$\lambda_{\max}(\mathbf{A})$ denotes the maximum eigenvalue of the
Hermitian matrix $\mathbf{A}$.
 \\\indent Treating the intra-cell and inter-cell interference at
each MT as noise, the maximum achievable rate of each user $i_{k}$
is 
\begin{equation}
R_{i_{k}}(\mathbf{Q})\triangleq\log\det\left(\mathbf{I}+\mathbf{H}_{i_{k}k}\mathbf{Q}_{i_{k}}\mathbf{H}_{i_{k}k}^{H}\bar{\mathbf{R}}_{i_{k}}(\mathbf{Q}_{-i_{k}})^{-1}\right)\label{eq:MIMO-rate-1}
\end{equation}
where $\mathbf{H}_{i_{k}l}\in\mathbb{C}^{M_{i_{k}}\times T_{l}}$
represents the channel matrix between BS $l$ and MT $i_{k}$; $\bar{\mathbf{R}}_{i_{k}}(\mathbf{Q}_{-i_{k}})\triangleq\sigma_{i_{k}}^{2}\mathbf{I}+\sum_{j\neq i}\mathbf{H}_{i_{k}k}\mathbf{Q}_{j_{k}}\mathbf{H}_{i_{k}k}^{H}+\sum_{l\neq k}\sum_{j=1}^{I_{l}}\mathbf{H}_{i_{k}l}\mathbf{Q}_{j_{l}}\mathbf{H}_{i_{k}l}^{H}$
is the covariance matrix of the Gaussian thermal noise (assumed to
be white w.l.o.g., otherwise one can always pre-whiten
the channel matrices) plus the intra-cell (second term) and inter-cell
(last term) interference; and we denoted $\mathbf{Q}_{-i_{k}}\triangleq(\mathbf{Q}_{j_{l}})_{(j,l)\neq(i,k)}$
and $\mathbf{Q}\triangleq(\mathbf{Q}_{k})_{k=1}^{K}.$ Accordingly,
with slight abuse of notation, we will write $\mathbf{Q}\in\mathcal{Q}$
to mean $\mathbf{Q}_{k}\in\mathcal{Q}_{k}$ for all $k\in\mathcal{K}_{\text{{BS}}}$.
\smallskip\\\noindent\textbf{ Rate profile maximization.} Max-min
fairness has long been considered an important design criterion for
wireless networks. Here we introduce the following more general rate-profile
maximization: given the profile ${\boldsymbol{\alpha}}\triangleq(\alpha_{i_{k}})_{i_{k}\in\mathcal{I}}$,
with each $\alpha_{i_{k}}>0$ and $\sum_{i_{k}\in\mathcal{I}}\alpha_{i_{k}}=1$,
let\vspace{-.1cm} 
\begin{equation}
\hspace{-1.8cm}\begin{array}{cl}
\underset{\mathbf{Q}}{\textrm{max}} & U(\mathbf{Q})\triangleq{\displaystyle \min_{i_{k}\in\mathcal{I}}}\,{\displaystyle \frac{R_{i_{k}}\left(\mathbf{Q}\right)}{\alpha_{i_{k}}}}\\
\textrm{\hspace{-0.06cm}s.t.} & \mathbf{Q}\in\mathcal{Q}.
\end{array}\tag{{\ensuremath{\mathcal{P}}}}\vspace{-.1cm}\label{eq:max-min-IBC}
\end{equation}

\noindent Special instances of this formulation have been proved to
be NP-hard (see, e.g., \citep{RazaviyayniHongLuo_maxminSP13}); therefore
in the following our focus is on computing efficiently (d-)stationary
solutions of \ref{eq:max-min-IBC}. 
\begin{defn}[d-stationarity]
\noindent  Given $\mathbf{Q}$
and $\mathbf{D}\triangleq(\mathbf{D}_{k}\mathbf{)}_{k=1}^{K}$, with
$\mathbf{D}_{k}\triangleq(\mathbf{D}_{i_{k}})_{i=1}^{I_{k}}$ and
$\mathbf{D}_{i_{k}}\in\mathbb{{C}}^{T_{k}\times T_{k}}$, let $U^{\prime}\!\left(\mathbf{Q};\mathbf{D}\right)$
denote the directional derivative of $U$ at $\mathbf{Q}\in\mathcal{Q}$
in the direction $\mathbf{D}$, defined as 
\[
U^{\prime}\!\left(\mathbf{Q};\mathbf{D}\right)\triangleq{\displaystyle \lim_{t\downarrow0}\dfrac{{U(\mathbf{Q}+t\mathbf{D})}-U(\mathbf{Q})}{t}.}
\]
 A tuple $\mathbf{Q^{\star}}$ is a d-stationary solution of \ref{eq:max-min-IBC} if 
\begin{equation}
U^{\prime}\!(\mathbf{Q}^{\star};\mathbf{Q}-\mathbf{Q}^{\star})\leq0,\quad\forall\mathbf{Q}\in\mathcal{Q}.\vspace{-.3cm}\label{eq:d-stationarity}
\end{equation}
\vspace{-0.4cm}
\end{defn}
\noindent Of course, (local/global) optimal solutions of \ref{eq:max-min-IBC}
satisfy (\ref{eq:d-stationarity}).

\noindent {\it Equivalent smooth reformulation:} To compute d-stationary
solutions of the nonconvex and nonsmooth problem \ref{eq:max-min-IBC},
we preliminarily rewrite \ref{eq:max-min-IBC} in an equivalent \emph{smooth}
(still nonconvex) form: introducing the slack variables $R\geq0,$
we have\smallskip
\noindent 
\begin{equation}
\begin{array}{cl}
\underset{\mathbf{Q},R\geq0}{\textrm{max}} & R\\
\textrm{\hspace{-0.15cm}\ s.t.} & \mathbf{Q}\in\mathcal{Q}\smallskip\\
 & \hspace{0.07cm}R_{i_{k}}(\mathbf{Q})\geq\alpha_{i_{k}}R,\;\,\forall i_{k}\in\mathcal{I}.\vspace{-0.1cm}\tag{\ensuremath{\ensuremath{\mathcal{P}}_{s}}}
\end{array}\label{eq:max-min-IBC-epi}
\end{equation}

We denote by $\mathcal{Z}$ the feasible set of \ref{eq:max-min-IBC-epi}.
Problems \ref{eq:max-min-IBC} and \ref{eq:max-min-IBC-epi} are equivalent
in the following sense. \vspace{-0.2cm}
\begin{prop}
\label{prop: eq-prob} Given \emph{\ref{eq:max-min-IBC}} and \emph{\ref{eq:max-min-IBC-epi}},
the following hold: 
\begin{description}
\item [{(i)}] Every feasible point of \emph{\ref{eq:max-min-IBC-epi}}
is regular {[}i.e., it satisfies the Mangasarian-Fromovitz Constraint
Qualification (MFCQ){]};
\item [{(ii)}] $\mathbf{Q}^{\star}$ is a d-stationary solution of \emph{\ref{eq:max-min-IBC}}
if and only if there exists $R^{\star}$ such that $\left(\mathbf{Q}^{\star},R^{\star}\right)$
is a stationary solution of \emph{\ref{eq:max-min-IBC-epi}}.
\end{description}	
\begin{IEEEproof}
See supporting material. 
\end{IEEEproof}
\end{prop}

Proposition \ref{prop: eq-prob} opens the way to the computation of d-stationary solutions of   \ref{eq:max-min-IBC} while designing algorithms for the smooth (nonconvex) formulation \ref{eq:max-min-IBC-epi}. We are not aware of any  algorithm with provable convergence to d-stationary solutions of   \ref{eq:max-min-IBC}, as documented next. \vspace{-0.3cm}

\subsection{Related works\label{sub:Related-works_IBC}}

Several resource allocation problems have been studied in the literature
for the vector Gaussian Interference Channel (IC), modeling multiuser
interference networks. Some representative examples corresponding
to different design criteria are: i) the sum-rate maximization problem
\citep{SchmidtShiBerryHonigUtschick-SPMag,KimGiannakisIT11,ScutFacchSonPalPang13,Wang-Krunz-Cui_JSTSP08};
ii) the minimization of the transmit power subject to QoS constraints
\citep{Chiang-WeiTan-PalomarOneil-Julian_ITWC-GP,Chiang_Hande_Lan_Tan_book_PC};
iii) the weighted Mean-Square-Error (MSE) minimization and the min-max
MSE fairness design \citep{shen2010mse,chen2012iterative}; and iv)
the rate profile optimization over MISO/SIMO \citep{ZhangCui_RateProfileTSP10,QiuZhangLuoCui_maxminTSP11,LiuZhangChuai_RateProfileTWC12,qiu2011optimal}
and MIMO (single-stream) \citep{MochaourabCaoJorswieck_RateProfile_arxiv13,Cai-Quek}
ICs. Since the IC is a special case of the IBC model, algorithms in
\citep{ZhangCui_RateProfileTSP10,MochaourabCaoJorswieck_RateProfile_arxiv13,LiuZhangChuai_RateProfileTWC12}
along with their convergence analysis cannot be applied to Problem
\ref{eq:max-min-IBC-epi} (and thus \ref{eq:max-min-IBC}). Moreover,
they are all\emph{ centralized}. \\ \indent Related to Problem \ref{eq:max-min-IBC}
is the max-min fairness formulation recently considered in \citep{RazaviyayniHongLuo_maxminSP13}.
There are however several differences between \citep{RazaviyayniHongLuo_maxminSP13}
and  our approach. First, the formulation in \citep{RazaviyayniHongLuo_maxminSP13}
is a special case of Problem \ref{eq:max-min-IBC}, corresponding
to equal $\alpha_{i_{k}}$ and standard power constrains (i.e., without
the additional constraints $\mathcal{W}_{k}$). Hence, the algorithm
in \citep{RazaviyayniHongLuo_maxminSP13} and its convergence analysis
do not apply to \ref{eq:max-min-IBC}. Second, even in the simplified
setting considered in \citep{RazaviyayniHongLuo_maxminSP13}, the
algorithm therein is not proved to converge to a (d-)stationary solution
of the max-min fairness formulation, but to critical points of an
auxiliary smooth problem that however might not be stationary for
the original problem. Our algorithmic framework instead is
guaranteed to converge to stationary solutions of \ref{eq:max-min-IBC-epi}
and thus, by Proposition \ref{prop: eq-prob}, to d-stationary solutions
of \ref{eq:max-min-IBC}. Third, the algorithm in \citep{RazaviyayniHongLuo_maxminSP13}
is \emph{centralized}. 

The analysis of the literature shows that Problem \ref{eq:max-min-IBC}
(and \ref{eq:max-min-IBC-epi}) remains unexplored in its generality.
The main contribution of this section is to propose the first (distributed)
algorithmic framework with provable convergence to d-stationary solutions
of \ref{eq:max-min-IBC}. More specifically, building on the iNner cOnVex Approximation (NOVA)  
framework developed in Part I \citep{ScuFacLamPartI}, we propose
next three alternative convexifications of the nonconvex constraints
in \ref{eq:max-min-IBC-epi}, which lead to different convex subproblems
{[}cf. (\ref{eq:convexified-subproblem}){]} and algorithms {[}cf.
(\ref{eq:main-iterate}){]}. We start in Sec. \ref{sub:Algorithmic-design-maxmin-centr}
with a centralized instance, while two alternative distributed implementations
are derived in Sec. \ref{sub:Distributed-implementation}. We remark
from the outset that all the convexifications we are going to introduce
satisfy Assumptions 2 and 3 (or 3 and 4) in \citep{ScuFacLamPartI},
implying {[}together with Proposition \ref{prop: eq-prob}(b){]} convergence
of the our algorithms. \vspace{-0.3cm}

\subsection{\label{sub:Algorithmic-design-maxmin-centr} Centralized solution
method}

\label{sec:max-min-fairness-IBC} The nonconvexity of Problem \ref{eq:max-min-IBC-epi}
is due to the nonconvex rate constraints $R_{i_{k}}(\mathbf{Q})\geq\alpha_{i_{k}}R.$
Exploiting the concave-convex structure of the rate function 
\begin{equation}
R_{i_{k}}(\mathbf{Q})=f_{i_{k}}^{+}(\mathbf{Q})-f_{i_{k}}^{-}(\mathbf{Q}_{-i_{k}}),\vspace{-0.1cm}\label{rate_user_ik}
\end{equation}
where $f_{i_{k}}^{+}(\mathbf{Q})=\log\det\left(\bar{\mathbf{R}}_{i_{k}}(\mathbf{Q}_{-i_{k}})+\mathbf{H}_{i_{k}k}\mathbf{Q}_{i_{k}}\mathbf{H}_{i_{k}k}^{H}\right)$
and \\
 $f_{i_{k}}^{-}(\mathbf{Q}_{-i_{k}})=\log\det(\bar{\mathbf{R}}_{i_{k}}(\mathbf{Q}_{-i_{k}}))$
are concave functions, a tight concave lower bound of $R_{i_{k}}(\mathbf{Q})$
(satisfying Assumptions 2-4 in \citep{ScuFacLamPartI}) is naturally
obtained by retaining in (\ref{rate_user_ik}) the concave part $f_{i_{k}}^{+}$
and linearizing the convex function $-f_{i_{k}}^{-}$, which leads
to the following rate approximation functions: given $\mathbf{Q}^{\nu}\triangleq(\mathbf{Q}_{i_{k}}^{\nu})_{i_{k}\in\mathcal{I}}$,
with each $\mathbf{Q}_{i_{k}}^{\nu}\succeq\mathbf{0}$,\vspace{-0.4cm}

\begin{equation}
{R}_{i_{k}}(\mathbf{Q})\geq\widetilde{R}_{i_{k}}(\mathbf{Q};\mathbf{Q}^{\nu})\triangleq{f_{i_{k}}^{+}}(\mathbf{Q})-\tilde{f}_{i_{k}}^{-}(\mathbf{Q}_{-i_{k}};\mathbf{Q}^{\nu})\vspace{-0.1cm}\label{eq:IBC-DC-rate}
\end{equation}
with \vspace{-0.5cm}

\begin{equation}
\begin{array}{l}
\tilde{f}_{i_{k}}^{-}(\mathbf{Q}_{-i_{k}};\mathbf{Q}^{\nu})\smallskip\\
{\displaystyle \triangleq{f_{i_{k}}^{-}}(\mathbf{Q}_{-i_{k}}^{\nu})+\sum_{(j,l)\neq(i,k)}\!}\!\!\!\!\left<\boldsymbol{\Pi}_{i_{k}\,j_{l}}^{\,-}(\mathbf{Q}_{-i_{k}}^{\nu}),\mathbf{Q}_{j_{l}}-\mathbf{Q}_{j_{l}}^{\nu}\right>
\end{array},\label{eq:f2tilde}
\end{equation}
\vspace{-0.1cm}

\noindent where $\left<\mathbf{A},\mathbf{B}\right>\!\triangleq\!\text{Re}\{\text{tr}(\mathbf{A}^{H}\!\mathbf{B})\}$,
and we denoted by $\boldsymbol{\Pi}_{i_{k}\,j_{l}}^{\,-}(\mathbf{Q}_{-i_{k}}^{\nu})\triangleq\nabla_{\mathbf{Q}_{j_{l}}^{*}}f_{i_{k}}^{-}(\mathbf{Q}_{-i_{k}}^{\nu})=\mathbf{H}_{i_{k}l}^{H}\bar{\mathbf{R}}_{i_{k}}^{-1}(\mathbf{Q}_{-i_{k}}^{\nu})\mathbf{H}_{i_{k}l}$
the conjugate gradient of $f_{i_{k}}^{-}$ w.r.t. $\mathbf{Q}_{jl}$
\citep{Scutari-Facchinei-Pang-Palomar_IT_PI}. \\\indent Given $\mathbf{Z}^{\nu}\triangleq(\mathbf{Q}^{\nu},R^{\nu})\in\mathcal{Z}$
and $\tau_{R},\,\tau_{\mathbf{Q}}>0$, the convex approximation of
problem \ref{eq:max-min-IBC-epi} {[}cf. (\ref{eq:convexified-subproblem}){]}
reads \vspace{-0.3cm}

\begin{equation}
\begin{array}{ccc}
\hat{\mathbf{Z}}(\mathbf{Z}^{\nu})\triangleq & \!\!\underset{\mathbf{Q},\,R\geq0}{\textrm{argmax}} & \!\!\!\!\!\left\{ R-{\displaystyle {\frac{\tau_{R}}{2}}(R-R^{\nu})^{2}}-\tau_{\mathbf{Q}}\left\Vert \mathbf{Q}-\mathbf{Q}^{\nu}\right\Vert _{F}^{2}\right\} \\
 & \textrm{\hspace{-0.4cm}\,\,s.t.} & \!\!\!\!\mathbf{Q}\in\mathcal{Q}\smallskip\hfill\\
 &  & \!\!\!\!\widetilde{R}_{i_{k}}(\mathbf{Q};\mathbf{Q}^{\nu})\geq\alpha_{i_{k}}R,\;\forall i_{k}\in\mathcal{I},\hfill\vspace{-0.1cm}
\end{array}\label{eq:max-min-IBC-approx-centr}
\end{equation}
where the the quadratic terms in the objective function are added to make
it strongly convex (see \citep[Assumption B1]{ScuFacLamPartI});
and we denoted by $\hat{\mathbf{Z}}(\mathbf{Z}^{\nu})$ the unique
solution of (\ref{eq:max-min-IBC-approx-centr}). Stationary solutions
of Problem \ref{eq:max-min-IBC-epi} can be then computed solving
the sequence of convexified problems (\ref{eq:max-min-IBC-approx-centr})
via (\ref{eq:main-iterate}); the formal description of the scheme
is given in Algorithm \ref{algoC}, whose convergence is stated in
Theorem \ref{th:conver_centralized}; the proof of the theorem follows
readily from Proposition \ref{prop: eq-prob} and \citep[Th.2]{ScuFacLamPartI}.\vspace{-0.1cm}

\begin{algorithm}[H]
\textbf{Data}: $\tau_{R},\tau_{\mathbf{Q}}>0$,
$\mathbf{Z}^{0}\triangleq(\mathbf{Q}^{0},R^{0})\,\in\,\mathcal{Z}$;
set $\nu=0$.

(\texttt{S.1}) If $\mathbf{Z}^{\nu}$ is a stationary solution of
\eqref{eq:max-min-IBC-epi}: \texttt{STOP}.

(\texttt{S.2}) Compute $\hat{\mathbf{Z}}(\mathbf{Z}^{\nu})$ by \eqref{eq:max-min-IBC-approx-centr}.

(\texttt{S.3}) Set $\mathbf{Z}^{\nu+1}=\mathbf{Z}^{\nu}+\gamma^{\nu}(\hat{\mathbf{Z}}(\mathbf{Z}^{\nu})-\mathbf{Z}^{\nu})$ for some $\gamma^{\nu}\in(0,1]$.

(\texttt{S.4}) $\nu\leftarrow\nu+1$ and go to step (\texttt{S.1}).

\protect\caption{\hspace{-3pt}\textbf{: }\label{algoC}NOVA Algorithm for Problem
\ref{eq:max-min-IBC-epi} (and \ref{eq:max-min-IBC}). }
\end{algorithm}
 \vspace{-0.3cm}
\begin{thm}
\label{th:conver_centralized} Let $\{\mathbf{Z}^{\nu}\triangleq(\mathbf{Q}^{\nu},R^{\nu})\}$
be the sequence generated by Algorithm \ref{algoC}. Choose any $\tau_{R},\tau_{\mathbf{Q}}>0$,
and the step-size sequence $\{\gamma^{\nu}\}$ such that $\gamma^{\nu}\in(0,1]$,
$\gamma^{\nu}\to0,$ and $\sum_{\nu}\gamma^{\nu}=+\infty.$ Then $\{\mathbf{Z}^{\nu}\}$
is bounded and every of its limit points $\mathbf{Z}^{\infty}\triangleq(\mathbf{Q}^{\infty},R^{\infty})$
is a stationary solution of Problem \ref{eq:max-min-IBC-epi}. Therefore,
$\mathbf{Q}^{\infty}$ is d-stationary for Problem \emph{\ref{eq:max-min-IBC}.
}Furthermore, if the algorithm does not stop after a finite number
of steps, none of the $\mathbf{Q}^{\infty}$ above is a local minimum
of U, and thus $R^{\infty}>0$. \vspace{-0.1cm}
\end{thm}
Theorem \ref{th:conver_centralized} offers some flexibility in the
choice of free parameters $(\tau_{R},\tau_{Q})$ and $\{\gamma^{\nu}\}$,
while guaranteeing convergence of Algorithm \ref{algoC}. Some effective
choices for $\{\gamma^{\nu}\}$ are discussed in Part I \citep{ScuFacLamPartI}.
Note also that the theorem guarantees that Algorithm \ref{algoC}
does not remain trapped in $R^{\infty}=0$, a ``degenerate'' stationary
solution of \ref{eq:max-min-IBC-epi} (the global minimizer of \ref{eq:max-min-IBC-epi}
and \emph{\ref{eq:max-min-IBC}}), at which some users do not receive
any service.

Algorithm \ref{algoC} is centralized because (\ref{eq:max-min-IBC-approx-centr})
cannot be decomposed across the base stations. This is due to the
lack of separability of the rate constraints in (\ref{eq:max-min-IBC-approx-centr}):
$f_{i_{k}}^{+}(\mathbf{Q})$ depends on the covariance matrices of
all the users. We introduce next an alternative valid convex approximation 
of the nonconvex rate constraints leading to distributed schemes.\vspace{-0.2cm}

\subsection{\label{sub:Distributed-implementation} Distributed implementation}

A centralized implementation might not be appealing in heterogeneous
multi-cell systems, where global information is not available at each
BS. Distributing the computation over the cells as well as alleviating
the communication overhead among the BSs is thus mandatory. This
subsection addresses this issue, and it is devoted to the design of a distributed algorithm  converging to d-stationary
solutions of Problem \emph{\ref{eq:max-min-IBC}}. 


  By keeping the concave part $f_{i_{k}}^{+}(\mathbf{Q})$
unaltered, the approximation $\widetilde{R}_{i_{k}}$ in (\ref{eq:IBC-DC-rate})
has the desired property of preserving the structure of the original
constraint function ${R}_{i_{k}}$ as much as possible. However, the
structure of ${R}_{i_{k}}$ is not suited to be decomposed across the users
due to the \emph{nonadditive} coupling among the variables $\mathbf{Q}_{i_{k}}$
in $f_{i_{k}}^{+}(\mathbf{Q})$. To cope with this issue, 
the proposed idea is to introduce in $\mathcal{P}$ slack variables whose purpose
is to decouple in each $f_{i_{k}}^{+}$ the covariance matrix $\mathbf{Q}_{i_{k}}$
of user $i_{k}$ from those of the other users$-$the interference
term $\bar{\mathbf{R}}_{i_{k}}(\mathbf{Q}_{-i_{k}})$. More specifically,
introducing the slack variables $\mathbf{Y}\triangleq(\mathbf{Y}_{k})_{k\in\mathcal{K}_{\mathtt{BS}}}$,
with $\mathbf{Y}_{k}=(\mathbf{Y}_{i_{k}})_{i_{k}\in\mathcal{I}_{k}}$,
and setting $\mathbf{I}_{i_{k}}(\mathbf{Q})\triangleq\sum_{l=1}^{K}\sum_{j=1}^{I_{l}}\mathbf{H}_{i_{k}l}\mathbf{Q}_{j_{l}}\mathbf{H}_{i_{k}l}^{H},$
we can write $f_{i_{k}}^{+}(\mathbf{Q})=\overline{f}_{i_{k}}^{\,+}(\mathbf{Y}_{i_{k}}),$
with $\overline{f}_{i_{k}}^{\,+}(\mathbf{Y}_{i_{k}})\triangleq\log\det(\sigma_{i_{k}}^{2}\mathbf{I}+\mathbf{Y}_{i_{k}})$
and $\mathbf{Y}_{i_{k}}=\mathbf{I}_{i_{k}}(\mathbf{Q})$. Then, in
view of (\ref{rate_user_ik}), Problem \ref{eq:max-min-IBC-epi} can
be rewritten in the following equivalent form:\vspace{-0.2cm} 
\begin{equation}
\hspace{-0.1cm}\begin{array}{cl}
\underset{\mathbf{Q},R\geq0,\mathbf{Y}}{\textrm{max}} & R\\
\textrm{\vspace{-1.5cm} s.t.} & \mathbf{Q}\in\mathcal{Q},\smallskip\\
 & \overline{f}_{i_{k}}^{\,+}(\mathbf{Y}_{i_{k}})-\!f_{i_{k}}^{-}(\mathbf{Q}_{-i_{k}})\!\geq\!\alpha_{i_{k}}R,\ \forall i_{k}\in\mathcal{I},\smallskip\\
 & \mathbf{0}\preceq\mathbf{Y}_{i_{k}}\preceq\mathbf{I}_{i_{k}}(\mathbf{Q}),\ \forall i_{k}\in\mathcal{I}.\tag{{\ensuremath{\mathcal{P}_{s}^{'}}}}\vspace{-0.1cm}
\end{array}\label{eq:max-min-IBC-distr}
\end{equation}
 The next proposition states the formal connection between \ref{eq:max-min-IBC-distr}
and \ref{eq:max-min-IBC-epi}, whose proof is omitted because of space
limitations. \vspace{-0.1cm}
\begin{prop}
\label{prop: eq-distributed} Given \emph{\ref{eq:max-min-IBC-epi}}
and \ref{eq:max-min-IBC-distr}, the following hold: 
\begin{description}
\item [{(i)}] Every feasible point $\left(\mathbf{Q},R,\mathbf{Y}\right)$
of \ref{eq:max-min-IBC-distr} satisfies the MFCQ;
\item [{(ii)}] $\left(\mathbf{Q}^{\star},R^{\star}\right)$ is a stationary
solution of \emph{\ref{eq:max-min-IBC-epi}} if and only if there
exists $\mathbf{Y}^{\star}$ such that $\left(\mathbf{Q}^{\star},R^{\star},\mathbf{Y}^{\star}\right)$
is a stationary solution of \ref{eq:max-min-IBC-distr}.
\vspace{-0.1cm}
\end{description}
\end{prop}
\noindent \indent It follows from Propositions \ref{prop: eq-prob}
and \ref{prop: eq-distributed} that, to compute d-stationary solutions
of \emph{\ref{eq:max-min-IBC}}, we can focus w.l.o.g on \ref{eq:max-min-IBC-distr}. Using (\ref{eq:f2tilde}), we can minorize
the left-hand side of the rate constraints in \ref{eq:max-min-IBC-distr}
as\vspace{-0.1cm} 
\[
\begin{array}{cc}
\overline{f}_{i_{k}}^{\,+}(\mathbf{Y}_{i_{k}})-\!f_{i_{k}}^{-}(\mathbf{Q}_{-i_{k}})\geq & \widehat{R}_{i_{k}}\left(\mathbf{Q}_{-i_{k}},\mathbf{Y}_{i_{k}};\mathbf{Q}^{\nu}\right)\hfill\\
\hfill\triangleq & \overline{f}_{i_{k}}^{\,+}(\mathbf{Y}_{i_{k}})-\widetilde{f}_{i_{k}}^{-}\left(\mathbf{Q}_{-i_{k}};\mathbf{Q}^{\nu}\right).
\end{array}
\]
 The approximation of Problem \ref{eq:max-min-IBC-distr} becomes
{[}cf.$\,$(\ref{eq:convexified-subproblem}){]}: given a feasible
$\mathbf{W}^{\nu}\triangleq\left(\mathbf{Q}^{\nu},R^{\nu},\mathbf{Y}^{\nu}\right)$,
and any $\tau_{\mathbf{Q}},\tau_{R},\tau_{\mathbf{Y}}>0$, 
\begin{equation}
\!\!\!\!\!\!\begin{array}{ccl}
\begin{array}{c}
\hat{\mathbf{W}}(\mathbf{W}^{\nu})\triangleq\\
\\
\end{array}\!\!\!\!\!\!\! & \!\!\!\begin{array}{c}
\underset{\mathbf{Q},\,R\geq0,\,\mathbf{Y}}{\textrm{argmax}}\\
\\
\end{array}\!\!\!\!\!\!\! & \!\!\!\begin{array}{c}
\left\{ R-{\displaystyle \frac{\tau_{R}}{2}}\left(R-R^{\nu}\right)^{2}\right.\hfill\\
\left.-\tau_{\mathbf{Q}}\left\Vert \mathbf{Q}-\mathbf{Q}^{\nu}\right\Vert _{F}^{2}-\tau_{\mathbf{Y}}\left\Vert \mathbf{Y}-\mathbf{Y}^{\nu}\right\Vert _{F}^{2}\right\} 
\end{array}\\
\!\!\!\!\!\!\! & \!\textrm{ \vspace{-1.6cm} s.t.}\!\!\!\!\!\!\! & \mathbf{Q}\in\mathcal{Q},\smallskip\\
\!\!\!\!\!\!\! & \!\!\!\!\!\!\!\! & \widehat{R}_{i_{k}}\left(\mathbf{Q}_{-i_{k}},\,\mathbf{Y}_{i_{k}};\mathbf{Q}^{\nu}\right)\geq\alpha_{i_{k}}R,\ \forall i_{k}\in\mathcal{I},\smallskip\\
\!\!\!\!\!\!\! & \!\!\!\!\!\!\!\! & \mathbf{0}\preceq\mathbf{Y}_{i_{k}}\preceq\mathbf{I}_{i_{k}}(\mathbf{Q}),\ \forall i_{k}\in\mathcal{I}. 
\end{array}\label{eq:max-min-IBC-approx-distr}
\end{equation}

\noindent To compute d-stationary solutions of Problem \emph{\ref{eq:max-min-IBC}}
via (\ref{eq:max-min-IBC-approx-distr}), we can invoke Algorithm
\ref{algoC} wherein the $\mathbf{Z}$ variables {[}resp. $\hat{\mathbf{Z}}(\mathbf{Z}^{\nu})$
in Step 2{]} are replaced by the $\mathbf{W}$ ones {[}resp. $\hat{\mathbf{W}}(\mathbf{W}^{\nu})$,
defined in (\ref{eq:max-min-IBC-approx-distr}){]}. Convergence of
this new algorithm is still given by Theorem \ref{th:conver_centralized}.
The difference with the centralized approach in Sec.$\,$\ref{sub:Algorithmic-design-maxmin-centr}
is that now, thanks to the additively separable structure of the objective
and constraint functions in (\ref{eq:max-min-IBC-approx-distr}),
one can compute $\hat{\mathbf{W}}$ in a distributed way, by leveraging
standard dual decomposition techniques; which is shown next. 

With $\mathbf{W}\triangleq\left(\mathbf{Q},R,\mathbf{Y}\right)$, denoting
$\boldsymbol{\lambda}\triangleq\left(\lambda_{i_{k}}\right)_{i_{k}\in\mathcal{I}}$
and $\boldsymbol{\Omega}\triangleq(\boldsymbol{\Omega}_{i_{k}})_{i_{k}\in\mathcal{I}}$
the multipliers associated to the rate constraints $\widehat{R}_{i_{k}}(\mathbf{Q}_{-i_{k}},\,\mathbf{Y}_{i_{k}};\mathbf{Q}^{\nu})\geq\alpha_{i_{k}}R$
and slack variable constraints $\mathbf{Y}_{i_{k}}\preceq\mathbf{I}_{i_{k}}(\mathbf{Q})$
respectively, let us define the (partial) Lagrangian of (\ref{eq:max-min-IBC-approx-distr})
as 
\[
\!\!\!\!\!\!\!\!\begin{array}{l}
\mathcal{L}(\mathbf{W},\boldsymbol{\lambda},\boldsymbol{\Omega};\mathbf{W}^{\nu})\triangleq\mathcal{L}_{R}(R,\boldsymbol{\lambda};R^{\nu})+\sum_{k=1}^{K}\mathcal{L}_{\mathbf{Q}_{k}}(\mathbf{Q}_{k},\boldsymbol{\lambda},\boldsymbol{\Omega};\mathbf{Q}^{\nu})\\
\hspace{3cm}+\sum_{k=1}^{K}\mathcal{L}_{\mathbf{Y}_{k}}(\mathbf{Y}_{k},\boldsymbol{\lambda},\boldsymbol{\Omega};\mathbf{Y}^{\nu})
\end{array}\smallskip
\]
\vspace{-0.2cm}where 
\[
\!\!\!\!\!\!\!\!\mathcal{L}_{R}(R,\boldsymbol{\lambda};R^{\nu})\triangleq-R+{\displaystyle \frac{\tau_{R}}{2}}(R-R^{\nu})^{2}+\boldsymbol{\boldsymbol{\lambda}}^{T}\boldsymbol{\alpha}R;
\]
\vspace{-0.2cm} 
\[
\begin{array}{l}
\!\!\!\!\!\!\mathcal{L}_{\mathbf{Q}_{k}}(\mathbf{Q}_{k},\boldsymbol{\boldsymbol{\lambda}},\boldsymbol{\Omega};\mathbf{Q}^{\nu})\triangleq\tau_{\mathbf{Q}}\left\Vert \mathbf{Q}_{k}-\mathbf{Q}_{k}^{\nu}\right\Vert _{F}^{2}+\sum_{i_{k}\in\mathcal{I}_{k}}\lambda_{i_{k}}f_{i_{k}}^{-}(\mathbf{Q}_{-i_{k}}^{\nu})\medskip\\
\qquad\qquad+\sum_{i_{k}\in\mathcal{I}_{k}}\sum_{(j,l)\neq(i,k)}\!\left\langle \lambda_{j_{l}}\boldsymbol{\Pi}_{j_{l}\,i_{k}}^{-}(\mathbf{Q}_{-j_{l}}^{\nu}),\mathbf{Q}_{i_{k}}-\mathbf{Q}_{i_{k}}^{\nu}\right\rangle \smallskip\\
\qquad\qquad-\sum_{i_{k}\in\mathcal{I}_{k}}\left\langle \sum_{j_{l}\in\mathcal{I}}\mathbf{H}_{j_{l}k}^{H}\boldsymbol{\Omega}_{j_{l}}\mathbf{H}_{j_{l}k},\mathbf{Q}_{i_{k}}\right\rangle ;
\end{array}
\]
 
\[
\!\!\!\!\!\begin{array}{l}
\mathcal{L}_{\mathbf{Y}_{k}}(\mathbf{Y}_{k},\boldsymbol{\lambda},\boldsymbol{\Omega};\mathbf{Y}^{\nu})\triangleq\tau_{\mathbf{Y}}\left\Vert \mathbf{Y}_{k}-\mathbf{Y}_{k}^{\nu}\right\Vert _{F}^{2}+\sum_{i_{k}\in\mathcal{I}_{k}}\left\langle \boldsymbol{\Omega}_{i_{k}},\mathbf{Y}_{i_{k}}\right\rangle \medskip\\
\qquad\qquad\qquad-\sum_{i_{k}\in\mathcal{I}_{k}}\lambda_{i_{k}}\log\det(\sigma_{i_{k}}^{2}\mathbf{I}+\mathbf{Y}_{i_{k}}),
\end{array}
\]
with $\boldsymbol{\alpha}\triangleq(\alpha_{i_{k}})_{i_{k}\in\mathcal{I}}$.
The additively separable structure of $\mathcal{L}(\mathbf{W},\boldsymbol{\lambda},\boldsymbol{\Omega};\mathbf{W}^{\nu})$
leads to the following decomposition of the dual function:\vspace{-0.1cm}
\begin{equation}
\begin{array}{lll}
D(\boldsymbol{\lambda},\boldsymbol{\Omega};\mathbf{W}^{\nu})\triangleq & {\displaystyle \min_{R\geq0}}\;\mathcal{L}_{R}(R,\boldsymbol{\lambda};R^{\nu})\\
 & +\sum_{k=1}^{K}{\displaystyle \min_{\mathbf{Q}_{k}\in\mathcal{Q}_{k}}}\mathcal{L}_{\mathbf{Q}_{k}}(\mathbf{Q}_{k},\boldsymbol{\lambda},\boldsymbol{\Omega};\mathbf{Q}^{\nu})\\
 & +\sum_{k=1}^{K}{\displaystyle \min_{(\mathbf{Y}_{i_{k}}\succeq\mathbf{0})_{i_{k}\in\mathcal{I}_{k}}}}\mathcal{L}_{\mathbf{Y}_{k}}(\mathbf{Y}_{k},\boldsymbol{\lambda},\boldsymbol{\Omega};\mathbf{Y}^{\nu}).\hfill
\end{array}\label{eq:dual-func-IBC-form1}
\end{equation}

The unique solutions of the above optimization problems are denoted
by $R^{\star}(\boldsymbol{\lambda};R^{\nu})={\displaystyle \text{{argmin}}_{R\geq0}}\mathcal{L}_{R}(R,\boldsymbol{\lambda};R^{\nu})$,
$\mathbf{Q}_{k}^{\star}(\boldsymbol{\lambda},\boldsymbol{\Omega};\mathbf{Q}^{\nu})={\displaystyle \text{{argmin}}_{\mathbf{Q}_{k}\in\mathcal{Q}_{k}}}$$\mathcal{L}_{\mathbf{Q}_{k}}(\mathbf{Q}_{k},\boldsymbol{\lambda},\boldsymbol{\Omega};\mathbf{Q}^{\nu})$,
and $\mathbf{Y}_{k}^{\star}(\boldsymbol{\lambda},\boldsymbol{\Omega};\mathbf{Y}^{\nu})={\displaystyle \text{{argmin}}_{(\mathbf{Y}_{i_{k}}\succeq\mathbf{0})_{i_{k}\in\mathcal{I}_{k}}}}$$\mathcal{L}_{\mathbf{Y}_{k}}(\mathbf{Y}_{k},\boldsymbol{\lambda},\boldsymbol{\Omega};\mathbf{Y}^{\nu})$.
We show next that $R^{\star}$ and $\mathbf{Y}_{k}^{\star}$ have
a closed form expression, and so does $\mathbf{Q}_{k}^{\star}$, when
the feasible sets $\mathcal{Q}_{k}$ contain only power budget constraints
[i.e., there is no set $\mathcal{W}_{k}$ in (\ref{eq:feasible-set-BS_k})],
a fact that will be tacitly assumed hereafter (the quite standard proof of Lemma \ref{lem:opt_R} is omitted because of space limitations). \vspace{-0.1cm}
\begin{lem}[Closed form of $R^{\star}(\boldsymbol{\lambda};R^{\nu})$]
\label{lem:opt_R} The optimal solution $R^{\star}(\boldsymbol{\lambda};R^{\nu})$
has the following expression\vspace{-0.1cm} 
\begin{equation}
R^{\star}(\boldsymbol{\lambda};R^{\nu})=\left[R^{\nu}-\frac{\boldsymbol{\lambda}^{T}\boldsymbol{\alpha}-1}{\tau_{R}}\right]_{+},\vspace{-0.1cm}\label{eq:bar-R-close-form-1}
\end{equation}
where $[\mathbf{x}]_{+}\triangleq\max(\mathbf{0},\mathbf{x})$ \emph{(}applied
component-wise\emph{).}
\begin{lem}[Closed form of $\mathbf{Q}_{k}^{\star}(\boldsymbol{\lambda},\boldsymbol{\Omega};\mathbf{Q}^{\nu})$]
\label{lem:opt_Q} Let $\mathbf{U}_{i_{k}}^{\nu}\mathbf{D}_{i_{k}}^{\nu}\mathbf{U}_{i_{k}}^{^{\nu}\,H}$
be the eigenvalue/eigenvector decomposition of\textcolor{red}{{} }$2\tau_{\mathbf{Q}}\mathbf{Q}_{i_{k}}^{\nu}-\sum_{(j,l)\neq(i,k)}\lambda_{j_{l}}\boldsymbol{\Pi}_{j_{l}\,i_{k}}^{-}\left(\mathbf{Q}_{-j_{l}}^{\nu}\right)+\sum_{j_{l}\in\mathcal{I}}\mathbf{H}_{j_{l}k}^{H}\boldsymbol{\Omega}_{j_{l}}\mathbf{H}_{j_{l}k}$
with $\mathbf{D}_{i_{k}}^{\nu}\triangleq\textrm{\emph{\text{{Diag}}}}(\mathbf{d}_{i_{k}}^{\nu})$.
Let us partition the optimal solution $\mathbf{Q}_{k}^{\star}(\boldsymbol{\lambda},\boldsymbol{\Omega};\mathbf{Q}^{\nu})\triangleq(\mathbf{Q}_{i_{k}}^{\star}(\boldsymbol{\lambda},\boldsymbol{\Omega};\mathbf{Q}^{\nu}))_{i_{k}\in\mathcal{I}_{k}}$.
Then each $\mathbf{Q}_{i_{k}}^{\star}(\boldsymbol{\lambda},\boldsymbol{\Omega};\mathbf{Q}^{\nu})$
has the following waterfilling-like expression \emph{{[}}we omit the
dependence on $(\boldsymbol{\lambda},\boldsymbol{\Omega};\mathbf{Q}^{\nu})$\emph{{]}}\vspace{-0.1cm}
\begin{equation}
\mathbf{Q}_{i_{k}}^{\star}=\mathbf{U}_{{i_{k}}}^{\nu}\textrm{\emph{\text{{Diag}}}}\!\left(\left[\frac{\mathbf{d}_{i_{k}}^{\nu}-\xi_{k}^{\star}}{2\tau_{\mathbf{Q}}}\right]_{+}\right)\mathbf{U}_{{i_{k}}}^{\nu\,H},\ \forall i_{k}\in\mathcal{I}_{k},\label{eq:Q-i-k-close-form1}
\end{equation}
where $\xi_{k}^{\star}\geq0$ is the water-level chosen to satisfy
the power constraint $\sum_{i_{k}\in\mathcal{I}_{k}}^{I_{k}}\text{tr}(\mathbf{Q}_{i_{k}}^{\star})\leq P_{k}$,
which can be found either exactly by the finite-step hypothesis method
\emph{(}see Algorithm$\,$\ref{algorithm:opt_multipliers}, Appendix$\,$\ref{app:clsoed_form_Q}\emph{)}\textcolor{red}{{}
}or approximately via bisection on the interval $\left[0,\sum_{i_{k}\in\mathcal{I}_{k}}\textrm{tr}\left(\textrm{\emph{\text{{Diag}}}}(\mathbf{d}_{i_{k}}^{\nu})\right)/(T_{k}\,I_{k})\right]$.
\vspace{-0.1cm}
\end{lem}
\end{lem}
\begin{IEEEproof}
See Appendix \ref{app:clsoed_form_Q}.\vspace{-0.1cm}\end{IEEEproof}
\begin{lem}[Closed form of $\mathbf{Y}_{k}^{\star}(\boldsymbol{\lambda},\boldsymbol{\Omega};\mathbf{Y}^{\nu})$]
 \label{lem:opt_Y} Let $\mathbf{V}_{i_{k}}^{\nu}\mathbf{D}_{i_{k}}^{\nu}\mathbf{V}_{i_{k}}^{\nu H}$
be the eigenvalue/eigenvector decomposition of $2\tau_{\mathbf{Y}}\mathbf{Y}_{i_{k}}^{\nu}-\boldsymbol{\Omega}_{i_{k}}^{\nu}$,
with $\mathbf{D}_{i_{k}}^{\nu}=\textrm{\textrm{\emph{\text{{Diag}}}}}(\mathbf{d}_{i_{k}}^{\nu})$.
Let us partition the optimal solution $\mathbf{Y}_{k}^{\star}(\boldsymbol{\lambda},\boldsymbol{\Omega};\mathbf{Y}^{\nu})\triangleq(\mathbf{Y}_{i_{k}}^{\star}(\boldsymbol{{\lambda}}_{i_{k}},\boldsymbol{\Omega}_{i_{k}};\mathbf{Y}^{\nu}))_{i_{k}\in\mathcal{I}_{k}}$.
Then, each $\mathbf{Y}_{i_{k}}^{\star}(\boldsymbol{{\lambda}}_{i_{k}},\boldsymbol{\Omega}_{i_{k}};\mathbf{Y}^{\nu})$
has the following expression \emph{{[}}we omit the dependence on $(\boldsymbol{\lambda}_{i_{k}},\boldsymbol{\Omega}_{i_{k}};\mathbf{Y}^{\nu})$\emph{{]}}\vspace{-0.1cm}
\begin{equation}
\mathbf{Y}_{i_{k}}^{\star}=\mathbf{V}_{i_{k}}^{\nu}\textrm{\emph{\text{{Diag}}}}\left(\mathbf{y}_{i_{k}}^{\nu}\right)\mathbf{V}_{i_{k}}^{\nu H},\ \forall i_{k}\in\mathcal{I}_{k},\label{eq:Y-i-k-close-form1}
\end{equation}
with\vspace{-0.3cm} 
\begin{equation}
\begin{array}{cc}
\mathbf{y}_{i_{k}}^{\nu}\triangleq & \left[-{\displaystyle \frac{1}{2}\left(\mathbf{\sigma}_{i_{k}}^{2}-\frac{1}{2\tau_{\mathbf{Y}}}\mathbf{d}_{i_{k}}^{\nu}\right)}\right.\hfill\smallskip\\
 & \left.+{\displaystyle \frac{1}{2}\sqrt{\left(\mathbf{\sigma}_{i_{k}}^{2}+\frac{1}{2\tau_{\mathbf{Y}}}\mathbf{d}_{i_{k}}^{\nu}\right)^{2}+\frac{2\lambda_{i_{k}}}{\tau_{\mathbf{Y}}}\,\mathbf{1}}}\,\right]_{+},\vspace{-0.1cm}
\end{array}\label{eq:y_opt}
\end{equation}
where $\mathbf{1}$ denotes the vector of all ones. \vspace{-0.2cm}\end{lem}
\begin{IEEEproof}
See Appendix \ref{app:clsoed_form_Y}. 
\end{IEEEproof}
\indent Finally, note that, since $\mathcal{L}\left(\mathbf{W},\boldsymbol{\lambda},\boldsymbol{\Omega};\mathbf{W}^{\nu}\right)$
is strongly convex for any given $\boldsymbol{\lambda}\geq\mathbf{0}$
and $\boldsymbol{\Omega}\triangleq\left(\boldsymbol{\Omega}_{i_{k}}\succeq\mathbf{0}\right)_{i_{k}\in\mathcal{I}}$,
the dual function $D(\boldsymbol{\lambda},\boldsymbol{\Omega};\mathbf{W}^{\nu})$
in (\ref{eq:dual-func-IBC-form1}) is ($\mathbb{{R}}$-)differentiable
on\textcolor{red}{{} }$\mathbb{{R}}_{+}^{I}\times\ensuremath{\prod_{i_{k}\in\mathcal{I}}\mathbb{C}^{M_{i_{k}}\times M_{i_{k}}}}$,
with (conjugate) gradient given by 
\begin{equation}
\begin{array}{c}
\begin{array}{cc}
\nabla_{\boldsymbol{\lambda}_{i_{k}}}D(\boldsymbol{\lambda},\boldsymbol{\Omega};\mathbf{W}^{\nu})= & \alpha_{i_{k}}R^{\star}-\widehat{R}_{i_{k}}\left(\mathbf{Q}_{-i_{k}}^{\star},\,\mathbf{Y}_{i_{k}}^{\star};\mathbf{Q}^{\nu}\right)\end{array}\\
\hspace{-0.7cm}\nabla_{\boldsymbol{\Omega}_{i_{k}}^{*}}D(\boldsymbol{\lambda},\boldsymbol{\Omega};\mathbf{W}^{\nu})=\mathbf{Y}_{i_{k}}^{\star}-\sum_{j_{l}\in\mathcal{I}}\mathbf{H}_{i_{k}l}\mathbf{Q}_{j_{l}}^{\star}\mathbf{H}_{i_{k}l}^{H},
\end{array}\label{eq:gradients_dual}
\end{equation}
where $R^{\star}$, $\mathbf{Q}_{i_{k}}^{\star}$, and $\mathbf{Y}_{i_{k}}^{\star}$
are given by (\ref{eq:bar-R-close-form-1}), (\ref{eq:Q-i-k-close-form1}),
and (\ref{eq:Y-i-k-close-form1}), respectively. Also, it can be shown
that the dual function is $\mathcal{C}^{2}$, with Lipschitz continuous
(augmented) Hessian with respect to $\mathbf{W}^{\nu}$. Then, the
dual problem \vspace{-0.3cm}

\begin{equation}
\max_{\underset{\boldsymbol{\Omega}=(\boldsymbol{\Omega}_{i_{k}}\succeq\mathbf{0})_{i_{k}\in\mathcal{I}}}{\boldsymbol{\lambda}\geq\mathbf{0}\smallskip}}D(\boldsymbol{\lambda},\boldsymbol{\Omega};\mathbf{W}^{\nu})\vspace{-0.1cm}\label{eq:dual-prob-IBC-form1}
\end{equation}
can be solved using either first or second order methods. A gradient-based
scheme with diminishing step-size is given in Algorithm \ref{alg:Dual-algorithm-form1},
whose convergence is stated in Theorem \ref{thm:2} (the proof of
the theorem follows from classical arguments and thus is omitted).
In Algorithm \ref{alg:Dual-algorithm-form1}, $P_{\mathbb{{H}}_{+}}(\mathbf{\bullet})$
denotes the orthogonal projection onto the set of complex positive
semidefinite (and thus Hermitian) matrices.
\begin{algorithm}[tbh]
\protect\protect\protect\caption{\label{alg:Dual-algorithm-form1}Distributed dual algorithm solving
(\ref{eq:dual-prob-IBC-form1})}

\textbf{Data:} $\mathbf{\boldsymbol{\lambda}}^{0}\triangleq(\lambda_{i_{k}}^{0})_{i_{k}\in\mathcal{I}}\geq\mathbf{0}$,
$\boldsymbol{\Omega}^{0}\triangleq(\boldsymbol{\Omega}_{i_{k}}^{0}\succeq\mathbf{0})_{i_{k}\in\mathcal{I}}$,
$\mathbf{W}^{\nu}=(\mathbf{Q}^{\nu},R^{\nu},\mathbf{Y}^{\nu})$.
Set $n=0$.

(\texttt{S.1}) If $\mathbf{\boldsymbol{{\lambda}}}^{n}$ and $\boldsymbol{\Omega}^{n}$
satisfy a suitable termination criterion: STOP.

(\texttt{S.2}) Compute each $R^{\star}(\boldsymbol{\lambda}^n;R^{\nu})$,
$\mathbf{Q}_{i_{k}}^{\star}(\boldsymbol{\lambda}^{n},\boldsymbol{\Omega}^{n};\mathbf{Q}^{\nu})$,
and $\mathbf{Y}_{i_{k}}^{\star}(\boldsymbol{\lambda}^{n},\boldsymbol{\Omega}^{n};\mathbf{Y}^{\nu})$
{[}cf. (\ref{eq:bar-R-close-form-1})-(\ref{eq:Y-i-k-close-form1}){]}.

(\texttt{S.3}) Update $\boldsymbol{\lambda}$ and $\boldsymbol{\Omega}$
according to 
\[
\begin{array}{c}
\lambda_{i_{k}}^{n+1}=\left[\lambda_{i_{k}}^{n}+\beta^{n}\nabla_{\lambda_{i_{k}}}D(\boldsymbol{\lambda},\boldsymbol{\Omega};\mathbf{W}^{\nu})\right]_{+}\\
\boldsymbol{\Omega}_{i_{k}}^{n+1}=P_{\mathbb{{H}}_{+}}\left(\boldsymbol{\Omega}_{i_{k}}^{n}+\beta^{n}\nabla_{\boldsymbol{\Omega}_{i_{k}}^{*}}D(\boldsymbol{\lambda},\boldsymbol{\Omega};\mathbf{W}^{\nu})\right)
\end{array},\ \forall i_{k}\in\mathcal{I}
\]
for some $\beta^n >0$.

(\texttt{S.4}) $n\leftarrow n+1$, go back to (\texttt{S.1}).
\end{algorithm}
\vspace{-0.1cm}
\begin{thm}
\label{thm:2} Let $\left\{ \left(\boldsymbol{\lambda}^{n},\boldsymbol{\Omega}^{n}\right)\right\} $
be the sequence generated by Algorithm \ref{alg:Dual-algorithm-form1}. Choose
the step-size sequence $\left\{ \beta^{n}\right\} $ such that $\beta^{n}>0$,
$\beta^{n}\rightarrow0$, $\sum_{n}\beta^{n}=+\infty$, and $\sum_{n}\left(\beta^{n}\right)^{2}<\infty$. Then, $\left\{ \left(\boldsymbol{\lambda}^{n},\boldsymbol{\Omega}^{n}\right)\right\}$ converges to a solution of (\ref{eq:dual-prob-IBC-form1}). Therefore, the sequence
$\left\{ \left(R^{\star}(\boldsymbol{\lambda}^{n};R^{\nu}),\,\mathbf{Q}^{\star}\left(\boldsymbol{{\lambda}}^{n},\boldsymbol{\Omega}^{n};\mathbf{Q}^{\nu}\right),\mathbf{Y}^{\star}\left(\boldsymbol{{\lambda}}^{n},\boldsymbol{\Omega}^{n};\mathbf{Y}^{\nu}\right)\right)\right\} $
converges to the unique solution of (\ref{eq:max-min-IBC-approx-distr}). 
\end{thm}
Algorithm \ref{alg:Dual-algorithm-form1} can be implemented in a
fairly distributed way across the BSs, with limited communication
overhead. More specifically, given the current value of $(\boldsymbol{\lambda},\boldsymbol{\Omega})$,
each BS $k$ updates individually the covariance matrices $\mathbf{Q}_{k}=(\mathbf{Q}_{i_{k}})_{i_{k}\in\mathcal{I}_{k}}$
of the users in the cell $k$ as well as the slack variables $\mathbf{Y}_{k}=(\mathbf{Y}_{i_{k}})_{i_{k}\in\mathcal{I}_{k}}$,
by computing the closed form solutions $\mathbf{Q}_{i_{k}}^{\star}(\boldsymbol{\lambda}^{n},\boldsymbol{\Omega}^{n};\mathbf{Q}^{\nu})$
{[}cf. (\ref{eq:Q-i-k-close-form1}){]} and $\mathbf{Y}_{i_{k}}^{\star}(\boldsymbol{\lambda}^{n},\boldsymbol{\Omega}^{n};\mathbf{Y}^{\nu})$
{[}cf. (\ref{eq:Y-i-k-close-form1}){]}. The update of the $R$-variable {[}using
$R^{\star}(\boldsymbol{\lambda};R^{\nu})${]} and the multipliers
$(\boldsymbol{\lambda},\boldsymbol{\Omega})$ require some coordination
among the BSs: it can be either carried out by a BS header or locally
by all the BSs if a consensus-like scheme is employed in order to obtain locally
the information required to compute $R^{\star}(\boldsymbol{\lambda};R^{\nu})$
and the gradients in (\ref{eq:gradients_dual}).

\indent The dual problem (\ref{eq:dual-prob-IBC-form1}) can be also
solved using a second order-based scheme, which is expected in practice
to be faster than a gradient-based one. It is sufficient to replace
Step 3 of Algorithm \ref{alg:Dual-algorithm-form1} with the following
updating rules for the multipliers:\vspace{-.3cm} 
\begin{equation}
\begin{split} & \boldsymbol{\lambda}^{n+1}=\boldsymbol{\boldsymbol{\lambda}}^{n}+s^{n}\left([\hat{\boldsymbol{\lambda}}^{n+1}]_{+}-\boldsymbol{\boldsymbol{\lambda}}^{n}\right),\\
 & \boldsymbol{\Omega}^{n+1}=\boldsymbol{\Omega}^{n}+s^{n}\left(P_{\mathbb{{H}}_{+}}(\hat{\boldsymbol{\Omega}}^{n+1})-\boldsymbol{\Omega}^{n}\right),
\end{split}
\label{Newton_algo}
\end{equation}
where $\hat{\boldsymbol{\lambda}}^{n+1}$ and $\hat{\boldsymbol{\Omega}}^{n+1}$
can be computed by $\hat{\boldsymbol{\Lambda}}^{n+1}\triangleq[\hat{\boldsymbol{\boldsymbol{\lambda}}}^{n+1},\text{vec}\!(\hat{{\boldsymbol{\Omega}}}^{n+1})^{T}]^{T}$,
with $\text{vec}\!(\hat{{\boldsymbol{\Omega}}})\triangleq(\text{vec}\!(\hat{{\boldsymbol{\Omega}}})_{i_{k}})_{i_{k}\in\mathcal{I}}$,
which is updated according to 
\begin{equation}
\begin{split}\hat{\boldsymbol{\Lambda}}^{n+1}\triangleq\hat{\boldsymbol{\Lambda}}^{n}+ & (\nabla_{\boldsymbol{\boldsymbol{\lambda}},\text{vec}(\boldsymbol{\Omega}^{*})}^{2}D(\boldsymbol{\boldsymbol{\lambda}}^{n},\boldsymbol{\Omega}^{n};\mathbf{W}^{\nu}))^{-1}\\
 & \cdot\nabla_{\boldsymbol{\boldsymbol{\lambda}},\text{vec}(\boldsymbol{\Omega}^{*})}D(\boldsymbol{\boldsymbol{\lambda}}^{n},\boldsymbol{\Omega}^{n};\mathbf{W}^{\nu}).
\end{split}
\label{lambda_new}
\end{equation}

In Appendix \ref{sec:Newton_Hessian} we provide the explicit expression
of the (augmented) Hessian matrices and gradients in (\ref{lambda_new}).\vspace{-0.3cm}

\subsection{Numerical results\label{sub:Numerical-results_IBC}\vspace{-0.1cm}}
\begin{figure}[t]
\center \vspace{-0.9cm}
\includegraphics[scale=0.45, ]{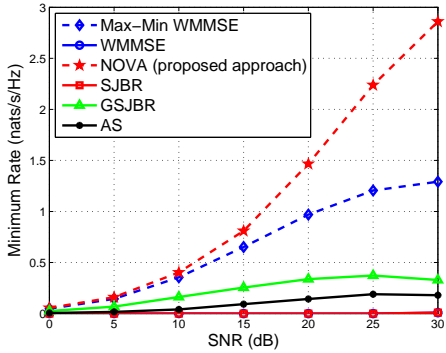} \vspace{-0.2cm}
 \protect\caption{Minimum rate vs. $\text{SNR}$: The proposed algorithm reaches better
solutions than the state of the art.}
\label{fig:rate_vs_snr_IBC} \vspace{-0.6cm}
\end{figure}
In this section we present some experiments assessing the effectiveness
of the proposed formulation and algorithms.

\noindent \emph{Example\,$\#\,1:\,$Centralized algorithm}.
We start comparing Algorithm \ref{algoC} with three other approaches
presented in the literature for the resource allocation over IBCs,
namely: 1) the Max-Min WMMSE \citep{RazaviyayniHongLuo_maxminSP13},
which aims at maximizing the minimum rate of the system (a special
case of \ref{eq:max-min-IBC-epi}); 2) the WMMSE algorithm \citep{ShiRazaviyaynLuoHe-TSP11}
and the partial linearization-based algorithm (termed SJBR) \citep{ScutFacchSonPalPang13},
which consider the maximization of the system sum-rate; and 3) the
partial linearization-based algorithm \citep{ScutFacchSonPalPang13},
applied to maximize the geometric mean of the rates (the proportional
fairness utility function), termed GSJBR. As benchmark,
we also report the results achieved using the standard nonlinear programming
solver in Matlab, specifically the active-set algorithm in 'fmincon' (among the other  options in fmincon, the active-set algorithm was the one that showed better performance);
we refer to it as ``AS'' algorithm. To allow the comparison,
we consider a special case of \ref{eq:max-min-IBC-epi} as in \citep{RazaviyayniHongLuo_maxminSP13}.
We simulated a $K=4$ cell IBC with $\mathcal{I}_{k}=3$ randomly
placed active MTs per cell; the BSs and MTs are equipped with 4 antennas.
Channels are Rayleigh fading, whose path-loss are generated using
the 3GPP(TR 36.814) methodology. We assume white zero-mean Gaussian
noise at each receiver, with variance $\sigma^{2}$, and same power
budget $P$ for all the BSs; the signal to noise ratio is then $\textrm{SNR}=P/\sigma^{2}$.
Algorithm \ref{algoC} is simulated using $\tau_{R}=1$e$-7$ and
the step-size rule $\gamma^{\nu}=\gamma^{\nu-1}(1-10^{-3}\gamma^{\nu-1})$,
with $\gamma^{0}=1$. The same step-size rule is used for SJBR and
GSJBR. The same random feasible initialization is used for all the
algorithms. All algorithms are terminated when the absolute value
of the difference between two consecutive values of the objective
function becomes smaller than $1e-3$. In Fig.$\,$\ref{fig:rate_vs_snr_IBC}
we plot the minimum rate versus $\text{SNR}$ achieved by the aforementioned
algorithms. All results are averaged over 300 independent channel/topology
realizations. The figures show that our algorithm yields
substantially more fair rate allocations (larger minimum rates) than
all the others. As expected, we observed that SJBR and WMMSE achieve
higher sum-rates (not reported in the figure) while sacrificing the
fairness: SJBR and WMMSE can shut off some users (the associated minimum
rate is zero). More specifically, for  $\text{SNR}\sim15$dB (resp. $\text{SNR}\sim30$dB) we observed the following   average losses on the sum-rate  w.r.t.  SJBR (and WMMSE):   $\sim$ \!$70\%$ (resp.  $\sim$ \!$45\%$) for our algorithm and   Max-Min WMMSE;  $\sim$ \!$50\%$ (resp.  $\sim$ \!$60\%$) for  GSJBR; and $\sim$ \!$80\%$ (resp.  $\sim$ \!$80\%$) for AS. Between Algorithm \ref{algoC} and the Max-Min WMMSE
\citep{RazaviyayniHongLuo_maxminSP13}, the former provides better
solutions, both in terms of minimum rate (cf.$\,$Fig.$\,$\ref{fig:rate_vs_snr_IBC})
and sum-rate (not reported). This might be due to the fact that the
Max-Min WMMSE converges to stationary solutions of an auxiliary nonconvex
problem (obtained lifting \ref{eq:max-min-IBC-epi}) that are not
proved to be stationary also for the original formulation \ref{eq:max-min-IBC};
our algorithm instead converges to d-stationary solutions of \ref{eq:max-min-IBC}.

\begin{figure}[t]
\vspace{-.5cm}\center \includegraphics[scale=0.245, ]{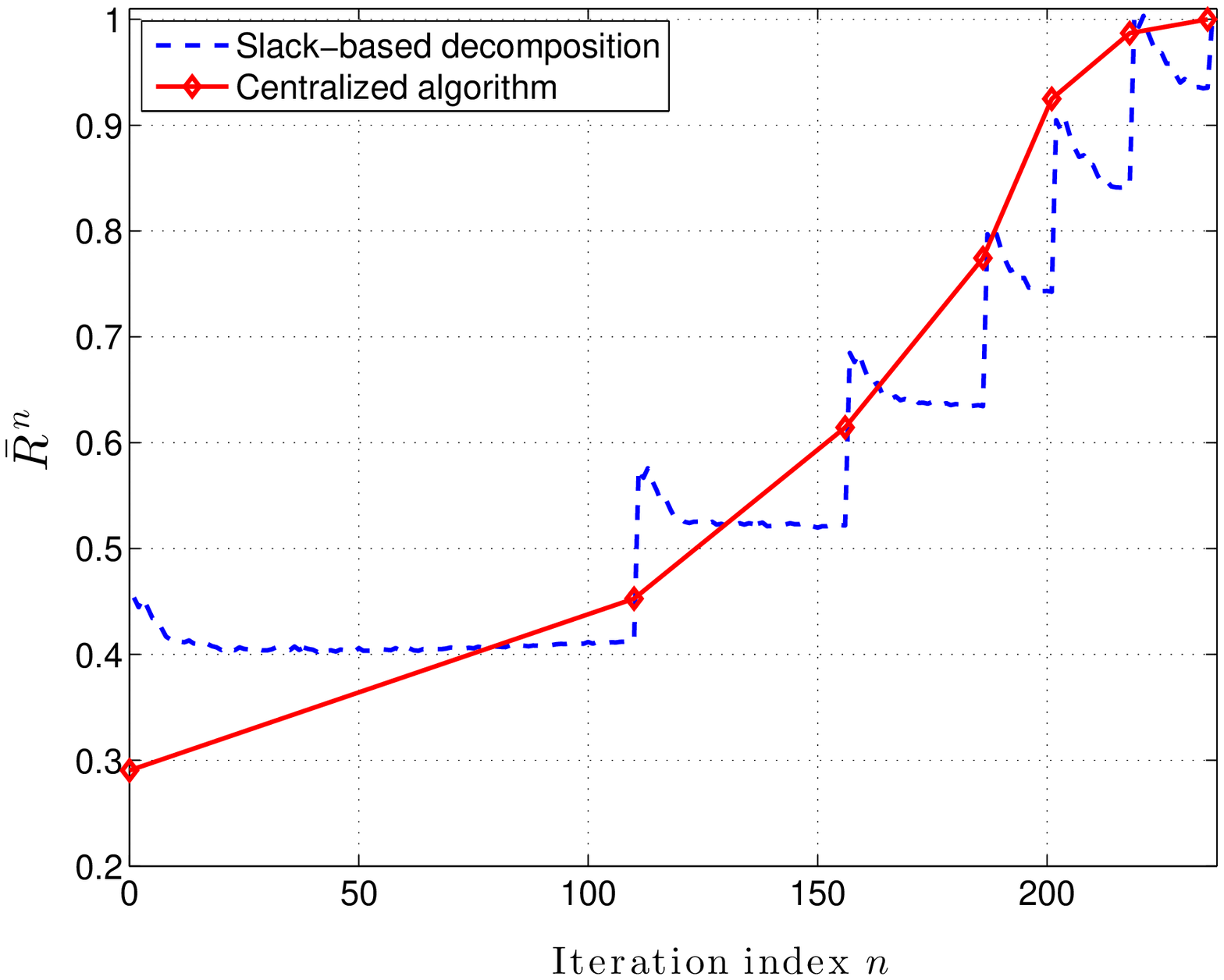}\,\includegraphics[scale=0.245, ]{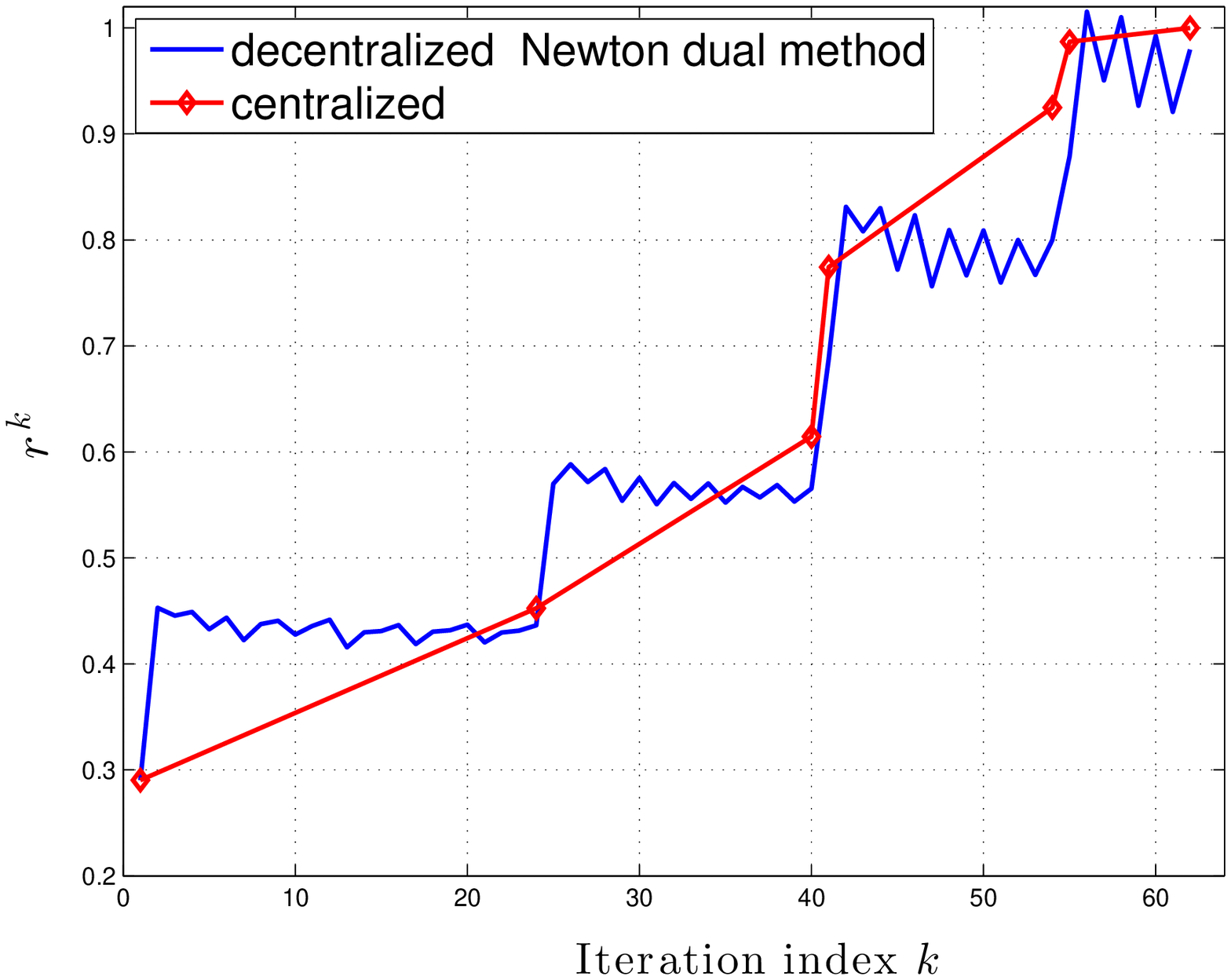}\caption{{\footnotesize{}Minimum rate vs. iterations. Left plot: Centralized
algorithm vs. Distributed, first order; Right plot: Centralized algorithm
vs. Distributed, second order.}}

\label{fig:rate_vs_snr_IBC-dual} \vspace{-0.6cm}
\end{figure}

\noindent \emph{Example$\,$$\#$$\,$2:$\,$Distributed algorithms.}
We test now the distributed algorithms in Sec. \ref{sub:Distributed-implementation}
and compare them with the centralized implementation. Specifically,
we simulate i) Algorithm \ref{algoC} based on the solution $\hat{{\mathbf{W}}}(\mathbf{W}^{\nu})$
(termed \emph{Centralized algorithm}); ii) the same algorithm as in
i) but with $\hat{{\mathbf{W}}}(\mathbf{W}^{\nu})$ computed in a
distributed way using Algorithm \ref{alg:Dual-algorithm-form1} (termed
\emph{Distributed, first-order}); and iii) the same algorithm as in
i) but with $\hat{{\mathbf{W}}}(\mathbf{W}^{\nu})$ computed solving
the dual problem (\ref{eq:dual-prob-IBC-form1}) using a second-order
method (termed \emph{Distributed, second-order}). The simulated scenario
as well as the tuning of the algorithms is as in Example$\,\#1$.
In Algorithm \ref{alg:Dual-algorithm-form1}, the step-size sequence
$\{\beta^{n}\}$ has been chosen as $\beta^{n}=\beta^{n-1}(1-0.8\cdot\beta^{n-1})$;
also the starting point $(\lambda^{0},\boldsymbol{{\Omega}}^{0})$
is set equal to the optimal solution $(\lambda^{\star},\boldsymbol{{\Omega}}^{\star})$
of the previous round. Fig. \ref{fig:rate_vs_snr_IBC-dual} shows
the normalized rate evolution achieved by the aforementioned algorithms versus the iteration index $n$. For the distributed algorithms,
the number of iterations counts both the inner and outer iterations.
Note that all the algorithms converge to the same stationary point
of Problem \ref{eq:max-min-IBC}, and they are quite fast. As expected,
exploiting second order information accelerates the practical convergence
but at the cost of extra signaling among the BSs.\vspace{-0.1cm}

\section{Multigroup Multicast Beamforming\label{sec:Multigroup-Multicast-Beamforming}}

\subsection{System model}

\begin{figure}[t]
 \center \includegraphics[scale=0.30]{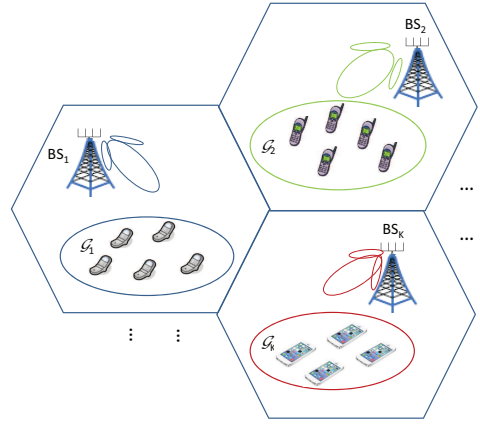}\vspace{-.2cm}
\caption{Multicell multi-group multicast network.}
\label{fig:multicast3} \vspace{-.2cm}
\end{figure}

We study the general Max-Min Fair (MMF) beamforming problem for \emph{multi}-group
multicasting \citep{KariSidiLuo08}, where different groups of subscribers
request different data streams from the transmitter; see Fig.$\,$\ref{fig:multicast3}.
\\\indent We assume that there are $K$ BSs (one per cell), each of them equipped
with $N_{t}$ transmit antennas and a total of $I$ active users,
which have a single receive antenna; let $\mathcal{I}\triangleq\{1,\ldots,I\}$.
For notational simplicity, we assume w.l.o.g. that each BS serves
a single multicast group; let $\mathcal{G}_{k}$ denote the group
of users served by the $k$-th BS, with $k\in\mathcal{K}_{\text{{BS}}}\triangleq\{1,\cdots,K\}$;
$\mathcal{G}_{1},\ldots,\mathcal{G}_{K}$ is a partition of $\mathcal{I}$.
We will denote by $i_{k}$ user $i$ belonging to group $\mathcal{G}_{k}$.
The extension of the algorithm to the multi-group case (i.e.,
multiple groups in each cell) is straightforward. Letting $\mathbf{w}_{k}\in\mathbb{C}^{N_{t}}$
be the beamforming vector for transmission at BS $k$ (to group $\mathcal{G}_{k}$),
the Max-Min Fair (MMF) beamforming problem reads\vspace{-0.1cm} 
\begin{equation}
\begin{array}{cl}
\underset{\mathbf{w}\triangleq(\mathbf{w}_{k})_{k\in\mathcal{K}_{\text{{BS}}}}}{\textrm{max}} & \!\!\!\!U(\mathbf{w})\triangleq\underset{i_{k}\in\mathcal{G}_{k},\,k\in\mathcal{K}_{\text{{BS}}}}{\textrm{min}}{\displaystyle {\frac{\mathbf{w}_{k}^{H}\mathbf{H}_{i_{k}k}\mathbf{w}_{k}}{\sum_{\ell\neq k}{\mathbf{w}_{\ell}^{H}\mathbf{H}_{i_{k}\ell}\mathbf{w}_{\ell}}+\sigma_{i_{k}}^{2}}}\bigskip}\\
\textrm{s. t.} & \!\!\!\!\!\!\!\!\begin{array}[t]{l}
\|\mathbf{w}_{k}\|_{2}^{2}\leq P_{k},\,\,\forall k\in\mathcal{K}_{\text{{BS}}},\end{array}
\end{array}\label{eq:MMF}
\end{equation}
where $\mathbf{H}_{i_{k}\ell}$ is a positive semidefinite (not all
zero) matrix modeling the channel between the $\ell$-th BS and user
$i_{k}\in\mathcal{G}_{k}$; specifically, $\mathbf{H}_{i_{k}\ell}=\mathbf{h}_{i_{k}\ell}\mathbf{h}_{i_{k}\ell}^{H}$
if instantaneous CSI is assumed, with $\mathbf{h}_{i_{k}\ell}\in\mathbb{C}^{N_{t}}$
being the frequency-flat quasi-static channel vector from BS $\ell$
to user $i_{k}$; and $\mathbf{H}_{i_{k}\ell}=\mathbb{{E}}(\mathbf{h}_{i_{k}\ell}\mathbf{h}_{i_{k}\ell}^{H})$
represents the spatial correlation matrix if only long-term CSI is
available (in the latter case, no special structure for $\mathbf{H}_{i_{k}\ell}$
is assumed); $\sigma_{i_{k}}^{2}$ is the variance of the AWGN at
receiver $i_{k}$; and $P_{k}$ is the power budget of cell $k$.
Note that different qualities of service among users can be readily
accommodated by multiplying each SINR in (\ref{eq:MMF}) by a predetermined
positive factor, which we will tacitly assume to be absorbed in the
channel matrices $\mathbf{H}_{i_{k}k}$. We denote by $\mathcal{W}$
the (convex) feasible set of (\ref{eq:MMF}). We remark that, differently
from the literature, one can add further (convex) constraints in $\mathcal{W}$,
such as per-antenna power, null or interference constraints; the algorithmic
framework we are going to introduce is still applicable.

A simplified instance of problem (\ref{eq:MMF}), where there exists
only one cell (and multiple groups), i.e., $K=1$, has been proved
to be NP-hard \citep{KariSidiLuo08}. Therefore, in the following,
we aim at computing efficiently (d-)stationary solutions of (\ref{eq:MMF}), the definition of which is analogous to the previous case (cf. \eqref{eq:d-stationarity} for Problem \ref{eq:max-min-IBC}). 

\noindent \textbf{Equivalent reformulation:} We start rewriting the
nonconvex and nonsmooth problem (\ref{eq:MMF}) in an equivalent \emph{smooth}
(still nonconvex) form: introducing the slack variables $t\geq0,$
and $\boldsymbol{{\beta}}\triangleq((\beta_{i_{k}})_{i_{k}\in\mathcal{G}_{k}})_{k\in\mathcal{K}_{\text{{BS}}}}$,
we have\vspace{-0.4cm}

\begin{equation}
\!\!\!\!\!\!\!\!\!\!\!\begin{array}{cl}
\underset{\begin{array}{c}
t\geq0,\,\boldsymbol{{\beta}},\,\mathbf{w}\end{array}}{\textrm{max}} & \!\!\!\!\!\!t\\
\textrm{s. t.} & \!\!\!\!\!\!\mbox{(a)}:\underset{\triangleq g_{i_{k}}(t,\beta_{i_{k}},\mathbf{w}_{k})}{\underbrace{t\cdot\beta_{i_{k}}-\mathbf{w}_{k}^{H}\mathbf{H}_{i_{k}k}\mathbf{w}_{k}}}\leq0,\\
 & \!\!\!\hspace{3cm}\forall i_{k}\in\mathcal{G}_{k},\,\forall k\in\mathcal{K}_{\text{{BS}}},\medskip\\
 & \!\!\!\!\!\!\mbox{(b)}:{\displaystyle \sum_{\ell\neq k}{\mathbf{w}_{\ell}^{H}\mathbf{H}_{i_{k}\ell}\mathbf{w}_{\ell}}+\sigma_{i_{k}}^{2}}\!\leq\beta_{i_{k}},\\
 & \!\!\!\hspace{3cm}\,\forall i_{k}\in\mathcal{G}_{k},\,\forall k\in\mathcal{K}_{\text{{BS}}},\medskip\\
 & \!\!\!\!\!\!\mbox{(c)}:\beta_{i_{k}}\leq\beta_{i_{k}}^{\max},\qquad\forall i_{k}\in\mathcal{G}_{k},\,\forall k\in\mathcal{K}_{\text{{BS}}},\medskip\\
 & \!\!\!\!\!\!\mbox{(d)}:\|\mathbf{w}_{k}\|_{2}^{2}\leq P_{k},\quad\,\,\,\,\forall k\in\mathcal{K}_{\text{{BS}}},\vspace{-.1cm}
\end{array}\label{eq:MMF_2}
\end{equation}
where condition (c) is meant to bound each $\beta_{i_{k}}$ by $\beta_{i_{k}}^{\max}\triangleq\underset{\ell\in\mathcal{K}_{\text{{BS}}}}{\max}{\;\lambda_{\max}\left(\mathbf{H}_{i_{k}\ell}\right)}\cdot\sum_{k=1}^{K}{P_{k}}+\sigma_{i_{k}}^{2}$,
so that the feasible set of (\ref{eq:MMF_2}), which we will denote by
$\mathcal{Z}$, is compact (a condition that is desirable to strengthen
the convergence properties of our algorithms, see \citep[Th. 2]{ScuFacLamPartI}).
Problems (\ref{eq:MMF}) and (\ref{eq:MMF_2}) are equivalent in the
following sense. \vspace{-0.1cm}
 
\begin{prop}
\label{prop: eq-prob-MMF}Given (\ref{eq:MMF}) and (\ref{eq:MMF_2}),
the following hold: 
\begin{description}
\item [{(i)}] Every stationary solution $\left(t^{\star},\boldsymbol{{\beta}}^{\star},\mathbf{w}^{\star}\right)$
of (\ref{eq:MMF_2}) satisfies the MFCQ;
\item [{(ii)}] $\mathbf{w}^{\star}$ is a d-stationary solution of (\ref{eq:MMF})
if and only if there exist $t^{\star}$ and $\boldsymbol{{\beta}}^{\star}$
such that $\left(t^{\star},\boldsymbol{{\beta}}^{\star},\mathbf{w}^{\star}\right)$
is a stationary solution of (\ref{eq:MMF_2}).\end{description}
\end{prop}
\begin{IEEEproof}
See supporting material. 
\end{IEEEproof}
Therefore, we can focus w.l.o.g. on (\ref{eq:MMF_2}).
We remark that the equivalence stated in Proposition \ref{prop: eq-prob-MMF}
is a new result in the literature. 

\subsection{Related works\label{sub:MFB_Related-works}}

The multicast beamforming problem has been widely studied in the literature,
under different channel models and settings (single-group vs. multi-group
and single-cell vs. multi-cell). While the general formulation is
nonconvex, special instances exhibit ad-hoc structures that allow
them to be solved efficiently, leveraging equivalent (quasi-)convex
reformulations; see, e.g., \citep{GopaSidi15,KariSidiLuo07,DartAsch13}.
In the case of general channel vectors, however, the (single-cell)
MMF beamforming problem {[}and thus also Problem (\ref{eq:MMF_2}){]}
was proved to be NP-hard \citep{KariSidiLuo08}. This has motivated
a lot of interest to pursuit approximate solutions that approach optimal
performance at moderate complexity. SemiDefinite Relaxations (SDR)
followed by Gaussian randomization (SDR-G) have been extensively studied
in the literature to obtain good suboptimal solutions \citep{KariSidiLuo08,ChriChatOtte14,HsuWangSuLin14,XianTaoWang13},
with theoretical bound guarantees \citep{Luo_So_et_al_alSPMag10,Chang_Luo_ChiTSP08}.
For a large number of antennas or users, however, the quality of the
approximation obtained by SDR-G methods deteriorates considerably.
In fact, SDR-based approaches return feasible points that in general
may not be even stationary for the original nonconvex problem. Moreover,
in a multi-cell scenario, \emph{SDR-G is not suitable for a} \emph{distributed}
implementation across the cells.

Two schemes based on heuristic convex approximations have been recently
proposed in \citep{SchaPesa12} and \citep{ChriChatOtte15} (the latter
based on earlier work \citep{MehaHuanGopaKonaSidi15}) for the \emph{single-cell}
multiple-group MMF beamforming problem. While extensive experiments
show that these schemes achieve better solutions than SDR-G approaches,
their theoretical convergence and guarantees remain an open question.
Finally, we are not aware of any \emph{distributed} scheme with provable
convergence for the \emph{multi-cell} MMF beamforming problem. 

Leveraging our NOVA framework, we propose next a novel centralized
algorithm and the first \emph{distributed} algorithm, both converging
to d-stationary solutions of Problem (\ref{eq:MMF}). Numerical results
(cf.$\,$Sec.$\,$\ref{sec:Simulations_MMFB}) show that  our
schemes reach better solutions than SDR-G approaches with high probability,
while having comparable computational complexity.\vspace{-0.3cm}

\subsection{Centralized solution method\label{sec:Centralized-MMF}\vspace{-0.1cm}}

Problem (\ref{eq:MMF_2}) is nonconvex due to the nonconvex constraint
functions $g_{i_{k}}(t,\beta_{i_{k}},\mathbf{w}_{k})$. Several valid
convexifications of $g_{i_{k}}$ are possible; two examples are given
next. 

\noindent \emph{Example \#1}: Note that $g_{i_{k}}$ is the sum of
a bilinear function and a concave one, namely: $g_{i_{k}}(t,\beta_{i_{k}},\mathbf{w}_{k})=g_{i_{k},1}(t,\beta_{i_{k}})+g_{i_{k},2}(\mathbf{w}_{k})$,
with 
\begin{equation}
g_{i_{k},1}(t,\beta_{i_{k}})\triangleq{\displaystyle t\cdot\beta_{i_{k}}},\text{ and }g_{i_{k},2}(\mathbf{w}_{k})\triangleq-\mathbf{w}_{k}^{H}\mathbf{H}_{i_{k}k}\mathbf{w}_{k}.\label{eq:g_constraints-1}
\end{equation}
A valid surrogate $\tilde{g}_{i_{k}}$ can be then obtained as follows:
i) linearize $g_{i_{k},2}(\mathbf{w}_{k})$ around $\mathbf{w}_{k}^{\nu}$,
that is,\vspace{-0.2cm}
 
\begin{equation}
\!\!\!\!\begin{array}{l}
\tilde{{g}}_{i_{k},2}\left(\mathbf{w}_{k};\mathbf{w}_{k}^{\nu}\right)\triangleq-\left(\mathbf{w}_{k}^{\nu}\right)^{H}\mathbf{H}_{i_{k}k}\mathbf{w}_{k}^{\nu}\\
\qquad\qquad\qquad\quad\,-\left\langle \nabla_{\mathbf{w}_{k}^{\ast}}g_{i_{k},2}\left(\mathbf{w}_{k}^{\nu}\right),\,\mathbf{w}_{k}-\mathbf{w}_{k}^{\nu}\right\rangle \geq g_{i_{k},2}(\mathbf{w}_{k}),
\end{array}\label{eq:linearization}
\end{equation}
with $\nabla_{\mathbf{w}_{k}^{\ast}}g_{i_{k},2}\left(\mathbf{w}_{k}^{\nu}\right)=\mathbf{H}_{i_{k}k}\mathbf{w}_{k}^{\nu}$
and $\left\langle \mathbf{a},\,\mathbf{b}\right\rangle \triangleq2\,\text{{Re}}\{\mathbf{a}^{H}\mathbf{b}\}$;
and ii) upper bound $g_{i_{k},1}(t,\beta_{i_{k}})$ around $(t^{\nu},\beta_{i_{k}}^{\nu})\neq(0,0)$
as 
\begin{equation}
\tilde{{g}}_{i_{k},1}(t,\beta_{i_{k}};t^{\nu},\beta_{i_{k}}^{\nu})\triangleq\frac{1}{2}\left(\frac{\beta_{i_{k}}^{\nu}}{t^{\nu}}\,t^{2}+\frac{t^{\nu}}{\beta_{i_{k}}^{\nu}}\,\beta_{i_{k}}^{2}\right)\geq g_{i_{k},1}(t,\beta_{i_{k}}).\label{eq:cvxify-bilinear-1}
\end{equation}
Overall, this results in the following surrogate function which satisfies
\citep[Assumptions 2-4]{ScuFacLamPartI}: 
\begin{equation}
\begin{array}{l}
\tilde{{g}}_{i_{k}}(t,\beta_{i_{k}},\mathbf{w}_{k};t^{\nu},\beta_{i_{k}}^{\nu},\mathbf{w}_{k}^{\nu})\smallskip\\
\triangleq\tilde{{g}}_{i_{k},1}(t,\beta_{i_{k}};t^{\nu},\beta_{i_{k}}^{\nu})+\tilde{{g}}_{i_{k},2}\left(\mathbf{w}_{k};\mathbf{w}_{k}^{\nu}\right).
\end{array}\label{eq:g_tilde-1}
\end{equation}
\noindent \emph{Example \#2}: Another valid approximation can be
readily obtained using a different bound for the bilinear term $g_{i_{k},1}(t,\beta_{i_{k}})$
in (\ref{eq:g_constraints-1}). Rewriting $g_{i_{k},1}(t,\beta_{i_{k}})$
as the difference of two convex functions, $g_{i_{k},1}(t,\beta_{i_{k}})=\frac{1}{2}((t+\beta_{i_{k}})^{2}-(t^{2}+\beta_{i_{k}}^{2})),$
the desired convex upper bound of $g_{i_{k},1}(t,\beta_{i_{k}})$
can be obtained by linearizing the concave part of $g_{i_{k},1}(t,\beta_{i_{k}})$
around $(t^{\nu},\beta_{i_{k}}^{\nu})$ while retaining the convex
part, which leads to 
\begin{equation}
\!\!\!\!\!\begin{array}{l}
\widehat{{g}}_{i_{k},1}(t,\beta_{i_{k}};t^{\nu},\beta_{i_{k}}^{\nu})\triangleq{\displaystyle \frac{1}{2}}\left(\left(t+\beta_{i_{k}}\right)^{2}-\left(t^{\nu}\right)^{2}-\left(\beta_{i_{k}}^{\nu}\right)^{2}\right)\smallskip\\
\qquad\qquad\qquad\qquad\,\,-\left(t^{\nu}\left(t-t^{\nu}\right)+\beta_{i_{k}}^{\nu}\left(\beta_{i_{k}}-\beta_{i_{k}}^{\nu}\right)\right).
\end{array}\label{eq:cvxify-bilinear-new-1}
\end{equation}
The resulting valid surrogate function is then 
\begin{equation}
\begin{array}{l}
\tilde{{g}}_{i_{k}}(t,\beta_{i_{k}},\mathbf{w}_{k};t^{\nu},\beta_{i_{k}}^{\nu},\mathbf{w}_{k}^{\nu})\\
\qquad\triangleq\widehat{{g}}_{i_{k},1}(t,\beta_{i_{k}};t^{\nu},\beta_{i_{k}}^{\nu})+\tilde{{g}}_{i_{k},2}\left(\mathbf{w}_{k};\mathbf{w}_{k}^{\nu}\right).
\end{array}\label{eq:g_hat-1}
\end{equation}

The strongly convex inner approximation of (\ref{eq:MMF_2}) {[}cf.$\,$(\ref{eq:convexified-subproblem}){]}
then reads: given a feasible $\mathbf{z}^{\nu}\triangleq\left(t^{\nu},\boldsymbol{{\beta}}^{\nu},\mathbf{w}^{\nu}\right)$,
\vspace{-0.4cm}

\begin{equation}
\!\!\!\!\!\!\!\!\!\begin{array}{cl}
\underset{\begin{array}{c}
t\geq0,\,\boldsymbol{{\beta}},\,\mathbf{w}\end{array}}{\textnormal{max}} & \!\!\!\!\!t-{\displaystyle \frac{\tau_{t}}{2}}\left(t-t^{\nu}\right)^{2}-\tau_{\mathbf{w}}\left\Vert \mathbf{w}-\mathbf{w}^{\nu}\right\Vert _{2}^{2}-\frac{\tau_{\boldsymbol{\beta}}}{2}\left\Vert \boldsymbol{\beta}-\boldsymbol{\beta}^{\nu}\right\Vert _{2}^{2}\\
\mbox{s. t.} & \!\!\!\!\!\!\tilde{{g}}_{i_{k}}(t,\beta_{i_{k}},\mathbf{w}_{k};t^{\nu},\beta_{i_{k}}^{\nu},\mathbf{w}_{k}^{\nu})\leq0,\,\forall i_{k}\in\mathcal{G}_{k},\,\forall k\in\mathcal{K}_{\text{{BS}}},\medskip\\
 & \!\!\!\!\!\!\mbox{(b), (c), and (d) of }(\ref{eq:MMF_2});
\end{array}\label{eq:SCA-cvx-frame-1}
\end{equation}
where $\tilde{{g}}_{i_{k}}$ is the surrogate defined either in (\ref{eq:g_tilde-1})
or in (\ref{eq:g_hat-1}). In the objective function of (\ref{eq:SCA-cvx-frame-1})
we added a proximal regularization to make it strongly convex; therefore,
problem (\ref{eq:SCA-cvx-frame-1}) has a unique solution, which we
denote by $\hat{\mathbf{z}}({\mathbf{z}}^{\nu})$.

Using \eqref{eq:SCA-cvx-frame-1}, the NOVA algorithm based on (\ref{eq:main-iterate})
is described in Algorithm \ref{algoC-1}, whose convergence is established
in Theorem \ref{th:conver}. Note that $t^{\nu}>0$, for all $\nu\geq1$
(provided that $t^{0}>0$), which guarantees that, if (\ref{eq:g_tilde-1})
is used in (\ref{eq:SCA-cvx-frame-1}), then $\tilde{{g}}_{i}$ is
always well defined. Also, the algorithm will never converge to a
degenerate stationary solution of (\ref{eq:MMF}) ($U=0$, i.e.,
$t^{\infty}=0$), at which some users will not receive any signal.
\vspace{-0.1cm}

\begin{algorithm}[H]
\textbf{Data}: $\mathbf{z}^{0}\triangleq(t^{0},\boldsymbol{{\beta}}^{0},\mathbf{w}^{0})\,\in\,\mathcal{Z}$,
with $t^{0}>0$, and $(\tau_{t},\tau_{\mathbf{w}},\tau_{\boldsymbol{\beta}})>\mathbf{0}$.
Set $\nu=0$.

$(\texttt{S.1})$ If $\mathbf{z}^{\nu}$ is a stationary solution of (\ref{eq:MMF}): \texttt{STOP}.

$(\texttt{S.2})$ Compute $\hat{\mathbf{z}}({\mathbf{z}}^{\nu})$.

$(\texttt{S.3})$ Set $\mathbf{z}^{\nu+1}=\mathbf{z}^{\nu}+\gamma^{\nu}\left(\hat{\mathbf{z}}({\mathbf{z}}^{\nu})-\mathbf{z}^{\nu}\right)$ for some $\gamma^{\nu}\in(0,1]$.

$(\texttt{S.4})$ $\nu\leftarrow\nu+1$ and go to step (\texttt{S.1}).

\caption{\hspace{-2.5pt}\textbf{:} \label{algoC-1}NOVA Algorithm for Problem
(\ref{eq:MMF})}
\end{algorithm}

\vspace{-0.4cm}

\begin{thm}
\label{th:conver} Let $\{\mathbf{z}^{\nu}=(t^{\nu},\boldsymbol{{\beta}}^{\nu},\mathbf{w}^{\nu})\}$ be the sequence generated by Algorithm \ref{algoC-1}. Choose any $\tau_{t}$, $\tau_{\mathbf{w}}$, $\tau_{\boldsymbol{\beta}} > 0$, and the step-size sequence $\{\gamma^{\nu}\}$ such that
$\gamma^{\nu}\in(0,1]$, $\gamma^{\nu}\to0$, and $\sum_{\nu}\gamma^{\nu}=+\infty$. Then, $\{\mathbf{z}^{\nu}\}$ is bounded (with $t^{\nu}>0$, for all $\nu\geq1$), and every of its limit points $(t^{\infty},\boldsymbol{{\beta}}^{\infty},\mathbf{w}^{\infty})$ is a stationary solution of  Problem (\ref{eq:MMF_2}), such that $t^{\infty}>0$.
Therefore, $\mathbf{w}^{\infty}$ is d-stationary for Problem (\ref{eq:MMF}). Furthermore, if the algorithm does not stop after a finite number of steps, none of the $\mathbf{w}^{\infty}$ above is a local minimum of U. \vspace{-.2cm}\end{thm} 
\begin{IEEEproof}
See Appendix \ref{sub:Proof_convergence_centralized_MMF}.\vspace{-0.7cm}

\end{IEEEproof}

\subsection{Distributed implementation \label{sub:Distributed-implementation_MMF}}

\vspace{-0.1cm}

Algorithm \ref{algoC-1} is centralized, because subproblems (\ref{eq:SCA-cvx-frame-1})
do not decouple across the BSs. In this section, we develop a distributed
solution method for (\ref{eq:SCA-cvx-frame-1}), using the surrogate
in Example$\:\#1$ {[}cf.$\,$(\ref{eq:g_tilde-1}){]}. We exploit
the additive separability in the BSs' variables of the objective function
and constraints in (\ref{eq:SCA-cvx-frame-1}), as outlined next. 

Denoting by $\boldsymbol{\lambda}\triangleq(\boldsymbol{{\lambda}}_{k}\triangleq(\lambda_{i_{k}})_{i_{k}\in\mathcal{G}_{k}})_{k\in\mathcal{K}_{\text{{BS}}}}$
and $\boldsymbol{\eta}\triangleq(\boldsymbol{\eta}_{k}\triangleq(\eta_{i_{k}})_{i_{k}\in\mathcal{G}_{k}})_{k\in\mathcal{K}_{{BS}}}$ the multipliers associated to the constraints $\tilde{{g}}_{i_{k}}\leq0$
and (b) in (\ref{eq:SCA-cvx-frame-1}), respectively, and introducing
$\boldsymbol{{\sigma}}^{2}\triangleq((\sigma_{i_{k}}^{2})_{i_{k}\in\mathcal{G}_{k}})_{k\in\mathcal{K}_{{BS}}}$,
$\boldsymbol{{\beta}}_{k}\triangleq(\beta_{i_{k}})_{i_{k}\in\mathcal{G}_{k}}$,
and $\boldsymbol{\beta}^{\max}\triangleq(\boldsymbol{\beta}_{k}^{\max}\triangleq(\beta_{i_{k}}^{\max})_{i_{k}\in\mathcal{G}_{k}})_{k\in\mathcal{K}_{\text{{BS}}}},$ the (partial) Lagrangian of (\ref{eq:SCA-cvx-frame-1}) can be shown
to have the following structure: $\mathcal{L}\left(t,\,\boldsymbol{\beta},\mathbf{w},\boldsymbol{\lambda},\boldsymbol{\eta}; \mathbf{z}^{\nu}\right)=\mathcal{L}_t\left(t,\boldsymbol{\lambda},\boldsymbol{\eta}; t^\nu,\boldsymbol{{\beta}}^{\nu}\right)+\sum_{k=1}^{K}\mathcal{L}_{\mathbf{w}_{k}}\left(\mathbf{w}_{k},\boldsymbol{\lambda},\boldsymbol{\eta}; \mathbf{w}^{\nu} \right)+\sum_{k=1}^{K}\mathcal{L}_{\boldsymbol{\beta}_{k}}\left(\boldsymbol{\beta}_{k},\boldsymbol{\lambda},\boldsymbol{\eta}; t^\nu,\boldsymbol{{\beta}}^{\nu}\right)$,
where \vspace{-0.1cm}
\[
\begin{array}{l}
\!\!\!\!\mathcal{L}_t\left(t,\boldsymbol{\lambda},\boldsymbol{\eta}; t^\nu,\boldsymbol{{\beta}}^{\nu}\right)\triangleq-t+\frac{\tau_{t}}{2}\left(t-t^{\nu}\right)^{2}+\frac{\boldsymbol{\lambda}^{T}\boldsymbol{\beta}^{\nu}}{2t^{\nu}}\,t^{2}+\boldsymbol{{\eta}}^{T}\boldsymbol{{\sigma}}^{2};\medskip\\
\!\!\!\!\mathcal{L}_{\mathbf{w}_{k}}\left(\mathbf{w}_{k},\boldsymbol{\lambda},\boldsymbol{\eta}; \mathbf{w}^{\nu} \right)\triangleq\tau_{\mathbf{w}}\|\mathbf{w}_{k}-\mathbf{w}_{k}^{\nu}\|_{2}^{2}\smallskip\\
\qquad\qquad\qquad\qquad\quad-\sum_{i_{k}\in\mathcal{G}_{k}}\lambda_{i_{k}}\mathbf{w}_{k}^{\nu H}\mathbf{H}_{i_{k}k}\mathbf{w}_{k}^{\nu}\smallskip\\
\qquad\qquad\qquad\qquad\quad-\sum_{i_{k}\in\mathcal{G}_{k}}\lambda_{i_{k}}\left\langle \mathbf{H}_{i_{k}k}\mathbf{w}_{k}^{\nu},\mathbf{w}_{k}-\mathbf{w}_{k}^{\nu}\right\rangle \smallskip\\
\qquad\qquad\qquad\qquad\quad+\sum_{\ell\neq k}\sum_{i_{\ell}\in\mathcal{G}_{\ell}}\eta_{i_{\ell}}\mathbf{w}_{k}^{H}\mathbf{H}_{i_{\ell}k}\mathbf{w}_{k};\medskip\\
\!\!\!\!\mathcal{L}_{\boldsymbol{\beta}_{k}}\left(\boldsymbol{\beta}_{k},\boldsymbol{\lambda},\boldsymbol{\eta}; t^\nu,\boldsymbol{{\beta}}^{\nu}\right)\triangleq\frac{\tau_{\boldsymbol{\beta}}}{2}\|\boldsymbol{\beta}_{k}-\boldsymbol{\beta}_{k}^{\nu}\|_{2}^{2}-\boldsymbol{\eta}_{k}^{T}\boldsymbol{\beta}_{k}\smallskip\\
\qquad\qquad\qquad\qquad\qquad+{\sum_{i_{k}\in\mathcal{G}_{k}}}\frac{\lambda_{i_{k}}\cdot\,t^{\nu}}{2\beta_{i_{k}}^{\nu}}\,\beta_{i_{k}}^{2}.\vspace{-0.1cm}
\end{array}
\]
The above structure of the Lagrangian leads naturally to the following
decomposition of the dual function: 
\begin{equation}
\!\!\!\!\!\begin{array}{l}
D\left(\boldsymbol{\lambda},\boldsymbol{\eta}; \mathbf{z}^{\nu}\right)=\underset{t\geq0}{\textrm{min}}\,\mathcal{L}_t\left(t,\boldsymbol{\lambda},\boldsymbol{\eta}; t^\nu,\boldsymbol{{\beta}}^{\nu}\right)\\
\qquad\quad\qquad\quad+{\displaystyle {\sum_{k\in\mathcal{K}_{\text{{BS}}}}}}\,\underset{\|\mathbf{w}_{k}\|_{2}^{2}\leq P_{k}}{\textrm{min}}\mathcal{L}_{\mathbf{w}_{k}}\left(\mathbf{w}_{k},\boldsymbol{\lambda},\boldsymbol{\eta}; \mathbf{w}^{\nu} \right)\\
\qquad\quad\qquad\quad+{\displaystyle {\sum_{k\in\mathcal{K}_{\text{{BS}}}}}}\underset{\mathbf{0}\leq\boldsymbol{\beta}_{k}\leq\boldsymbol{\beta}_{k}^{\max}}{\textrm{min}}\mathcal{L}_{\boldsymbol{\beta}_{k}}\left(\boldsymbol{\beta}_{k},\boldsymbol{\lambda},\boldsymbol{\eta}; t^\nu,\boldsymbol{{\beta}}^{\nu}\right).
\end{array}\vspace{-.1cm}\label{eq:dual_MMF}
\end{equation}
The unique solutions of the above optimization problems can be computed
in closed form (the proof is omitted because of space limitations):\vspace{-0.4cm}

\begin{equation}
\!\!\!\!\!\begin{array}{l}
\hat{t}\left(\boldsymbol{\lambda}; t^\nu,\boldsymbol{{\beta}}^{\nu}\right)\triangleq\underset{t\geq0}{\textrm{argmin}}\,\mathcal{L}_t\left(t,\boldsymbol{\lambda},\boldsymbol{\eta}; t^\nu,\boldsymbol{{\beta}}^{\nu}\right)=\left[\dfrac{{1+\tau_{t}\cdot t^{\nu}}}{\tau_{t}+\boldsymbol{\lambda}^{T}\boldsymbol{\beta}^{\nu}/t^{\nu}}\right]_{+}\smallskip\\
\hat{\boldsymbol{\beta}}_{k}\left(\boldsymbol{\lambda},\boldsymbol{\eta}; t^\nu,\boldsymbol{{\beta}}^{\nu}\right)\triangleq\underset{\mathbf{0}\leq\boldsymbol{\beta}_{k}\leq\boldsymbol{\beta}_{k}^{\max}}{\textrm{argmin}}\,\mathcal{L}_{\boldsymbol{\beta}_{k}}\left(\boldsymbol{\beta}_{k},\boldsymbol{\lambda},\boldsymbol{\eta}; t^\nu,\boldsymbol{{\beta}}^{\nu}\right)\smallskip\\
\qquad\quad\qquad\qquad=\left(\left[\dfrac{{\tau_{\boldsymbol{\beta}}\cdot\beta_{i_{k}}^{\nu}+\eta_{i_{k}}}}{\tau_{\boldsymbol{\beta}}+\lambda_{i_{k}}\cdot t^{\nu}/\beta_{i_{k}}^{\nu}}\right]_{0}^{\beta_{i_{k}}^{\max}}\right)_{i_{k}\in\mathcal{G}_{k}}\smallskip\\
\hat{{\mathbf{w}}}_{k}\left(\boldsymbol{\lambda},\boldsymbol{\eta}; \mathbf{w}^{\nu}\right)\triangleq\underset{\|\mathbf{w}_{k}\|_{2}^{2}\leq P_{k}}{\textrm{argmin}}\,\mathcal{L}_{\mathbf{w}_{k}}\left(\mathbf{w}_{k},\boldsymbol{\lambda},\boldsymbol{\eta}; \mathbf{w}^{\nu} \right)\smallskip\\
\qquad\qquad\qquad=(\mathbf{\xi}_{k}^{\star}\mathbf{I}+\mathbf{A}_{k})^{-1}\mathbf{b}_{k}^{\nu},
\end{array}\label{eq:closed_form}
\end{equation}
where $[x]_{a}^{b}\triangleq\min(b,\max(a,x))$, $\mathbf{A}_{k}$
and $\mathbf{b}_{k}^{\nu}$ are defined as 
\[
\begin{array}{l}
\mathbf{A}_{k}\triangleq\tau_{\mathbf{w}}\mathbf{I}+\sum_{\ell\neq k}\sum_{i_{\ell}\in\mathcal{G}_{\ell}}\eta_{i_{\ell}}\mathbf{H}_{i_{\ell}k}\smallskip\\
\mathbf{b}_{k}^{\nu}\triangleq\left(\tau_{\mathbf{w}}\mathbf{I}+\sum_{i\in\mathcal{G}_{k}}\lambda_{i_{k}}\mathbf{H}_{i_{k}k}\right)\mathbf{w}_{k}^{\nu},
\end{array}
\]
and $\mathbf{\xi}_{k}^{\star}$, which is such that $0\leq\mathbf{\xi}_{k}^{\star}\perp\|\hat{{\mathbf{w}}}_{k}\left(\boldsymbol{\lambda},\boldsymbol{\eta},\mathbf{\xi}_{k}^{\star}\right)\|_{2}^{2}-P_{k}\leq0$,
can be efficiently computed as follows. Denoting by $\mathbf{U}_{k}\mathbf{D}_{k}\mathbf{U}_{k}^{H}$
the eigendecomposition of $\mathbf{A}_{k}$, we have $f_{k}(\xi_{k})\triangleq\|\hat{{\mathbf{w}}}_{k}^{\nu}\left(\boldsymbol{\lambda},\boldsymbol{\eta}; \mathbf{w}^{\nu}\right)\|_{2}^{2}-P_{k}=\sum_{j=1}^{N_{t}}\frac{[\mathbf{U}_{k}^{H}\mathbf{b}_{k}^{\nu}\mathbf{b}_{k}^{\nu H}\mathbf{U}_{k}]_{jj}}{(\xi_{k}+\left[\mathbf{D}_{k}\right]_{jj})^{2}}-P_{k}$.
Therefore, $\mathbf{\xi}_{k}^{\star}=0$ if $f_{k}(0)<0$; otherwise
$\mathbf{\xi}_{k}^{\star}$ is such that $f_{k}(\xi_{k}^{\star})=0$,
which can be computed using bisection on $[0,\sqrt{\textrm{tr}(\mathbf{U}_{k}^{H}\mathbf{b}_{k}^{\nu}\mathbf{b}_{k}^{\nu H}\mathbf{U}_{k})/P_{k}}-\underset{j}{\textrm{min}}[\mathbf{D}_{k}]_{jj}]$. 

Finally, note that the dual function $D\left(\boldsymbol{\lambda},\boldsymbol{\eta}; \mathbf{z}^{\nu}\right)$
is differentiable on $\mathbb{R}^I_+\times \mathbb{R}^I_+$ ,with gradient given by  \vspace{-0.2cm}

\begin{equation}
\!\!\!\!\begin{array}{l}
\nabla_{\lambda_{i_{k}}}D^{\nu}\left(\boldsymbol{\lambda},\boldsymbol{\eta};\mathbf{z}^{\nu}\right)=\tilde{g}_{i_{k}}\big(\hat{t}(\boldsymbol{\lambda}; t^\nu,\boldsymbol{{\beta}}^{\nu}),\hat{\boldsymbol{\beta}}(\boldsymbol{\lambda},\boldsymbol{\eta}; t^\nu,\boldsymbol{{\beta}}^{\nu}),\\
\qquad\qquad\qquad\qquad\qquad \hat{\mathbf{w}}(\boldsymbol{\lambda},\boldsymbol{\eta};\mathbf{{w}}^{\nu});\mathbf{z}^{\nu}\big),\\
\nabla_{\eta_{i_{k}}}D^{\nu}\left(\boldsymbol{\lambda},\boldsymbol{\eta};\mathbf{z}^{\nu}\right)=\sum_{\ell\neq k}\hat{\mathbf{w}}_{\ell}(\boldsymbol{\lambda},\boldsymbol{\eta};\mathbf{{w}}^{\nu})^{H}\mathbf{H}_{i_{k}\ell}\hat{\mathbf{w}}_{\ell}(\boldsymbol{\lambda},\boldsymbol{\eta};\mathbf{{w}}^{\nu})\\
\qquad\qquad\qquad\qquad+\sigma_{i_{k}}^{2}-\hat{\beta}_{i_{k}}\left(\boldsymbol{\lambda},\boldsymbol{\eta}; t^\nu,\boldsymbol{{\beta}}^{\nu}\right).\vspace{-0.2cm}
\end{array}\label{eq:gradient_dual_function}
\end{equation}
Using (\ref{eq:gradient_dual_function}), the dual problem $\max_{\boldsymbol{{\lambda},{\eta}}\geq\mathbf{0}}D\left(\boldsymbol{\lambda},\boldsymbol{\eta}; \mathbf{z}^{\nu}\right)$ can be solved in a distributed way with convergence guarantees
using, e.g., a gradient-based scheme with diminishing step-size; we omit
further details. 
Overall, the proposed algorithm consists in updating $\mathbf{z}^{\nu}$
via $\mathbf{z}^{\nu+1}=\mathbf{z}^{\nu}+\gamma^{\nu}\left(\hat{\mathbf{z}}({\mathbf{z}}^{\nu})-\mathbf{z}^{\nu}\right)$
{[}cf.$\,$(\ref{eq:main-iterate}){]} wherein $\hat{\mathbf{z}}({\mathbf{z}}^{\nu})$
is computed in a distributed way solving the dual problem (\ref{eq:dual_MMF}),
e.g., using a first or second order method. The algorithm is thus
a double-loop scheme. The inner loop deals with the update of the
multipliers $\left(\boldsymbol{\lambda},\boldsymbol{\eta}\right)$,
given $\mathbf{z}^{\nu}$;
let $(\boldsymbol{\lambda}^{\infty},\boldsymbol{\eta}^{\infty})$
be the limit point (within the desired accuracy) of the sequence $\left\{ (\boldsymbol{\lambda}^{n},\boldsymbol{\eta}^{n})\right\} $
generated by the algorithm solving the dual problem (\ref{eq:dual_MMF}).
In the outer loop the BSs update locally their $t$, $\boldsymbol{\beta}_{k}$s and $\mathbf{w}_{k}$s using the closed form solutions $\hat{\mathbf{z}}({\mathbf{z}}^{\nu}) = (\hat{t}(\boldsymbol{\lambda}^{\infty}; t^\nu,\boldsymbol{{\beta}}^{\nu}),\hat{\boldsymbol{\beta}}(\boldsymbol{\lambda}^{\infty},\boldsymbol{\eta}^{\infty}; t^\nu,\boldsymbol{{\beta}}^{\nu}),\hat{\mathbf{w}}(\boldsymbol{\lambda}^{\infty},\boldsymbol{\eta}^{\infty};\mathbf{{w}}^{\nu}))$ {[}cf.$\,$(\ref{eq:closed_form}){]}. The inner and outer updates
can be performed in a fairly distributed way among the cells. Indeed,
 to compute the closed form solutions $\hat{\mathbf{w}}_{k}(\boldsymbol{\lambda}^{n},\boldsymbol{\eta}^{n};\mathbf{{w}}^{\nu})$
and $\hat{\boldsymbol{\beta}_{k}}(\boldsymbol{\lambda}^{n},\boldsymbol{\eta}^{n};t^\nu,\boldsymbol{{\beta}}^{\nu})$,
the BSs need only information within their cell. The update of the
$t-$variable {[}using $\hat{t}(\boldsymbol{\lambda}^{n};t^\nu,\boldsymbol{{\beta}}^{\nu})${]}
and multipliers $\left(\boldsymbol{\lambda}^{n+1},\boldsymbol{\eta}^{n+1}\right)$
require some coordination among the BSs: it can be either carried
out by a BS header or locally by all the BSs if a consensus-like scheme
is employed to obtain locally the information required to compute
$\hat{t}(\boldsymbol{\lambda}^{n};t^\nu,\boldsymbol{{\beta}}^{\nu})$ and the gradients in (\ref{eq:gradient_dual_function}). 

\vspace{-0.2cm}

\subsection{Numerical Results \label{sec:Simulations_MMFB}}\vspace{-0.1cm}

\begin{figure}[t]
\vspace{-0.8cm}
\begin{minipage}[b]{1\linewidth}%
\centering \centerline{\includegraphics[width=6cm]{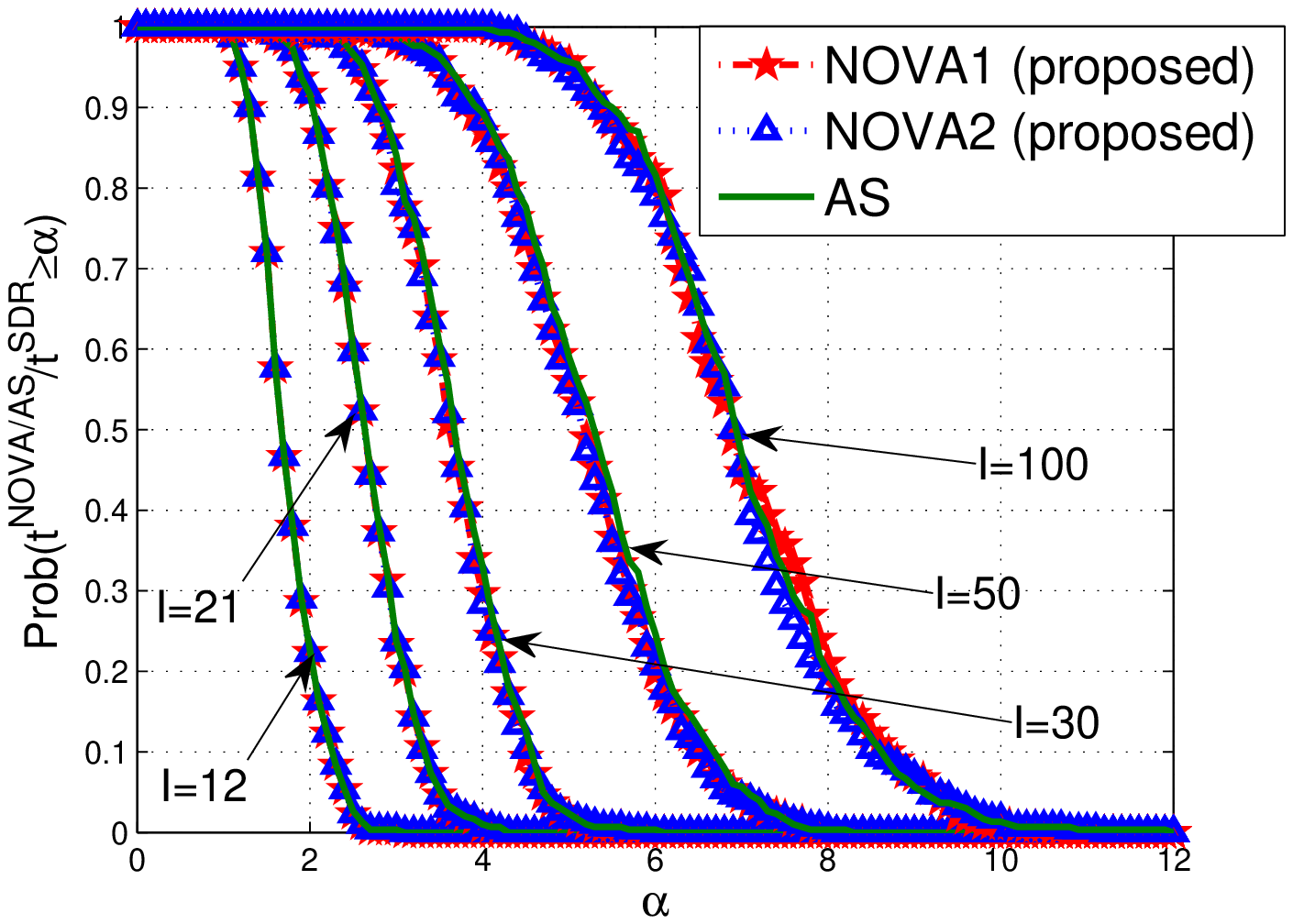}}\vspace{-0.2cm}
 (a)%
\end{minipage}

\begin{minipage}[b]{1\linewidth}%
\centering \centerline{\includegraphics[width=6cm]{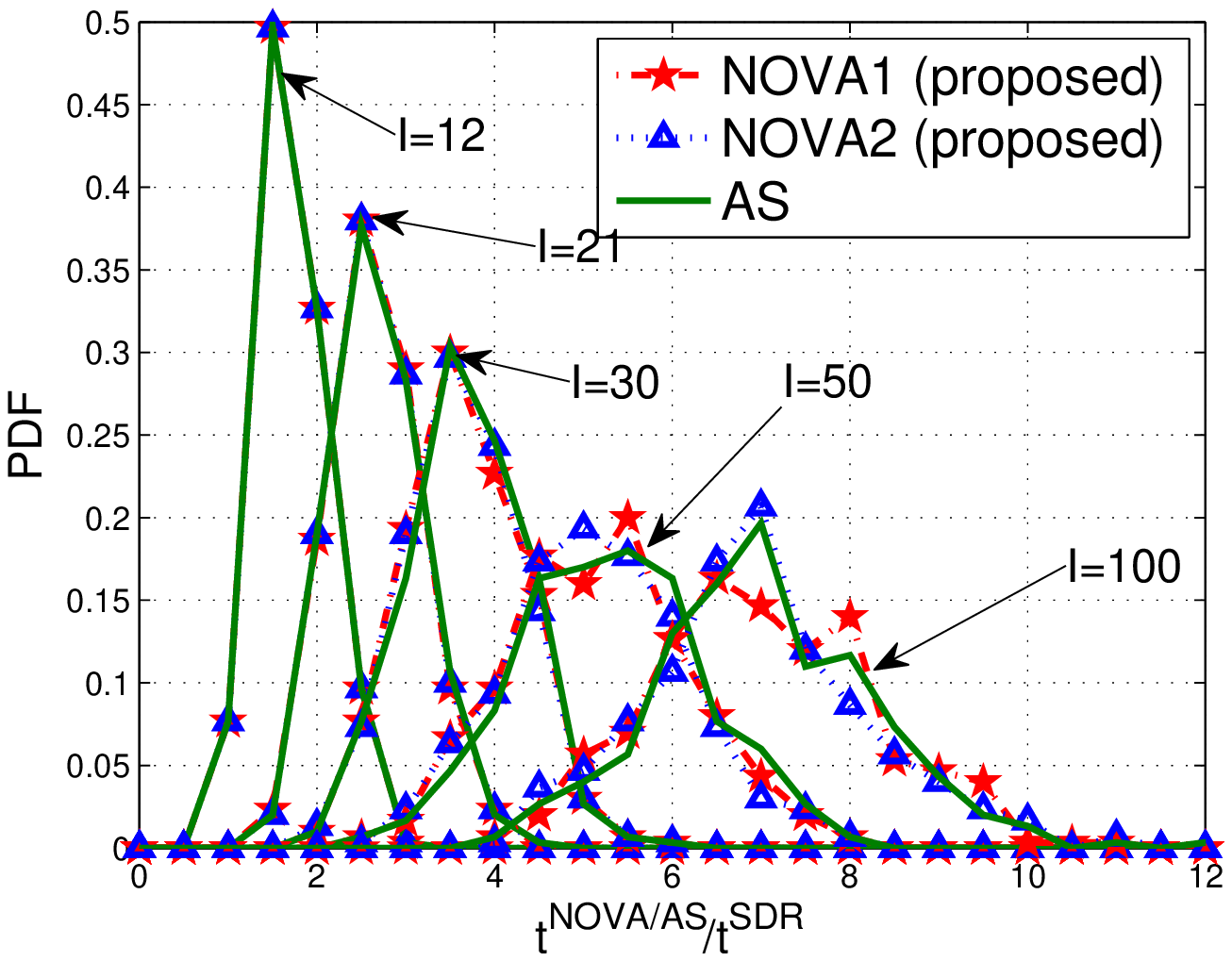}}
(b)\vspace{-0.2cm}
\end{minipage}

\caption{\label{fig:s-prob-CDFnPDF}(a): $\text{{Prob}}(t^{\text{NOVA}/\text{AS}}/t^{\text{SDR}}\geq\alpha)$
versus $\alpha$, for $I=12,\,21,\,30,\,50,\,100$; (b): Estimated
p.d.f. of $t^{\text{NOVA}}/t^{\text{SDR}}$ and $t^{\text{AS}}/t^{\text{SDR}}$.
}\vspace{-.4cm}
\end{figure}

In this section, we present some numerical results validating the
proposed approach and algorithmic framework.\emph{ \\\noindent Example\,$\#\,1:$\,Centralized
algorithm}. We compare our NOVA algorithm with the renowned SDR-G
scheme in \citep{KariSidiLuo08}. For the NOVA algorithm, we considered two instances,
corresponding to the two approximation strategies introduced in Sec.
\ref{sec:Centralized-MMF} (see Examples 1 and 2); we will term them
``NOVA1'' and ``NOVA2'', respectively. The setup of our experiment
is the following. We simulated a single BS system; the transmitter
is equipped with $N_{t}=8$ transmit antennas and serves $K=2$ multicast
groups, each with $I$ single-antenna users. Different numbers of
users per group are considered, namely: $I=12,\ 24,\ 30,\ 50,\ 100$.
NOVA algorithms are simulated using the step-size rule
$\gamma^{\nu}=\gamma^{\nu-1}\left(1-10^{-2}\gamma^{\nu-1}\right)$,
with $\gamma^{0}=1$; the proximal gain
is set to $\tau=1$e$-5$. The iterate is terminated when the absolute
value of the difference of the objective function in two consecutive
iterations is less than $1$e$-3$. For the SDR-G in \citep{KariSidiLuo08},
300 Gaussian samples are taken during the randomization phase, where
the principal component of the relaxed SDP solution is also included
as a candidate; the best value of the resulting objective function
is denoted by $t^{\text{SDR}}$. To be fair, for our
schemes, we considered 300 random feasible starting points and kept
the best value of the objective function at convergence, denoted by
$t^{\text{NOVA}}$. We then compared the performance of the two algorithms
in terms of the ratio $t^{\text{NOVA}}/t^{\text{SDR}}$. As benchmark,
we also report the results achieved using the standard nonlinear programming
solver in Matlab, specifically the active-set algorithm in 'fmincon';
we refer to it as ``AS'' algorithm and denote by $t^{\text{AS}}$
the best value of the objective function at convergence (obtained
over the same random initializations of the NOVA schemes). In Fig.
\ref{fig:s-prob-CDFnPDF} a) we plot the probability that $t^{\text{NOVA/AS}}/t^{\text{SDR}}\geq\alpha$
versus $\alpha$, for different values of $I$ (number of users per
group), and $\text{SNR}\triangleq P/\sigma^{2}=3\text{dB}$; this
probability is estimated taking 300 independent channel realizations.
The figures show a significant gain of the proposed NOVA methods.
For instance, when $I=30$, the minimum achieved SINR of all NOVA
methods is about \emph{at least} three times and \textit{at most}
5 times the one achieved by SDR-G, with probability one. It seems
that the gap tends to grow in favor of the NOVA methods, as the number
of users increases. In Fig. \ref{fig:s-prob-CDFnPDF} b) we plot the
distribution of $t^{\text{NOVA}/\text{AS}}/t^{\text{SDR}}$. For instance,
when $I=30$, the minimum achieved SINR of all NOVA methods is on
average about four times the one achieved by SDR-G; the variance is
about 1.

\begin{figure}[t]
\vspace{-0.1cm}
\centering \centerline{\includegraphics[width=5cm]{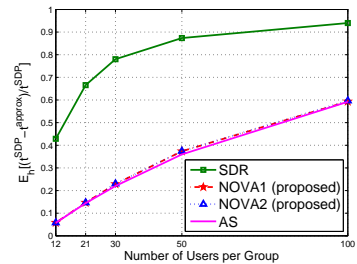}}\vspace{-0.1cm}
 \caption{\label{fig:s-prob-Dist} Average normalized distance of the achieved
minimum SINR from the SDP upper bound versus $I$.\vspace{-0.5cm}
}
\end{figure}

\begin{figure}[!h]
\vspace{-0.1cm}
\center \includegraphics[scale=0.31]{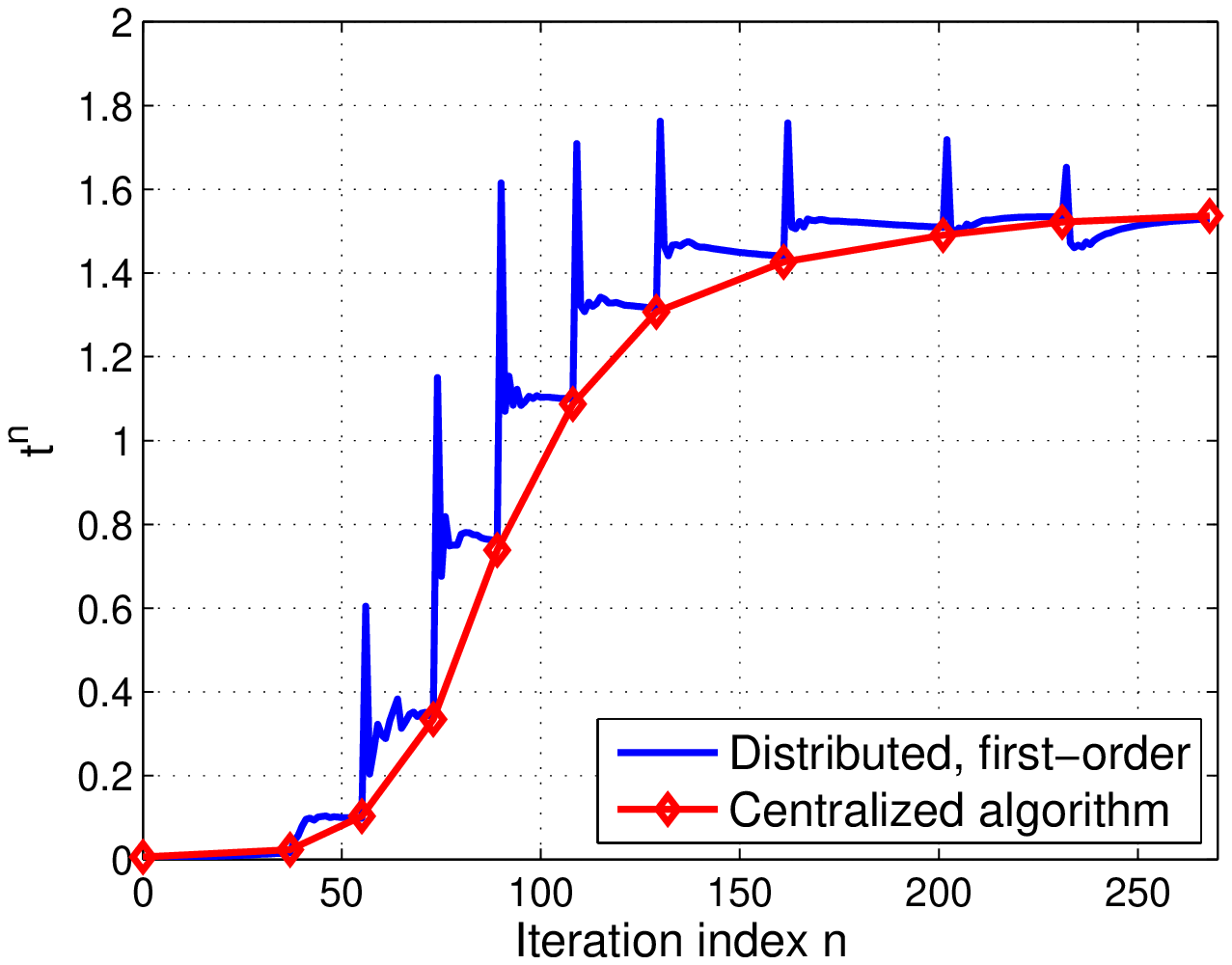}\,\includegraphics[scale=0.31]{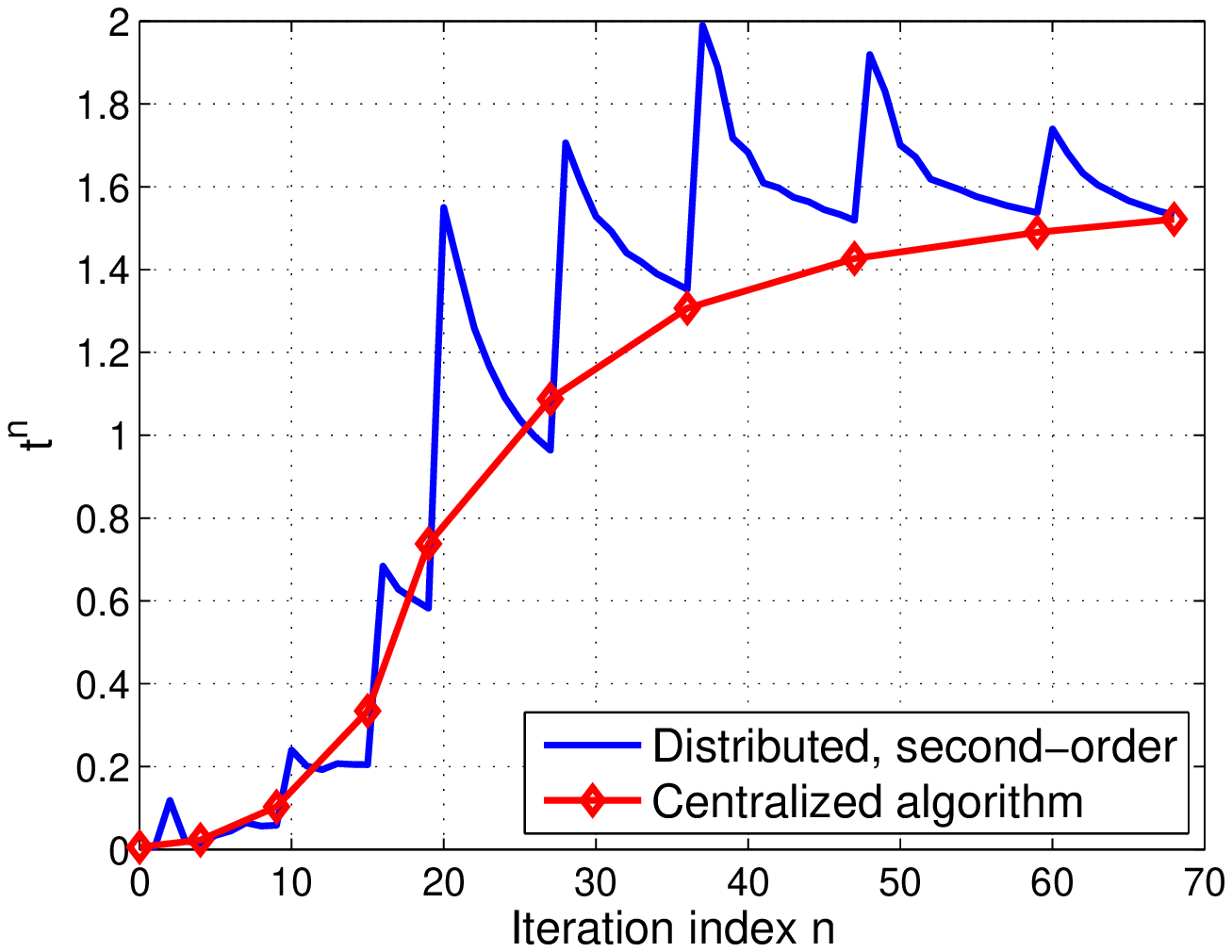}\caption{{\footnotesize{}{}Minimum rate vs. iterations. Left plot: Centralized
algorithm vs. Distributed, first order; Right plot: Centralized algorithm
vs. Distributed, second order.}}

\label{fig:tate_vs_iter-MMF} \vspace{-0.4cm}
 
\end{figure}

In Fig. \ref{fig:s-prob-Dist} we plot the average (normalized) distance
of the objective value achieved at convergence of the aforementioned
algorithms from the upper bound obtained solving the SDP relaxation
(denoted by $t^{\text{SDP}}$). More specifically, we plot the average
of $1-t^{\text{approx}}/t^{\text{SDP}}$ versus $I$ (the average
is taken over 300 independent channel realizations), where $t^{\text{approx}}=t^{\text{SDR}}$
for the SDR-G algorithm, $t^{\text{approx}}=t^{\text{NOVA}}$ for
our methods, and $t^{\text{approx}}=t^{\text{AS}}$ for the
AS algorithm in Matlab, with $t^{\text{SDR}}$, $t^{\text{NOVA}}$
and $t^{\text{AS}}$ defined as in\textcolor{blue}{{} }Fig.\textcolor{blue}{$\,$}\ref{fig:s-prob-CDFnPDF}.
The figure shows that the objective value reached by our methods is
much closer to the SDP bound than the one obtained by SDR-G. For instance,
when $I=30$, the solution of our NOVA methods are within 25\% the
upper bound.\emph{\\\noindent Example$\,\#\,2:\,$ Distributed
algorithms.} The previous example shows that the proposed schemes
compare favorably with the commercial off-the-shelf software and outperform
SDR-based scheme (in terms of quality of the solution and convergence
speed). However, differently from off-the-shelf softwares, our schemes
allow for a distributed implementation in a multi-cell scenario with
convergence guarantees. We test the distributed implementation
of the algorithms, as described in Sec. \ref{sub:Distributed-implementation_MMF},
and compare it with the centralized one. More specifically, we simulate
i) Algorithm \ref{algoC-1} based on the solution $\hat{{\mathbf{z}}}(\mathbf{z}^{\nu})$
(termed \emph{Centralized algorithm}); ii) the same algorithm as in
i) but with $\hat{{\mathbf{z}}}(\mathbf{z}^{\nu})$ computed in a
distributed way using the heavy ball method (termed \emph{Distributed,
first-order}); and iii) the same algorithm as in i) but with $\hat{{\mathbf{z}}}(\mathbf{z}^{\nu})$
computed by solving the dual problem $\max_{\boldsymbol{{\lambda},{\eta}}\geq\mathbf{0}}D^{\nu}\left(\boldsymbol{\lambda},\boldsymbol{\eta};\mathbf{z}^{\nu}\right)$\textcolor{blue}{{}
}using the damped Netwon method (termed \emph{Distributed, second-order}).
The simulated scenario of our experiment is the following. We simulated
a system comprising $K=4$ BSs, each equipped with $N_{t}=4$ transmit
antennas and serving $G=1$ multicast group. Each group has $I=3$
single-antenna users. In both loops (inner and outer), the iterate
is terminated when the absolute value of the difference of the objective
function in two consecutive iterations is less than $1$e$-2$. Fig.
\ref{fig:tate_vs_iter-MMF} shows the evolution of the objective function
$t$ of (\ref{eq:MMF_2}) versus the iterations. For the distributed algorithms, the number of iterations
counts both the inner and outer iterations. Note that all the algorithms
converge to the same stationary point of Problem (\ref{eq:MMF}),
and they are quite fast. As expected, exploiting second order information
accelerates the practical convergence but with the cost of extra signaling
among the BSs.

\section{Concluding Remarks \label{sec:Conclusions}}

In this two-part paper we introduced and analyzed a new algorithmic
framework for the distributed minimization of nonconvex functions,
subject to nonconvex constraints. Part I \citep{Larsson-Jorswieck-Lindblom-Mochaourab_SPMag_sub09}
developed the general framework and studied its convergence properties.
In this Part II, we customized our general results to two challenging
and timely problems in communications, namely: 1) the rate profile
maximization in MIMO IBCs; and 2) the max-min fair multicast multigroup
beamforming problem in multi-cell systems. Our algorithms i) were
proved to converge to d-stationary solutions of the aforementioned
problems; ii) represent the first attempt to design distributed solution
methods for 1) and 2); and iii) were shown numerically to reach better
local optimal solutions than ad-hoc schemes proposed in the literature
for special cases of the considered formulations. 

We remark that, although we considered in details only problems 1)
and 2) above, the applicability of our framework goes much
beyond these two specific formulations and, more generally, applications
in communications. Moreover, even within the network systems considered
in 1) and 2), one can consider alternative objective functions and
constraints. Two examples of unsolved problems to which our framework
readily applies (with convergence guarantees) are: 1) the distributed
minimization of the BS's (weighted) transmit power over MIMO IBCs,
subject to rate constraints; and 2) the maximization of the (weighted)
sum of the multi-cast multigroup capacity in multi-cell systems. 

Our framework finds applications also in other areas, such as signal
processing, smart grids, and machine learning.

\appendix{}

\vspace{-0.1cm}

\subsection{Proof of Lemma \ref{lem:opt_Q} \label{app:clsoed_form_Q}}

Let $\bar{\mathbf{Q}}_{i_{k}}\triangleq\mathbf{U}_{{i_{k}}}^{\nu H}\mathbf{Q}_{i_{k}}\mathbf{U}_{{i_{k}}}^{\nu}$,
and $\bar{\mathbf{Q}}_{k}\triangleq(\bar{\mathbf{Q}}_{i_{k}})_{i_{k}\in\mathcal{I}_{k}}$.
Then, each problem $\min_{\mathbf{Q}_{k}\in\mathcal{Q}_{k}}\mathcal{L}_{\mathbf{Q}_{k}}(\mathbf{Q}_{k},\boldsymbol{\lambda},\boldsymbol{\Omega};\mathbf{Q}^{\nu})$
in (\ref{eq:dual-func-IBC-form1}) can be rewritten as 
\begin{equation}
\begin{array}{cl}
{\displaystyle \min_{\bar{\mathbf{Q}}_{k}=(\bar{\mathbf{Q}}_{i_{k}}\succeq\mathbf{0})_{i_{k}\in\mathcal{I}_{k}}}} & {\displaystyle \sum_{i_{k}\in\mathcal{I}_{k}}}\text{tr}\left(\left(\tau_{\mathbf{Q}}\bar{\mathbf{Q}}_{i_{k}}^{H}-\mathbf{D}_{i_{k}}^{\nu}\right)\,\bar{\mathbf{Q}}_{i_{k}}\right)\\
\text{s.t.} & {\displaystyle \sum_{i_{k}\in\mathcal{I}_{k}}}\text{tr}(\bar{\mathbf{Q}}_{i_{k}})\leq P_{k}.
\end{array}\label{eq:diag-prob-Q-k}
\end{equation}

We claim that the optimal solution of (\ref{eq:diag-prob-Q-k}) must
be diagonal. Indeed, denoting by $\texttt{{diag}}(\bar{\mathbf{Q}}_{i_{k}})$
the diagonal matrix having the same diagonal entries of $\bar{\mathbf{Q}}_{i_{k}}$,
and $\texttt{{off}}(\bar{\mathbf{Q}}_{i_{k}})\triangleq\bar{\mathbf{Q}}_{i_{k}}-\texttt{{diag}}(\bar{\mathbf{Q}}_{i_{k}})$,
(each term in the sum of) the objective function of (\ref{eq:diag-prob-Q-k})
can be lower bounded as
\begin{equation}
\begin{array}{l}
\text{tr}\left(\left(\tau_{\mathbf{Q}}\bar{\mathbf{Q}}_{i_{k}}^{H}-\mathbf{D}_{i_{k}}^{\nu}\right)\bar{\mathbf{Q}}_{i_{k}}\right)\medskip\\
\qquad=\text{tr}\left(\left(\tau_{\mathbf{Q}}\texttt{{diag}}(\bar{\mathbf{Q}}_{i_{k}})-\mathbf{D}_{i_{k}}^{\nu}\right)\texttt{{diag}}(\bar{\mathbf{Q}}_{i_{k}})\right)\smallskip\\
\qquad\,\,\,\,\,+\text{tr}\left(\tau_{\mathbf{Q}}\texttt{{off}}(\bar{\mathbf{Q}}_{i_{k}})^{H}\texttt{{off}}(\bar{\mathbf{Q}}_{i_{k}})\right)\medskip\\
\qquad\geq\text{tr}\left(\left(\tau_{\mathbf{Q}}\texttt{{diag}}(\bar{\mathbf{Q}}_{i_{k}})-\mathbf{D}_{i_{k}}^{\nu}\right)\,\texttt{{diag}}(\bar{\mathbf{Q}}_{i_{k}})\right).
\end{array}\label{ineq_tr}
\end{equation}
The claim follows from the fact that the lower bound in (\ref{ineq_tr})
is achieved if and only if $\texttt{{off}}(\bar{\mathbf{Q}}_{i_{k}})=\mathbf{0}$
and that constraints in (\ref{eq:diag-prob-Q-k}) depend only on $\texttt{{diag}}(\bar{\mathbf{Q}}_{i_{k}})$.

Setting $\bar{\mathbf{Q}}_{i_{k}}=\texttt{{Diag}}(\bar{{\mathbf{q}}}_{i_{k}})$
to be diagonal with entries the elements of $\bar{{\mathbf{q}}}_{i_{k}}\triangleq(\bar{{q}}_{i_{k},t})_{t=1}^{T_{k}}$,
(\ref{eq:diag-prob-Q-k}) becomes\vspace{-.3cm} 
\begin{equation}
\begin{array}{cl}
{\displaystyle {\min_{\bar{{\mathbf{q}}}_{i_{k}}\geq\mathbf{0}}}} & {\displaystyle \sum_{i_{k}\in\mathcal{I}_{k}}\sum_{t=1}^{T_{k}}}\left(\tau_{\mathbf{Q}}\bar{q}_{i_{k},t}^{2}-d_{i_{k},t}^{\nu}\cdot\bar{q}_{i_{k},t}\right)\\
\mbox{s.t.} & {\displaystyle \sum_{i_{k}\in\mathcal{I}_{k}}\sum_{t=1}^{T_{k}}}\bar{q}_{i_{k},t}\leq P_{k}.
\end{array}\label{eq:sub-prob-bar-q-i-k-j}
\end{equation}
Problem (\ref{eq:sub-prob-bar-q-i-k-j}) has a closed form solution
$\bar{\mathbf{q}}_{i_{k}}^{\star}$ (up to the multiplier $\xi_{k}^{\star}$),
given by\vspace{-.1cm} 
\begin{equation}
\bar{\mathbf{q}}_{i_{k}}^{\star}=\left[\frac{\mathbf{d}_{i_{k}}^{\nu}-\xi_{k}^{\star}}{2\tau_{\mathbf{Q}}}\right]_{+},\label{eq:q_star}
\end{equation}
where $\xi_{k}^{\star}$ needs to be chosen so that  $\sum_{i_{k}\in\mathcal{I}_{k}}\mathbf{1}^{T}\bar{\mathbf{q}}_{i_{k}}^{\star}\leq P_{k}$.
This can be done, e.g., using Algorithm \ref{algorithm:opt_multipliers},
which converges in a finite number of steps. 
\begin{algorithm}[H]
\textbf{Data:} $\mathbf{d}_{k}^{\nu}\triangleq(d_{k,j}^{\nu})_{j=1}^{T_{k}\cdot I_{k}}=(\mathbf{d}_{i_{k}}^{\nu})_{i_{k}\in\mathcal{I}_{k}}$
(arranged in decreasing order). 

(\texttt{S.0}) Set $\mathcal{J}\triangleq\{j:\,d_{k,j}^{\nu}>0\}$.

(\texttt{S.1}) If $\sum_{i\in\mathcal{J}}\,\frac{d_{k,j}^{\nu}}{2\tau_{\mathbf{Q}}}\leq P_{k}$:
set $\xi_{k}^{\star}=0$ and \texttt{STOP}.

(\texttt{S.2}) Repeat

\quad{}\quad{}\quad{}\,\,(a) {Set} $\xi_{k}^{\star}=\left[\frac{\sum_{j\in\mathcal{J}}d_{k,j}^{\nu}-2\tau_{\mathbf{Q}}P_{k}}{\left|\mathcal{J}\right|}\right]_{+};$

\quad{}\quad{}\quad{}\,\,(b) {If} $\frac{d_{k,j}^{\nu}-\xi_{k}^{\star}}{2\tau_{\mathbf{Q}}}>0$,
$\forall j\in\mathcal{J}$: \texttt{STOP}; 

\quad{}\quad{}\quad{}\quad{}\quad{}\,\,\,{else}\, $\mathcal{J}\triangleq\mathcal{J}/\left\{ j=\left|\mathcal{J}\right|\right\} $; 

\quad{}\quad{}\quad{}{until} $\left|\mathcal{J}\right|=1$. \caption{Efficient computation of $\xi_{k}^{\star}$ in (\ref{eq:q_star}) }

\label{algorithm:opt_multipliers} 
\end{algorithm}

\noindent The optimal solution $\mathbf{Q}_{i_{k}}^{\star}$ is
thus given by $\mathbf{Q}_{i_{k}}^{\star}\triangleq\mathbf{U}_{{i_{k}}}^{\nu}\texttt{{Diag}}(\bar{{\mathbf{q}}}_{i_{k}}^{\star})\mathbf{U}_{{i_{k}}}^{\nu H}$,
with $\bar{{\mathbf{q}}}_{i_{k}}^{\star}$ defined in (\ref{eq:q_star}),
which completes the proof. \vspace{-.2cm}

\subsection{Proof of Lemma \ref{lem:opt_Y} \label{app:clsoed_form_Y}}

\textit{\emph{Let $\mathbf{V}_{i_{k}}^{\nu}\mathbf{D}_{i_{k}}^{\nu}\mathbf{V}_{i_{k}}^{\nu H}$
be the eigenvalue/eigenvector decomposition of $2\tau_{\mathbf{Y}}\mathbf{Y}_{i_{k}}^{\nu}-\boldsymbol{\Omega}_{i_{k}}^{\nu}$,
with $\mathbf{D}_{i_{k}}^{\nu}=\texttt{{Diag}}(\mathbf{d}_{i_{k}}^{\nu})$;
define $\bar{\mathbf{Y}}_{i_{k}}\triangleq\mathbf{V}_{i_{k}}^{\nu H}\mathbf{Y}_{i_{k}}\mathbf{V}_{i_{k}}^{\nu}$.}}
Each $\min_{(\mathbf{Y}_{i_{k}}\succeq\mathbf{0})_{i_{k}\in\mathcal{I}_{k}}}\mathcal{L}_{\mathbf{Y}_{k}}(\mathbf{Y}_{k},\boldsymbol{\lambda},\boldsymbol{\Omega};\mathbf{Y}^{\nu})$
in (\ref{eq:dual-func-IBC-form1}) can be decomposed in $I_{k}$ separate
subproblems, the $i_{k}$-th of which is 
\begin{equation}
\begin{split}{\displaystyle \min_{\bar{\mathbf{Y}}_{i_{k}}\succeq\mathbf{0}}}\text{tr}\left(\left(\tau_{\mathbf{Y}}\bar{\mathbf{Y}}_{i_{k}}^{H}-\mathbf{D}_{i_{k}}^{\nu}\right)\,\bar{\mathbf{Y}}_{i_{k}}\right)- & \lambda_{i_{k}}^{\nu}\log\det\left(\sigma_{i_{k}}^{2}\mathbf{I}+\bar{\mathbf{Y}}_{i_{k}}\right)\end{split}
.\label{eq:Y_problem}
\end{equation}

We claim that the optimal solution of (\ref{eq:Y_problem}) must be
diagonal. This is a consequence of the following two inequalities:
Denoting by $\texttt{{diag}}(\bar{\mathbf{Y}}_{i_{k}})$ the diagonal
matrix having the same diagonal entries of $\bar{\mathbf{Y}}_{i_{k}}$,
and $\texttt{{off}}(\bar{\mathbf{Y}}_{i_{k}})\triangleq\bar{\mathbf{Y}}_{i_{k}}-\texttt{{diag}}(\bar{\mathbf{Y}}_{i_{k}})$,
we have {[}note that $\texttt{{diag}}(\bar{\mathbf{Y}}_{i_{k}})\succeq\mathbf{0}${]}
\[
\begin{array}{l}
\text{tr}\left(\left(\tau_{\mathbf{Y}}\bar{\mathbf{Y}}_{i_{k}}^{H}-\mathbf{D}_{i_{k}}^{\nu}\right)\bar{\mathbf{Y}}_{i_{k}}\right)\smallskip\\
\quad\geq\text{tr}\left(\left(\tau_{\mathbf{Y}}\texttt{{diag}}(\bar{\mathbf{Y}}_{i_{k}})-\mathbf{D}_{i_{k}}^{\nu}\right)\texttt{{diag}}(\bar{\mathbf{Y}}_{i_{k}})\right),\medskip\\
\log\det\left(\sigma_{i_{k}}^{2}\mathbf{I}+\bar{\mathbf{Y}}_{i_{k}}\right)\leq\log\det\left(\sigma_{i_{k}}^{2}\mathbf{I}+\texttt{{diag}}(\bar{\mathbf{Y}}_{i_{k}})\right),
\end{array}
\]
where the first inequality has been proved in (\ref{ineq_tr}), while
the second one is the Hadamard's inequality. Both inequalities are
satisfied with equality if and only if $\bar{\mathbf{Y}}_{i_{k}}=\texttt{{diag}}(\bar{\mathbf{Y}}_{i_{k}})$.

Replacing in (\ref{eq:Y_problem}) $\bar{\mathbf{Y}}_{i_{k}}=\texttt{{diag}}(\bar{\mathbf{Y}}_{i_{k}})\triangleq\texttt{{Diag}}(\mathbf{y}_{i_{k}})$
and checking that $\mathbf{y}_{i_{k}}$ given by (\ref{eq:y_opt})
satisfies the KKT system of the resulting convex optimization problem,
one get the closed form expression (\ref{eq:y_opt}).

\subsection{\label{sec:Newton_Hessian}Augmented Hessian and Gradient Expressions
in (\ref{Newton_algo})}

In this section, we provide the closed form expressions for the augmented
Hessian matrices and gradients of the updating rules given in (\ref{Newton_algo}).
More specifically, we have 
\begin{equation}
\begin{split}\nabla_{\boldsymbol{\lambda},\text{vec}(\boldsymbol{\Omega}^{*})}D(\boldsymbol{\lambda},\boldsymbol{\Omega}; & \mathbf{W}^{\nu})\triangleq[\nabla_{\boldsymbol{\lambda}}D(\boldsymbol{\lambda},\boldsymbol{\Omega};\mathbf{W}^{\nu});\\
 & \text{vec}(\nabla_{\boldsymbol{\Omega}^{*}}D(\boldsymbol{\lambda},\boldsymbol{\Omega};\mathbf{W}^{\nu}))].
\end{split}
\end{equation}
Define 
\[
\mathbf{\bar{W}}^{n}\triangleq(R^{\star}(\boldsymbol{\lambda}^{n};R^{\nu}),\mathbf{Q}^{\star}(\boldsymbol{\lambda}^{n},\boldsymbol{\Omega}^{n};\mathbf{Q}^{\nu}),\mathbf{Y}^{\star}(\boldsymbol{\lambda}^{n},\boldsymbol{\Omega}^{n};\mathbf{Y}^{\nu}))
\]
and 
\[
\begin{array}{lll}
\nabla_{\lambda_{i_{k}}}D(\boldsymbol{\lambda}^{n},\boldsymbol{\Omega}^{n};\mathbf{W}^{\nu}) & = & \hat{h}_{i_{k}}(\mathbf{W};\mathbf{Q}^{\nu})\mid_{\mathbf{W}=\mathbf{\bar{W}}^{n}}\medskip\\
\nabla_{\boldsymbol{\Omega}_{i_{k}}^{*}}D(\boldsymbol{\lambda}^{n},\boldsymbol{\Omega}^{n};\mathbf{W}^{\nu}) & = & {h}_{i_{k}}(\mathbf{W})|_{\mathbf{W}=\mathbf{\bar{W}}^{n}}
\end{array}
\]
where 
\[
\begin{array}{lll}
\hat{h}_{i_{k}}(\mathbf{W};\mathbf{Q}^{\nu}) & \triangleq & \alpha_{i_{k}}{R}-\widehat{R}_{i_{k}}\left(\mathbf{Q}_{-i_{k}},\mathbf{Y}_{i_{k}};\mathbf{Q}^{\nu}\right)\medskip\\
{h}_{i_{k}}(\mathbf{W}) & \triangleq & \mathbf{Y}_{i_{k}}-\mathbf{I}_{i_{k}}(\mathbf{Q}).
\end{array}
\]
Then, we have 
\begin{equation}
\begin{aligned}\mathbf{G}\triangleq\, & [(\text{vec}(\nabla_{\mathbf{W}^{*}}\hat{h}_{i_{k}}(\bar{\mathbf{W}}^{n};\mathbf{Q}^{\nu}))^{H})_{\forall i_{k}\in\mathcal{I}};\\
 & \,(\nabla_{\mathbf{W}^{*}}{h}_{i_{k}}(\bar{\mathbf{W}}^{n};\mathbf{Q}^{\nu}))_{\forall i_{k}\in\mathcal{I}}],
\end{aligned}
\end{equation}
with 
\[
\begin{split} & \nabla_{\boldsymbol{\lambda},\text{vec}(\boldsymbol{\Omega}^{*})}^{2}D(\boldsymbol{\lambda}^{n},\boldsymbol{\Omega}^{n};\mathbf{W}^{\nu})=\mathbf{G}\cdot\boldsymbol{\mathcal{H}}(\bar{\mathbf{W}}^{n};\mathbf{W}^{\nu})^{-1}\cdot\mathbf{G}^{H}\end{split}
,
\]
and 
\[
\boldsymbol{\mathcal{H}}(\bar{\mathbf{W}}^{n};\mathbf{W}^{\nu})=\text{bdiag}(\nabla_{\mathbf{Q}^{*}}^{2}\mathcal{L},{\displaystyle {\frac{d^{2}\mathcal{L}}{d{R}^{2}}},\nabla_{\mathbf{Y}^{*}}^{2}\mathcal{L}),}
\]
where, for notation simplicity, we assume $\mathcal{L}=\mathcal{L}(\bar{\mathbf{W}}^{n},\boldsymbol{\lambda}^{n},\boldsymbol{\Omega}^{n};\mathbf{W}^{\nu})$.

It follows that 
\begin{equation}
\begin{split} & \nabla_{\mathbf{Q}^{*}}^{2}\mathcal{L}=\text{bdiag}\left((\boldsymbol{\mathcal{H}}_{\mathbf{Q}_{j_{m}}}(\bar{\mathbf{W}}^{n},\boldsymbol{\lambda}^{n},\boldsymbol{\Omega}^{n};\mathbf{W}^{\nu}))_{j_{m}\in\mathcal{I}}\right),\\
 & \nabla_{\mathbf{Y}^{*}}^{2}\mathcal{L}=\text{bdiag}\left((\boldsymbol{\mathcal{H}}_{\mathbf{Y}_{j_{m}}}(\bar{\mathbf{W}}^{n},\boldsymbol{\lambda}^{n},\boldsymbol{\Omega}^{n};\mathbf{W}^{\nu}))_{j_{m}\in\mathcal{I}}\right),\\
 & {\displaystyle {\frac{d^{2}\mathcal{L}}{d{R}^{2}}}=\tau_{R},}
\end{split}
\end{equation}
with 
\begin{equation}
\begin{split} & \begin{split}\boldsymbol{\mathcal{H}}_{\mathbf{Q}_{j_{m}}}(\bar{\mathbf{W}}^{n},\boldsymbol{\lambda}^{n},\boldsymbol{\Omega}^{n};\mathbf{W}^{\nu})= & 2\,\tau_{\mathbf{Q}}(\mathbf{I}_{T_{k}}\otimes\mathbf{I}_{T_{k}}),\end{split}
\\
 & \begin{split}\boldsymbol{\mathcal{H}}_{\mathbf{Y}_{j_{m}}}(\bar{\mathbf{W}}^{n}, & \boldsymbol{\lambda}^{n},\boldsymbol{\Omega}^{n};\mathbf{W}^{\nu})=2\,\tau_{\mathbf{Y}}(\mathbf{I}_{M_{i_{k}}}\otimes\mathbf{I}_{M_{i_{k}}})+\lambda_{i_{k}}^{n}\cdot\\
 & \left[{(\sigma_{i_{k}}^{2}\mathbf{I}+\mathbf{Y}_{i_{k}}^{n})^{-1}}^{H}\otimes(\sigma_{i_{k}}^{2}\mathbf{I}+\mathbf{Y}_{i_{k}}^{n})^{-1}\right].
\end{split}
\end{split}
\end{equation}
Finally, we obtain 
\[
\begin{array}{l}
\text{vec}(\nabla_{\mathbf{W}^{*}}\hat{h}_{i_{k}}(\bar{\mathbf{W}}^{n};\mathbf{Q}^{\nu}))=[(\mathbf{M}_{j_{l}}^{i_{k}})_{j_{l}\in\mathcal{I}}^{H},\alpha_{i_{k}},(\mathbf{N}_{j_{l}}^{i_{k}})_{j_{l}\in\mathcal{I}}^{H}]^{H},\medskip\\
\nabla_{\mathbf{W}^{*}}{h}_{i_{k}}(\bar{\mathbf{W}}^{n})=[{(\mathbf{F}_{j_{l}}^{i_{k}})}_{j_{l}\in\mathcal{I}},\mathbf{0}_{M_{i_{k}}^{2}\times1},{(\mathbf{G}_{j_{l}}^{i_{k}})}_{j_{l}\in\mathcal{I}}],
\end{array}
\]
where 
\[
\begin{split} & \mathbf{M}_{j_{l}}^{i_{k}}=\left\{ \begin{array}{lll}
(\mathbf{I}_{T_{k}}\otimes\mathbf{\Pi}_{i_{k},j_{l}}^{-})\text{vec}(\mathbf{I}_{T_{k}}),\quad\;\text{if}\;j_{l}\neq i_{k}\\
(\mathbf{0}_{T_{k}}\otimes\mathbf{0}_{T_{k}})\text{vec}(\mathbf{0}_{T_{k}}),\hspace{0.7cm}\text{otherwise;}
\end{array}\right.\\
 & \mathbf{N}_{j_{l}}^{i_{k}}=\left\{ \begin{array}{lll}
(\mathbf{I}_{M_{i_{k}}}\otimes(\sigma_{i_{k}}^{2}\mathbf{I}+\mathbf{Y}_{i_{k}})^{-1})\text{vec}(\mathbf{I}_{M_{i_{k}}}),\quad\;\text{if}\;j_{l}=i_{k};\\
(\mathbf{0}_{M_{i_{k}}}\otimes\mathbf{0}_{M_{i_{k}}})\text{vec}(\mathbf{0}_{M_{i_{k}}}),\hspace{2.05cm}\mbox{otherwise;}
\end{array}\right.\\
 & \mathbf{F}_{j_{l}}^{i_{k}}=-(\mathbf{H}_{i_{k}l}^{*}\otimes\mathbf{H}_{i_{k}l})\hspace{0.81cm}\forall j_{l}\in\mathcal{I};\\
 & \mathbf{G}_{j_{l}}^{i_{k}}=\left\{ \begin{array}{lll}
(\mathbf{I}_{M_{i_{k}}}\otimes\mathbf{I}_{M_{i_{k}}}),\quad\;\text{if}\;j_{l}=i_{k};\\
(\mathbf{0}_{M_{i_{k}}}\otimes\mathbf{0}_{M_{i_{k}}}),\hspace{0.35cm}\text{otherwise}.
\end{array}\right.
\end{split}
\]

\subsection{Proof of Theorem \ref{th:conver}\label{sub:Proof_convergence_centralized_MMF} }

The proof of convergence follows readily from Proposition \ref{prop: eq-prob-MMF}
and \citep[Th.2]{ScuFacLamPartI}, and thus is omitted. So does $t^{\infty}>0$,
when the algorithm does not converge in a finite number of steps.
Then, we only need to prove that, if $t^{0}>0$, then $\hat{t}(\boldsymbol{\lambda};t^\nu,\boldsymbol{\beta}^\nu)>0$,
for all $\nu\geq1$ and   $\boldsymbol{\lambda} \ge 0$. Because of space limitations, we consider (\ref{eq:SCA-cvx-frame-1})
with surrogate $\tilde{g}_{i}$ given by (\ref{eq:g_tilde-1}) only.

It is not difficult to see that, given the current iterate $(t^{\nu},\boldsymbol{\beta}^{\nu},\mathbf{w}^{\nu})$,
$\hat{t}(\boldsymbol{\lambda};t^\nu,\boldsymbol{\beta}^\nu)$ has the following expression:
\begin{equation}
\hat{t}(\boldsymbol{\lambda};t^\nu,\boldsymbol{\beta}^\nu)=\frac{t^{\nu}+\tau_{t}\cdot(t^{\nu})^{2}}{\tau_{t}\cdot t^{\nu}+\boldsymbol{\lambda}^{T}\boldsymbol{\beta}^{\nu}},\label{eq:closed_form_t_hat}
\end{equation}
where $\boldsymbol{\lambda}$ and $\boldsymbol{\eta}$ are the (nonnegative) multipliers associated with the constraints $\tilde{g}_{i_{k}}(t,\beta_{i_{k}},\mathbf{w}_{k};t^{\nu},\beta_{i_{k}}^{\nu},\mathbf{w}_{k}^{\nu})\leq0$
and (b), respectively. It follows from (\ref{eq:closed_form_t_hat}),
that if $t^{0}\neq0$, then ${t}^{1}=t^{0}+\gamma^{0}(\hat{t}(\boldsymbol{\lambda};t^0,\boldsymbol{\beta}^0)-t^{0})>0$.
Therefore, so is $t^{\nu}$, for $\nu\geq2$. \vspace{-0.1cm}

{\small{}\bibliographystyle{IEEEtran}
\bibliography{scutari_refs}

\begin{thebibliography}{10}
\providecommand{\url}[1]{#1}
\csname url@samestyle\endcsname
\providecommand{\newblock}{\relax}
\providecommand{\bibinfo}[2]{#2}
\providecommand{\BIBentrySTDinterwordspacing}{\spaceskip=0pt\relax}
\providecommand{\BIBentryALTinterwordstretchfactor}{4}
\providecommand{\BIBentryALTinterwordspacing}{\spaceskip=\fontdimen2\font plus
\BIBentryALTinterwordstretchfactor\fontdimen3\font minus
  \fontdimen4\font\relax}
\providecommand{\BIBforeignlanguage}[2]{{%
\expandafter\ifx\csname l@#1\endcsname\relax
\typeout{** WARNING: IEEEtran.bst: No hyphenation pattern has been}%
\typeout{** loaded for the language `#1'. Using the pattern for}%
\typeout{** the default language instead.}%
\else
\language=\csname l@#1\endcsname
\fi
#2}}
\providecommand{\BIBdecl}{\relax}
\BIBdecl

\bibitem{ScuFaccLampSonICASSP14}
G.~Scutari, F.~Facchinei, L.~Lampariello, and P.~Song, ``Parallel and
  distributed methods for nonconvex optimization,'' in \emph{Proc. of IEEE
  International Conference on Acoustics, Speech, and Signal Processing (ICASSP
  14)}, Florence, Italy, May 4-9 2014.

\bibitem{ScuFaccLampSonICASSP16}
P.~Song, G.~Scutari, F.~Facchinei, and L.~Lampariello, ``D3m: Distributed
  multi-cell multigroup multicasting,'' in \emph{Proc. of IEEE International
  Conference on Acoustics, Speech, and Signal Processing (ICASSP 16)}, Shangai,
  China, March 20-25 2016.

\bibitem{ScuFacLamPartI}
\BIBentryALTinterwordspacing
G.~Scutari, F.~Facchinei, L.~Lampariello, and P.~Song, ``Parallel and
  distributed methods for nonconvex optimization$-${P}art {I}: {T}heory,''
  \emph{{IEEE} Trans. Signal Process.}, (submitted). [Online]. Available:
  \url{arXiv:1410.4754}
\BIBentrySTDinterwordspacing

\bibitem{DC-BranchAndBound-1}
K.~T. Phan, S.~A. Vorobyov, C.~Telambura, and T.~Le-Ngoc, ``Power control for
  wireless cellular systems via {D.C.} programming,'' in \emph{IEEE/SP 14th
  Workshop on Statistical Signal Processing}, 2007, pp. 507--511.

\bibitem{DC-BranchAndBound-5}
H.~Al-Shatri and T.~Weber, ``Achieving the maximum sum rate using {D.C.}
  programming in cellular networks,'' \emph{IEEE Trans. Signal Process.},
  vol.~60, no.~3, pp. 1331--1341, Mar. 2012.

\bibitem{DC-Linearization-1}
N.~Vucic, S.~Shi, and M.~Schubert, ``{DC} programming approach for resource
  allocation in wireless networks,'' in \emph{8th International Symposium on
  Modeling and Optimization in Mobile, Ad Hoc and Wireless Networks (WiOpt)},
  2010, pp. 380--386.

\bibitem{Chiang-WeiTan-PalomarOneil-Julian_ITWC-GP}
M.~Chiang, C.~W. Tan, D.~P. Palomar, D.~O. Neill, and D.~Julian, ``Power
  control by geometric programming,'' \emph{{IEEE} Trans. Wireless Commun.},
  vol.~6, no.~7, pp. 2640--2651, Jul. 2007.

\bibitem{DC-Polynomial}
A.~Khabbazibasmenj, F.~Roemer, S.~A. Vorobyov, and M.~Haardt, ``Sum-rate
  maximization in two-way {AF MIMO} relaying: Polynomial time solutions to a
  class of {DC} programming problems,'' \emph{IEEE Trans. Signal Process.},
  vol.~60, no.~10, pp. 5478--5493, 2012.

\bibitem{DC-BranchAndBound-2}
Y.~Xu, T.~Le-Ngoc, and S.~Panigrahi, ``Global concave minimization for optimal
  spectrum balancing in multi-user {DSL} networks,'' \emph{IEEE Trans. Signal
  Process.}, vol.~56, no.~7, pp. 2875--2885, Jul. 2008.

\bibitem{CendrillonHuangChiangMoonen_TSP06}
R.~Cendrillon, J.~Huang, M.~Chiang, and M.~Moonen, ``Autonomous spectrum
  balancing for digital subscriber lines,'' \emph{{IEEE} Trans. Signal
  Process.}, vol.~55, no.~8, pp. 4241--4257, Aug. 2007.

\bibitem{SchmidtShiBerryHonigUtschick-SPMag}
D.~Schmidt, C.~Shi, R.~Berry, M.~Honig, and W.~Utschick, ``Distributed resource
  allocation schemes: Pricing algorithms for power control and beamformer
  design in interference networks,'' \emph{{IEEE} Signal Process. Mag.},
  vol.~26, no.~5, pp. 53--63, Sept. 2009.

\bibitem{KimGiannakisIT11}
S.-J. Kim and G.~B. Giannakis, ``Optimal resource allocation for {MIMO} ad hoc
  cognitive radio networks,'' \emph{{IEEE} Trans. on Information Theory},
  vol.~57, no.~5, pp. 3117--3131, May 2011.

\bibitem{ScutFacchSonPalPang13}
\BIBentryALTinterwordspacing
G.~Scutari, F.~Facchinei, P.~Song, D.~P. Palomar, and J.-S. Pang,
  ``Decomposition by partial linearization: Parallel optimization of
  multi-agent systems,'' \emph{IEEE Trans. Signal Process.}, vol.~62, no.~3,
  pp. 641--656, Feb. 2014. [Online]. Available:
  \url{http://arxiv.org/abs/1302.0756.}
\BIBentrySTDinterwordspacing

\bibitem{ZhangCui_RateProfileTSP10}
R.~Zhang and S.~Cui, ``Cooperative interference management with miso
  beamforming,'' \emph{{IEEE} Trans. Signal Process.}, vol.~58, no.~18, pp.
  5450--5458, Oct. 2010.

\bibitem{MochaourabCaoJorswieck_RateProfile_arxiv13}
R.~Mochaourab, P.~Cao, and E.~Jorswieck, ``Alternating rate profile
  optimization in single stream mimo interference channels,'' \emph{{IEEE}
  Signal Process. Lett.}, no.~21, Feb. 2014.

\bibitem{QiuZhangLuoCui_maxminTSP11}
J.~Qiu, R.~Zhang, Z.-Q. Luo, and S.~Cui, ``Optimal distributed beamforming for
  miso interference channels,'' \emph{{IEEE} Trans. Signal Process.}, vol.~59,
  no.~11, pp. 5638--5643, Nov. 2011.

\bibitem{LiuZhangChuai_RateProfileTWC12}
L.~Liu, R.~Zhang, and K.-C. Chua, ``Achieving global optimality for weighted
  sum-rate maximization in the k-user gaussian interference channel with
  multiple antennas,'' \emph{{IEEE} Trans. Wireless Commun.}, vol.~11, no.~5,
  pp. 1933--1945, May 2012.

\bibitem{RazaviyayniHongLuo_maxminSP13}
M.~Razaviyayn, M.~Hong, and Z.-Q. Luo, ``Linear transceiver design for a mimo
  interfering broadcast channel achieving max-min fairness,'' \emph{Signal
  Processing}, vol.~93, pp. 3327--3340, Dec. 2013.

\bibitem{DallAnese2012}
Y.~Zhang, E.~Dall'Anese, and G.~B. Giannakis, ``Distributed optimal beamformers
  for cognitive radios robust to channel uncertainties,'' \emph{IEEE Trans.
  Signal Process.}, vol.~60, no.~12, pp. 6495--6508, Dec. 2012.

\bibitem{YangScutariPalomar_JSAC13}
Y.~Yang, G.~Scutari, P.~Song, and D.~P. Palomar, ``Robust mimo cognitive radio
  systems under interference temperature constraints,'' \emph{{IEEE} J. Sel.
  Areas Commun.}, vol.~31, no.~11, pp. 2465--2482, Nov. 2013.

\bibitem{Wang-Krunz-Cui_JSTSP08}
F.~Wang, M.~Krunz, and S.~Cui, ``Price-based spectrum management in cognitive
  radio networks,'' \emph{{IEEE} J. Sel. Topics Signal Process.}, vol.~2,
  no.~1, pp. 74--87, Feb. 2008.

\bibitem{KariSidiLuo08}
E.~Karipidis, N.~D. Sidiropoulos, and Z.-Q. Luo, ``Quality of service and
  max-min fair transmit beamforming to multiple cochannel multicast groups,''
  \emph{IEEE Trans. on Signal Process.}, vol.~56, no.~3, pp. 1268 -- 1279, Mar.
  2008.

\bibitem{ChriChatOtte14}
D.~Christopoulos, S.~Chatzinotas, and B.~Ottersten, ``Weighted fair multicast
  multigroup beamforming under per-antenna power constraints,'' \emph{IEEE
  Trans. on Signal Process.}, vol.~62, no.~19, pp. 5132 -- 5142, Oct. 2014.

\bibitem{HsuWangSuLin14}
G.-W. Hsu, H.-H. Wang, H.-J. Su, and P.~Lin, ``Joint beamforming for multicell
  multigroup multicast with per-cell power constraints,'' in \emph{2014 IEEE
  25th Annual Int. Symposium on Personal, Indoor, and Mobile Radio
  Communication (PIMRC)}, Washington, DC, Sept. 2-5 2014, pp. 527 -- 532.

\bibitem{XianTaoWang13}
Z.~Xiang, M.~Tao, and X.~Wang, ``Coordinated multicast beamforming in multicell
  networks,'' \emph{IEEE Trans. on Wireless Commun.}, vol.~12, no.~1, pp. 12 --
  21, Jan. 2013.

\bibitem{ConvexSumSeparable-2}
M.~Chiang, S.~H. Low, A.~R. Calderbank, and J.~C. Doyle, ``Layering as
  optimization decomposition: A mathematical theory of network architectures,''
  \emph{Proc. IEEE}, vol.~95, no.~1, pp. 255--312, Jan. 2007.

\bibitem{Chiang_Hande_Lan_Tan_book_PC}
M.~Chiang, P.~Hande, T.~Lan, and C.~W. Tan, \emph{Power control in wireless
  cellular networks}.\hskip 1em plus 0.5em minus 0.4em\relax Foundations and
  Trends in Networking, Now Publishers, Jul. 2008, vol.~2, no.~4.

\bibitem{Palomar-Chiang_ACTran07-Num}
D.~P. Palomar and M.~Chiang, ``Alternative distributed algorithms for network
  utility maximization: Framework and applications,'' \emph{{IEEE} Trans. on
  Automatic Control}, vol.~52, no.~12, pp. 2254--2269, Dec. 2007.

\bibitem{SidiDaviLuo06}
S.~N. D., D.~T. N., and Z.-Q. Luo, ``Transmit beamforming for physical-layer
  multicasting,'' \emph{IEEE Trans. on Signal Process.}, vol.~54, no.~6, pp.
  2239--2251, Jun. 2006.

\bibitem{shen2010mse}
H.~Shen, B.~Li, M.~Tao, and X.~Wang, ``Mse-based transceiver designs for the
  mimo interference channel,'' \emph{{IEEE} Trans. Wireless Commun.}, vol.~9,
  no.~11, pp. 3480--3489, Sept. 2010.

\bibitem{chen2012iterative}
C.-E. Chen and W.-H. Chung, ``An iterative minmax per-stream mse transceiver
  design for mimo interference channel,'' \emph{IEEE Wireless Commun. Lett.},
  vol.~1, no.~3, pp. 229--232, Apr. 2012.

\bibitem{qiu2011optimal}
J.~Qiu, R.~Zhang, Z.-Q. Luo, and S.~Cui, ``Optimal distributed beamforming for
  miso interference channels,'' \emph{{IEEE} Trans. Signal Process.}, vol.~59,
  no.~11, pp. 5638--5643, Jul. 2011.

\bibitem{Cai-Quek}
D.~W.~H. Cai, T.~Q.~S. Quek, and C.~W. Tan, ``A unified analysis of max-min
  weighted sinr for mimo downlink system,'' \emph{IEEE Trans. on Signal
  Process.}, vol.~59, pp. 3850--3862, Aug. 2011.

\bibitem{Scutari-Facchinei-Pang-Palomar_IT_PI}
G.~Scutari, F.~Facchieni, J.-S. Pang, and D.~P. Palomar, ``Real and complex
  monotone communication games,'' \emph{{IEEE} Trans. on Information Theory},
  vol.~60, no.~7, pp. 4197--4231, July 2014.

\bibitem{ShiRazaviyaynLuoHe-TSP11}
Q.~Shi, M.~Razaviyayn, Z.-Q. Luo, and C.~He, ``An iteratively weighted {MMSE}
  approach to distributed sum-utility maximization for a {MIMO} interfering
  broadcast channel,'' \emph{{IEEE} Trans. Signal Process.}, vol.~59, no.~9,
  pp. 4331--4340, Sept. 2011.

\bibitem{GopaSidi15}
B.~Gopalakrishnan and N.~D. Sidiropoulos, ``High performance adaptive
  algorithms for single-group multicast beamforming,'' \emph{IEEE Trans. on
  Signal Process.}, vol.~63, no.~16, pp. 4373--4384, Aug. 2015.

\bibitem{KariSidiLuo07}
E.~Karipidis, N.~D. Sidiropoulos, and Z.-Q. Luo, ``Far-field multicast
  beamforming for uniform linear antenna arrays,'' \emph{IEEE Trans. on Signal
  Process.}, vol.~55, no.~10, pp. 4916--4927, Oct. 2007.

\bibitem{DartAsch13}
G.~Dartmann and G.~Ascheid, ``Equivalent quasi-convex form of the multicast
  max-min beamforming problem,'' \emph{IEEE Trans. on Vehicular Technology},
  vol.~62, no.~9, pp. 4643--4648, Nov. 2013.

\bibitem{Luo_So_et_al_alSPMag10}
Z.-Q. Luo, W.-K. Ma, A.~M.-C. So, Y.~Ye, and S.~Zhang, ``Semidefinite
  relaxation of quadratic optimization problems,'' \emph{IEEE Signal Process.
  Mag.}, vol.~27, no.~3, pp. 20--34, May 2010.

\bibitem{Chang_Luo_ChiTSP08}
T.-H. Chang, Z.-Q. Luo, and C.-Y. Chi, ``Approximation bounds for semidefinite
  relaxation of max-min-fair multicast transmit beamforming problem,''
  \emph{IEEE Trans. on Signal Process.}, vol.~56, no.~8, pp. 3932--3943, Aug.
  2008.

\bibitem{SchaPesa12}
A.~Schad and M.~Pesavento, ``Max-min fair transmit beamforming for multi-group
  multicasting,'' in \emph{2012 International ITG Workshop on Smart Antennas
  (WSA)}, Dresden, Germany, Mar. 7-8 2012, pp. 115 -- 118.

\bibitem{ChriChatOtte15}
D.~Christopoulos, S.~Chatzinotas, and B.~Ottersten, ``Multicast multigroup
  beamforming for per-antenna power constrained large-scale arrays,''
  \emph{arXiv preprint}, Mar. 2015.

\bibitem{MehaHuanGopaKonaSidi15}
O.~Mehanna, K.~Huang, B.~Gopalakrishnan, A.~Konar, and N.~D. Sidiropoulos,
  ``Feasible point pursuit and successive approximation of non-convex qcqps,''
  \emph{IEEE Signal Process. Lett.}, vol.~22, no.~7, pp. 804--808, Jul. 2015.

\bibitem{Larsson-Jorswieck-Lindblom-Mochaourab_SPMag_sub09}
E.~Larsson, E.~Jorswieck, J.~Lindblom, and R.~Mochaourab, ``Game theory and the
  flat-fading gaussian interference channel,'' \emph{{IEEE} Signal Process.
  Mag.}, vol.~26, no.~5, pp. 18--27, Sept. 2009.

\bibitem{Morduk06}
B.~S. Mordukhovich, \emph{Variational Analysis and Generalized Differentiation:
  Basic Theory}.\hskip 1em plus 0.5em minus 0.4em\relax Springer, 2006.

\bibitem{RockWets98}
R.~Rockafellar and J.~Wets, \emph{Variational Analysis}.\hskip 1em plus 0.5em
  minus 0.4em\relax Springer, 1998.

\end{thebibliography}
}{\small \par}
\newpage

\section{Supporting Material}

\subsection{\label{sec:sol-equivalence}Proof of Proposition \ref{prop: eq-prob}}
\noindent (i): The proof follows readily by inspection of the MFCQ,
written for Problem \emph{\ref{eq:max-min-IBC-epi}}, and thus
is omitted. \\ \noindent (ii):  We prove the statement in a more general setting.
Consider the following two optimization problems 
\begin{equation}
\begin{array}{cl}
\underset{\mathbf{x}\in\mathcal{K}}{\min} & F\left(\mathbf{x}\right)\triangleq\max\left\{ f_{1}(\mathbf{x}),\cdots,f_{I}(\mathbf{x})\right\} \end{array}\label{eq:max-min-general}
\end{equation}
and\vspace{-0.1cm} 
\begin{equation}
\begin{array}{cl}
\underset{\mathbf{x},\,t\leq0}{\textrm{min}} & t\\
\textrm{\,\,s. t.} & f_{i}(\mathbf{x})\leq t,\ \forall i\in\mathcal{I},\\
 & \mathbf{x}\in\mathcal{K},\vspace{-0.1cm}
\end{array}\label{eq:max-min-general-smooth}
\end{equation}
where each $f_{i}:\mathcal{K}\rightarrow\mathbb{{R}}_{-}$ is a nonpositive
(possibly nonconvex) differentiable function with Lipschitz gradient
on $\mathcal{K}$; $\mathcal{K}\subseteq\mathbb{{R}}^{n}$ is a (nonempty)
closed and convex set (with nonempty relative interior); and $\mathcal{I}\triangleq\{1,\cdots,I\}$.
We also assume that (\ref{eq:max-min-general}) has a solution. 

In order to prove Proposition \ref{prop: eq-prob}(a), it is sufficient
to show the following equivalence between (\ref{eq:max-min-general})
and (\ref{eq:max-min-general-smooth}): $\mathbf{x}^{\star}$ is a
d-stationary solution of (\ref{eq:max-min-general}) if and only if
there exists a $t^{\star}$ such that $\left(\mathbf{x}^{\star},t^{\star}\right)$
is a regular stationary point of (\ref{eq:max-min-general-smooth}).

We first introduce some preliminary results. Denoting by $\partial F(\mathbf{x})$
the set of subgradients of $F$ at $\mathbf{x}$ (see \citep[Definition 1.77]{Morduk06}),
we recall that, since each $f_{i}$ is continuously differentiable,
we can write \citep[Corollary 10.51]{RockWets98} 
\[
\partial F(\mathbf{x})=\mathtt{con}\left(\left\{ \nabla f_{i}(\mathbf{x}):\,i\in\mathcal{I}(\mathbf{x})\right\} \right),
\]
where $\mathcal{I}\left(\mathbf{x}\right)\triangleq\{i\in\mathcal{I}:\,{\displaystyle F(\mathbf{x})}=f_{i}(\mathbf{x})\}$
and $\mathtt{con}(\mathcal{A})$ denotes the convex hull of the set
$\mathcal{A}$. Moreover, invoking again \citep[Corollary 10.51]{RockWets98}
and thanks to \citep[Theorem 9.16]{RockWets98}, the directional derivative
of $F^{'}$ at $\mathbf{x}$ in the direction $\mathbf{d}$, $F^{\prime}(\mathbf{x};\mathbf{d})$,
exists for every $\mathbf{x}\in{\mathcal{K}}$ and $\mathbf{d}\in\mathbb{R}^{n}$,
and it is given by 
\begin{equation}
F^{\prime}(\mathbf{x};\mathbf{d})=\max\left\{ \left\langle \boldsymbol{\xi},\mathbf{d}\right\rangle :\boldsymbol{\xi}\in\partial F\left(\mathbf{x}\right)\right\} .\label{eq:dirdersub}
\end{equation}
Using the above definitions, if $\mathbf{x}^{\star}$ is
a d-stationary solution of (\ref{eq:max-min-general}), there exists
$\boldsymbol{\alpha^{\star}}\in\mathbb{R}_{+}^{I}$ such that the
following holds: 
\begin{equation}
\begin{array}{rcl}
{\displaystyle {\sum_{i\in\mathcal{I}}}\alpha_{i}^{\star}\nabla f_{i}\left(\mathbf{x}^{\star}\right)^{T}\left(\mathbf{x}-\mathbf{x}^{\star}\right)} & \geq & 0,\quad\forall\mathbf{x}\in{\mathcal{K}},\\[5pt]
{\displaystyle {\sum_{i\in{\mathcal{I}}}}\alpha_{i}^{\star}} & = & 1,\\[5pt]
\alpha_{i}^{\star} & \geq & 0,\,\forall i\in\mathcal{I}\left(\mathbf{x}^{\star}\right),\\[5pt]
\alpha_{i}^{\star} & = & 0,\,\forall i\in{\mathcal{I}}\setminus\mathcal{I}\left(\mathbf{x}^{\star}\right).
\end{array}\label{eq:d-stationary}
\end{equation}
 On the other hand, if $\left(\bar{\mathbf{x}},\bar{t}\right)$ is
regular stationary for (\ref{eq:max-min-general-smooth}), there exists
$\bar{\boldsymbol{\lambda}}\in\mathbb{R}_{+}^{I}$ such that 
\begin{equation}
\begin{array}{rcl}
{\displaystyle {\sum_{i\in\mathcal{I}}}}\bar{\lambda}_{i}\nabla f_{i}(\bar{\mathbf{x}})^{T}(\mathbf{x}-\bar{\mathbf{x}}) & \geq & 0,\quad\forall\mathbf{x}\in{\mathcal{K}},\\[5pt]
0\leq\,{\displaystyle {\sum_{i\in\mathcal{I}}}\bar{\lambda}_{i}}-1 & \,\perp\, & \bar{t}\leq0,\\[5pt]
0\leq\,\bar{\lambda}_{i}\, & \perp\, & \,f_{i}(\bar{\mathbf{x}})-\bar{t}\,\leq0,\quad\forall i\in\mathcal{I}.
\end{array}\label{eq:kkt}
\end{equation}
 We are now ready to prove the desired result. 

\noindent $\left(\boldsymbol{\Rightarrow}\right)$ Let $\mathbf{x}^{\star}$
be a d-stationary point of (\ref{eq:max-min-general}). By taking
$\left(\bar{\mathbf{x}},\bar{t},\bar{\boldsymbol{\lambda}}\right)\triangleq\left(\mathbf{x}^{\star},F(\mathbf{x}^{\star}),\boldsymbol{{\alpha}}^{\star}\right)$
and using \eqref{eq:d-stationary}, it is not difficult to check that
the tuple $\left(\bar{\mathbf{x}},\bar{t},\bar{\boldsymbol{\lambda}}\right)$
satisfies the KKT conditions \eqref{eq:kkt}. Therefore $\left(\bar{\mathbf{x}},\bar{t}\right)$
is stationary for (\ref{eq:max-min-general-smooth}).

\noindent $\left(\boldsymbol{\Leftarrow}\right)$ Let $\left(\bar{\mathbf{x}},\bar{t}\right)$
be a regular stationary point of (\ref{eq:max-min-general-smooth}).
Then, there exists $\bar{\boldsymbol{\lambda}}\in\mathbb{R}_{+}^{I}$
such that $\left(\bar{\mathbf{x}},\bar{t},\bar{\boldsymbol{\lambda}}\right)$
satisfies KKT conditions \eqref{eq:kkt}. Since ${\sum_{i\in\mathcal{I}}}\bar{\lambda}_{i}>0$,
it must be $\bar{t}=f_{i_{0}}(\bar{\mathbf{x}})$, for some $i_{0}\in\mathcal{I}$.
Therefore, $\bar{t}=F(\bar{\mathbf{x}})$. Furthermore, since $\mathcal{I}\left(\bar{\mathbf{x}}\right)=\left\{ i\in\mathcal{I}:\,\bar{t}=f_{i}\left(\bar{\mathbf{x}}\right)\right\} $,
we have $\bar{\lambda}_{j}=0$ for every $j\in{\mathcal{I}}\setminus\mathcal{I}\left(\bar{\mathbf{x}}\right)$.
Define $\tilde{{\lambda}}_{i}\triangleq\bar{\lambda}_{i}/\bar{\lambda}^{\text{{ave}}}$,
for all $i\in\mathcal{I}$, where $\bar{\lambda}^{\text{{ave}}}\triangleq\sum_{i\in\mathcal{I}\left(\bar{\mathbf{x}}\right)}\bar{\lambda}_{i}>0$.
It follows that $\sum_{i\in\mathcal{I}\left(\bar{\mathbf{x}}\right)}\tilde{{\lambda}}_{i}=1$,
and 
\[
\sum_{i\in\mathcal{I}}\tilde{{\lambda}}_{i}\nabla f_{i}(\bar{\mathbf{x}})=\sum_{i\in\mathcal{I}\left(\bar{\mathbf{x}}\right)}\tilde{{\lambda}}_{i}\nabla f_{i}(\bar{\mathbf{x}})\in\partial F\left(\bar{\mathbf{x}}\right).
\]
Therefore, for every $\mathbf{x}\in\mathcal{K}$, 
\[
\begin{array}{rcl}
F^{\prime}\left(\bar{\mathbf{x}};\mathbf{x}-\bar{\mathbf{x}}\right) & = & \max\left\{ \left\langle \boldsymbol{\xi},\mathbf{x}-\bar{\mathbf{x}}\right\rangle :\boldsymbol{\xi}\in\partial F\left(\bar{\mathbf{x}}\right)\right\} \smallskip\\
 & \geq & {\displaystyle {\sum_{i\in\mathcal{I}\left(\bar{\mathbf{x}}\right)}}\tilde{{\lambda}}_{i}\nabla f_{i}(\bar{\mathbf{x}})^{T}(\mathbf{x}-\bar{\mathbf{x}})\smallskip}\\
 & = & \left(1/\bar{\lambda}^{\text{{ave}}}\right)\cdot{\displaystyle {\sum_{i\in\mathcal{I}}}\bar{\lambda}_{i}\nabla f_{i}(\bar{\mathbf{x}})^{T}(\mathbf{x}-\bar{\mathbf{x}})\geq0,}
\end{array}
\]
where the last inequality follows from \eqref{eq:kkt} and $\bar{\lambda}^{\text{{ave}}}>0$.
Thus, $\bar{\mathbf{x}}$ is d-stationary for (\ref{eq:max-min-general}).
This completes the proof of statement (ii).

\subsection{\label{sec:MMF-sol-equiv-1}Proof of Proposition \ref{prop: eq-prob-MMF}}

\subsubsection{Preliminaries\label{sub:Preliminaries}}

\noindent Let us start by introducing some intermediate results needed
to prove the proposition. For notation simplicity, let define $y_{i_{k}}(\mathbf{w})\triangleq\sum_{\ell\neq k}{\mathbf{w}_{\ell}^{H}\mathbf{H}_{i_{k}\ell}\mathbf{w}_{\ell}}+\sigma_{i_{k}}^{2}$,
for all $i_{k}\in\mathcal{G}_{k}$ and $k\in\mathcal{K}_{\text{{BS}}}$.
\\
 \indent Following the same arguments as in Appendix \ref{sec:sol-equivalence}
{[}cf.$\,$(\ref{eq:d-stationary}){]}, if $\mathbf{w}^{\star}$
is a d-stationary solution of (\ref{eq:MMF}), then there exists $\boldsymbol{\alpha}^{\star}\in\mathbb{R}_{+}^{I}$
such that $(\mathbf{w}^{\star},\boldsymbol{\alpha}^{\star})$ satisfies
\begin{equation}
\begin{array}{l}
(a):\,\,{\displaystyle \sum_{k\in\mathcal{K}_{\text{{BS}}}}\sum_{i_{k}\in\mathcal{G}_{k}}\alpha_{i_{k}}^{\star}\left\langle \nabla_{\mathbf{w}^{*}}u_{i_{k}}(\mathbf{w}^{\star}),\mathbf{w}-\mathbf{w}^{\star}\right\rangle }\leq0,\,\,\,\forall\mathbf{w}\in{\mathcal{W}},\smallskip\\
({b)}:\,\,{\displaystyle \sum_{k\in\mathcal{K}_{\text{{BS}}}}\sum_{i_{k}\in\mathcal{G}_{k}}\alpha_{i_{k}}^{\star}}=1,\smallskip\\
({c)}:\,\,\alpha_{i_{k}}^{\star}\geq0,\qquad\forall i_{k}\in\mathcal{I}\left(\mathbf{w}^{\star}\right),\smallskip\\
({d)}:\,\,\alpha_{i_{k}}^{\star}=0,\qquad\forall i_{k}\notin\mathcal{I}\left(\mathbf{w}^{\star}\right),
\end{array}\label{d_stat_MMF}
\end{equation}
where $u_{i_{k}}(\mathbf{w})\triangleq\mathbf{w}_{k}^{H}\mathbf{H}_{i_{k}k}\mathbf{w}_{k}/y_{i_{k}}(\mathbf{w})$,
$\nabla_{\mathbf{w}^{*}}u_{i_{k}}(\mathbf{w})=(\nabla_{\mathbf{w}_{\ell}^{*}}u_{i_{k}}(\mathbf{w}))_{\ell\in\mathcal{K}_{\text{{BS}}}}$,
with 
\[
\nabla_{\mathbf{w}_{\ell}^{*}}u_{i_{k}}(\mathbf{w})\triangleq\left\{ \begin{array}{ll}
\frac{1}{y_{i_{k}}(\mathbf{w})}\,\mathbf{H}_{i_{k}k}\mathbf{w}_{k}, & \ \ell=k\smallskip\\
-\frac{\mathbf{w}_{k}^{H}\mathbf{H}_{i_{k}k}\mathbf{w}_{k}}{\left(y_{i_{k}}(\mathbf{w})\right)^{2}}\,\nabla_{\mathbf{w}_{\ell}^{*}}y_{i_{k}}(\mathbf{w}), & \ \ell\neq k;
\end{array}\right.
\]
and $\mathcal{I}(\mathbf{w}^{\star})\!\triangleq\!\left\{ i_{k}:U(\mathbf{w}^{\star})=u_{i_{k}}(\mathbf{w}^{\star})\right\} $,$\,$with
$U$$\,$defined$\,$in$\,$(\ref{eq:MMF}).

On the other hand, if $(\bar{{t}},\bar{{\boldsymbol{\beta}}},\bar{{\mathbf{w}}})$
is a regular stationary solution of (\ref{eq:MMF_2}), there exist
multipliers $(\bar{{\boldsymbol{{\lambda}}}},\bar{{\boldsymbol{{\eta}}}},\bar{{\boldsymbol{{\rho}}}})=((\bar{{\lambda}}_{i_{k}}$,
$\bar{{\eta}}_{i_{k}}$, $\bar{{\rho}}_{i_{k}})_{i_{k}\in\mathcal{G}_{k}})_{k\in\mathcal{K}_{\text{{BS}}}}$
and $\bar{{\zeta}}$ such that the following KKT conditions (referred, to be precise, to problem (\ref{eq:MMF_2}) where both sides of constraints (a) are divided by ${\beta}_{i_{k}}>0$) are satisfied 
\begin{equation}
\!\!\!\!\!\!\!\begin{array}{l}
(a^{'}):\!\!{\displaystyle \sum_{k\in\mathcal{K}_{\text{{BS}}}}\!\sum_{i_{k}\in\mathcal{G}_{k}}}\!\!\left\langle \nabla_{\mathbf{w}^{*}}\!\!\left(\frac{\bar{{\lambda}}_{i_{k}}}{\bar{{\beta}}_{i_{k}}}\bar{{\mathbf{w}}}_{k}^{H}\mathbf{H}_{i_{k}k}\bar{{\mathbf{w}}}_{k}\!-\bar{{\eta}}_{i_{k}}\,y_{i_{k}}(\bar{{\mathbf{w}}})\right)\!,\mathbf{w}-\bar{{\mathbf{w}}}\right\rangle \smallskip\\
\hfill\leq0,\,\,\forall\mathbf{w}\in\mathcal{W},\\
({b^{'})}:\!\!{\displaystyle \sum_{k\in\mathcal{K}_{\text{{BS}}}}\sum_{i_{k}\in\mathcal{G}_{k}}}\bar{{\lambda}}_{i_{k}}=1+\bar{{\zeta}},\smallskip\\
({c^{'})}:\,\,\bar{\eta}_{i_{k}}={\displaystyle {\frac{\bar{{\lambda}}_{i_{k}}\bar{{\mathbf{w}}}_{k}^{H}\mathbf{H}_{i_{k}k}\bar{{\mathbf{w}}}_{k}}{\bar{{\beta}}_{i_{k}}^{2}}}+\bar{{\rho}}_{i_{k}},\qquad\quad\;\forall i_{k}\in\mathcal{G}_{k},k\in\mathcal{K}_{\text{{BS}}},\smallskip}\\
({d^{'})}:\,\,0\leq\bar{{\lambda}}_{i_{k}}\perp \Big(\bar{{t}} -{\displaystyle {\frac{\bar{{\mathbf{w}}}_{k}^{H}\mathbf{H}_{i_{k}k}\bar{{\mathbf{w}}}_{k}}{\bar{{\beta}}_{i_{k}}}}\Big)\leq0,\;\,\, \forall i_{k}\in\mathcal{G}_{k},k\in\mathcal{K}_{\text{{BS}}},\smallskip}\\
({e^{'})}:\,\,0\leq\bar{{\eta}}_{i_{k}}\perp{(y_{i_{k}}(\bar{{\mathbf{w}}})-\bar{{\beta}}_{i_{k}})}\leq0,\hspace{0.9cm}\forall i_{k}\in\mathcal{G}_{k},k\in\mathcal{K}_{\text{{BS}}},\smallskip\\
({f^{'})}:\,\,0\leq\bar{{\rho}}_{i_{k}}\perp(\bar{{\beta}}_{i_{k}} - \beta_{i_{k}}^{\max})\leq0,\hspace{1.1cm}\forall i_{k}\in\mathcal{G}_{k},k\in\mathcal{K}_{\text{{BS}}},\smallskip\\
({g^{'})}:\,\,0\leq\bar{{\zeta}}\perp\bar{{t}}\geq0,
\end{array}\label{eq:kkt-smth}
\end{equation}
where, with a slight abuse of notation, we denoted by $\nabla_{\mathbf{w}^{*}}(\bar{{\mathbf{w}}}_{k}^{H}\mathbf{H}_{i_{k}k}\bar{{\mathbf{w}}}_{k})$
and $\nabla_{\mathbf{w}^{*}}(y_{i_{k}}(\bar{{\mathbf{w}}}))$ the conjugate
gradient of ${\mathbf{w}}_{k}^{H}\mathbf{H}_{i_{k}k}\mathbf{w}_{k}$
and $y_{i_{k}}(\mathbf{w})$, respectively, evaluated at $\mathbf{w}=\bar{{\mathbf{w}}}$. 

Note that it must always be $\bar{{\boldsymbol{{\rho}}}}=\mathbf{0}$,
as shown next. Suppose that there exists a $i_{k}$ such that $\bar{{\rho}}_{i_{k}}>0$.
Then, invoking the complementarity condition in $(f^{'})$, we have
$\bar{{\beta}}_{i_{k}}=\beta_{i_{k}}^{\max}$. Also, by the definition
of $\beta_{i_{k}}^{\max}$, it must be $\bar{{\beta}}_{i_{k}}>{y_{i_{k}}(\bar{\mathbf{w}})}>0$,
and thus $\bar{{\eta}}_{i_{k}}=0$ {[}by complementarity in $(e^{'})${]}.
It follows from $(c^{'})$ that $-\bar{{\rho}}_{i_{k}}=\bar{{\lambda}}_{i_{k}}\bar{{\mathbf{w}}}_{k}^{H}\mathbf{H}_{i_{k}k}\bar{{\mathbf{w}}}_{k}/\bar{{\beta}}_{i_{k}}^{2}<0$,
which contradicts the fact that $\bar{{\lambda}}_{i_{k}}\bar{{\mathbf{w}}}_{k}^{H}\mathbf{H}_{i_{k}k}\bar{{\mathbf{w}}}_{k}/\bar{{\beta}}_{i_{k}}^{2}\geq0$
(recall that $\mathbf{H}_{i_{k}k}$ is a positive semidefinite matrix).

\subsubsection{Proof of Proposition \ref{prop: eq-prob-MMF}}

We prove only statement (ii).

\noindent $(\Rightarrow)$\textcolor{blue}{: }Let ${\mathbf{w}}^{\star}$
be a d-stationary solution of (\ref{eq:MMF}). Partition the set of
users $\mathcal{I}$ according to $\mathcal{I}=\mathcal{I}_{\text{{a}}}(\mathbf{w}^{\star})\cup\overline{{\mathcal{I}}}_{\text{{a}}}(\mathbf{w}^{\star})$,
where $\mathcal{I}_{\text{{a}}}(\mathbf{w}^{\star})$ is the set of
active users at $\mathbf{w}^{\star}$ (i.e., the users served by a
BS), defined as
\begin{equation}
\mathcal{I}_{\text{{a}}}(\mathbf{w}^{\star})\triangleq\left\{ i_{k}\,:\,\mathbf{H}_{i_{k}k}\mathbf{w}_{k}^{\star}\neq\mathbf{0}\right\} ,\label{eq:I_active_at_d_stationarity}
\end{equation}
and $\overline{{\mathcal{I}}}_{\text{{a}}}(\mathbf{w}^{\star})$ is
its complement. Note that, since all $\mathbf{H}_{i_{k}k}$ are nonzero
and positive semidefinite, there must exists a feasible $\mathbf{w}=(\mathbf{w}_{k})_{k\in\mathcal{K}_{\text{{BS}}}}$
such that $\mathbf{H}_{i_{k}k}\mathbf{w}_{k}\neq\mathbf{0}$ for some
$i_{k}\in\mathcal{G}_{k}$ and $k\in\mathcal{{K}}_{\text{{BS}}}$.
We distinguish the following two complementary cases: either 1) $\overline{{\mathcal{I}}}_{\text{{a}}}(\mathbf{w}^{\star})\neq\emptyset$;
or 2) $\overline{{\mathcal{I}}}_{\text{{a}}}(\mathbf{w}^{\star})=\emptyset$.
The former is a degenerate (undesired) case: the d-stationary
point $\mathbf{w}^{\star}$ is a global \emph{minimizer} of $U$,
since $U(\mathbf{w}^{\star})=0$. The latter case, implying $U(\mathbf{w}^{\star})>0$,
is instead the interesting one, which the global optimal solution
of (\ref{eq:MMF}) belongs to. 

\noindent  \emph{Case 1:} $\overline{{\mathcal{I}}}_{\text{{a}}}(\mathbf{w}^{\star})\neq\emptyset$.
Choose $(\bar{t},\bar{\boldsymbol{\beta}},\bar{\mathbf{w}},\bar{\boldsymbol{\lambda}},\bar{\boldsymbol{\eta}},\bar{\boldsymbol{\rho}},\bar{\zeta})$
as follows: $\bar{{t}}=0$; $\bar{\boldsymbol{\beta}}$ is any vector
$\bar{\boldsymbol{\beta}}\triangleq((\bar{\beta}_{i_{k}})_{i_{k}\in\mathcal{G}_{k}})_{k\in\mathcal{K}_{\text{{BS}}}}$
satisfying the feasibility conditions in ($e^{'}$) and ($f^{'}$),
that is, $y_{i_{k}}(\mathbf{w}^{\star})\leq\bar{\beta}_{i_{k}}\leq\beta_{i_{k}}^{\max}$,
for all $i_{k}\in\mathcal{G}_{k}$ and $k\in\mathcal{K}_{\text{{BS}}}$;
$\bar{\mathbf{w}}=\mathbf{w}^{\star}$; $\bar{\boldsymbol{\lambda}}=((\alpha_{i_{k}}^{\star})_{i_{k}\in\mathcal{G}_{k}})_{k\in\mathcal{K}_{\text{{BS}}}}$\textcolor{blue}{,}
$\bar{\boldsymbol{\eta}}=\bar{\boldsymbol{\rho}}=\mathbf{0}$; and
$\bar{\zeta}=0$. We show next that such a tuple satisfies the KKT
conditions (\ref{eq:kkt-smth}).

Observe preliminary that $u_{i_{k}}(\mathbf{w}^{\star})=0$ if and
only if $i_{k}\in\overline{{\mathcal{I}}}_{\text{{a}}}(\mathbf{w}^{\star})$,
implying $U(\mathbf{w}^{\star})=u_{i_{k}}(\mathbf{w}^{\star})=0$,
for all $i_{k}\in\overline{{\mathcal{I}}}_{\text{{a}}}(\mathbf{w}^{\star})$
{[}recall that all $u_{i_{k}}(\mathbf{w}^{\star})\geq0${]}; therefore,
it must be 
\begin{equation}
{\mathcal{I}}(\mathbf{w}^{\star})\equiv\overline{{\mathcal{I}}}_{\text{{a}}}(\mathbf{w}^{\star}).\label{eq:I_equal_to_nonactive}
\end{equation}
Invoking now ($d$) in (\ref{d_stat_MMF}) and using (\ref{eq:I_equal_to_nonactive}),
we have $\bar{{\lambda}}_{i_{k}}=0$, for all $i_{k}\in{\mathcal{I}}_{\text{{a}}}(\mathbf{w}^{\star})$.
It follows that
\[
\bar{{\lambda}}_{i_{k}}\cdot\nabla_{\mathbf{w}^{*}}\left(\bar{{\mathbf{w}}}_{k}^{H}\mathbf{H}_{i_{k}k}\bar{{\mathbf{w}}}_{k}\right)=\mathbf{0},\quad\forall i_{k}\in\mathcal{G}_{i_{k}},\,\forall k\in\mathcal{K}_{\text{{BS}}},
\]
which, together with $\bar{\boldsymbol{\eta}}=\mathbf{0}$, make ($a^{'}$)
in (\ref{eq:kkt-smth}) satisfied, with the LHS equal to zero.

Condition ($b^{'}$) follows readily from $\sum_{k\in\mathcal{K}_{\text{{BS}}}}\sum_{i_{k}\in\mathcal{G}_{k}}\bar{{\lambda}}_{i_{k}}=1$
{[}cf. ($b$) in (\ref{d_stat_MMF}){]} and $\bar{\zeta}=0$.

Condition ($c^{'}$) can be checked observing that $\bar{{\lambda}}_{i_{k}}\cdot\bar{{\mathbf{w}}}_{k}^{H}\mathbf{H}_{i_{k}k}\bar{{\mathbf{w}}}_{k}=\mathbf{0}$,
for all $i_{k}\in\mathcal{G}_{i_{k}}$ and $k\in\mathcal{K}_{\text{{BS}}}$. 

Conditions ($d^{'}$)-($g^{'}$) follow readily by inspection. 

\noindent  \emph{Case 2:} $\overline{{\mathcal{I}}}_{\text{{a}}}(\mathbf{w}^{\star})=\emptyset$.
Choose now $(\bar{t},\bar{\boldsymbol{\beta}},\bar{\mathbf{w}},\bar{\boldsymbol{\lambda}},\bar{\boldsymbol{\eta}},\bar{\boldsymbol{\rho}},\bar{\zeta})$
as follows: $\bar{{t}}=U(\mathbf{w}^{\star})$; $\bar{\boldsymbol{\beta}}=(y_{i_{k}}(\mathbf{w}^{\star}))_{i_{k}\in\mathcal{G}_{k},k\in\mathcal{K}_{\text{{BS}}}}$;
$\bar{\mathbf{w}}=\mathbf{w}^{\star}$; $\bar{\boldsymbol{\lambda}}=((\alpha_{i_{k}}^{\star})_{i_{k}\in\mathcal{G}_{k}})_{k\in\mathcal{K}_{\text{{BS}}}}$;
$\bar{\boldsymbol{\eta}}\triangleq(\bar{\eta}_{i_{k}})_{i_{k}\in\mathcal{G}_{k},k\in\mathcal{K}_{\text{{BS}}}}$,
with each $\bar{\eta}_{i_{k}}=\bar{\lambda}_{i_{k}}\cdot u_{i_{k}}(\mathbf{w}^{\star})/y_{i_{k}}(\mathbf{w}^{\star})$;
$\bar{\boldsymbol{\rho}}=\mathbf{0}$; and $\bar{\zeta}=0$. 

By substitution, one can see that 
\[
\begin{array}{l}
{\displaystyle \sum_{k\in\mathcal{K}_{\text{{BS}}}}\sum_{i_{k}\in\mathcal{G}_{k}}}\!\!\left\langle \nabla_{\mathbf{w}^{*}}\left(\frac{\bar{{\lambda}}_{i_{k}}}{\bar{{\beta}}_{i_{k}}}\bar{{\mathbf{w}}}_{k}^{H}\mathbf{H}_{i_{k}k}\bar{{\mathbf{w}}}_{k}\!-\bar{{\eta}}_{i_{k}}\,y_{i_{k}}(\bar{{\mathbf{w}}})\right),\mathbf{w}-\bar{{\mathbf{w}}}\right\rangle \\
={\displaystyle \sum_{k\in\mathcal{K}_{\text{{BS}}}}\sum_{i_{k}\in\mathcal{G}_{k}}}\alpha_{i_{k}}^{\star}\left\langle \nabla_{\mathbf{w}^{*}}u_{i_{k}}(\mathbf{w}^{\star}),\mathbf{w}-\bar{{\mathbf{w}}}\right\rangle \leq0,\,\,\forall\mathbf{w}\in\mathcal{W},
\end{array}
\]
where the last implication follows from ($a$) in (\ref{d_stat_MMF});
therefore, ($a^{'}$) in (\ref{eq:kkt-smth}) is satisfied. 

Condition ($b^{'}$) follows readily from $\sum_{k\in\mathcal{K}_{\text{{BS}}}}\sum_{i_{k}\in\mathcal{G}_{k}}\bar{{\lambda}}_{i_{k}}=1$
{[}cf. ($b$) in (\ref{d_stat_MMF}){]} and $\bar{\zeta}=0$. 

Condition ($c^{'}$) follows by inspection. 

Condition ($d^{'}$) can be checked as follows. Since by ($b^{'}$)
there exists a $\bar{{\lambda}}_{i_{k}}>0$, it must be $\bar{{t}}={\displaystyle \bar{{\mathbf{w}}}_{k}^{H}\mathbf{H}_{i_{k}k}\bar{{\mathbf{w}}}_{k}/\bar{{\beta}}_{i_{k}},}$
which together with $\bar{{\beta}}_{i_{k}}=y_{i_{k}}(\mathbf{w}^{\star})$,
can be rewritten as $\bar{{t}}=U(\mathbf{w}^{\star})$.

Finally, conditions ($e^{'}$)-($g^{'}$) follow by inspection. 

\noindent (a)\textcolor{blue}{{} }$(\Leftarrow)$\textcolor{blue}{:}
Let us prove now the converse. Let $(\bar{t},\bar{\boldsymbol{\beta}},\bar{\mathbf{w}})$
be a regular stationary point of (\ref{eq:MMF_2}). One can show that, if $\overline{{\mathcal{I}}}_{\text{{a}}}(\bar{\mathbf{w}})\neq\emptyset$,
$\bar{\mathbf{w}}$ is a d-stationary solution of (\ref{eq:MMF})
with $U^{\prime}(\bar{\mathbf{w}};\mathbf{w}-\bar{\mathbf{w}})=0$,
for all $\mathbf{w}\in\mathcal{W}$; we omit further details.

Let us consider now the nondegenerate case $\overline{{\mathcal{I}}}_{\text{{a}}}(\bar{\mathbf{w}})=\emptyset$.
Let $(\bar{\boldsymbol{\lambda}},\bar{\boldsymbol{\eta}},\bar{\boldsymbol{\rho}},\bar{\zeta})$
be the multipliers such that $(\bar{t},\bar{\boldsymbol{\beta}},\bar{\mathbf{w}},\bar{\boldsymbol{\lambda}},\bar{\boldsymbol{\eta}},\bar{\boldsymbol{\rho}},\bar{\zeta})$
satisfies (\ref{eq:kkt-smth}). It follows from $(b^{'})$ that there
exists a $\bar{\lambda}_{i_{k}}>0$. Then, we have the following chain
of implication (each of them due to the condition reported on top
of the implication) 
\begin{equation}
\begin{array}{lll}
\bar{\lambda}_{i_{k}}>0 & \stackrel{(d^{'})}{\Rightarrow} & \bar{t}={\frac{\bar{{\mathbf{w}}}_{k}^{H}\mathbf{H}_{i_{k}k}\bar{{\mathbf{w}}}_{k}}{\bar{{\beta}}_{i_{k}}}}>0\\
 & \stackrel{(c^{'})}{\Rightarrow} & \bar{\eta}_{i_{k}}>0\\
 & \stackrel{(e^{'})}{\Rightarrow} & \bar{{\beta}}_{i_{k}}=y_{i_{k}}(\bar{\mathbf{w}})>0\\
 & \stackrel{(d^{'})}{\Rightarrow} & \bar{t}=u_{i_{k}}(\bar{{\mathbf{w}}}).
\end{array}\label{eq:chain_ineq}
\end{equation}

Denoting by $\mathcal{I}(\bar{\mathbf{w}})\triangleq\left\{ i_{k}\,:\,\bar{t}=u_{i_{k}}(\bar{\mathbf{w}})\right\} $
and $\alpha_{i_{k}}^{\star}\triangleq\frac{\bar{\lambda}_{i_{k}}}{1+\bar{\zeta}},$
for all $i_{k}\in\mathcal{G}_{k}$ and $k\in\mathcal{K}_{{BS}}$,
it follows from (\ref{eq:chain_ineq}) and ($b^{'}$) that 
\begin{equation}
\begin{array}{l}
\sum_{k\in\mathcal{K}_{{BS}}}\sum_{i_{k}\in\mathcal{G}_{k}}\alpha_{i_{k}}^{\star}=1,\smallskip\\
\alpha_{i_{k}}^{\star}\geq0,\quad\forall i_{k}\in\mathcal{I}(\bar{{\mathbf{w}}}),\smallskip\\
\alpha_{i_{k}}^{\star}=0,\quad\forall i_{k}\notin\mathcal{I}(\bar{{\mathbf{w}}}).
\end{array}\label{eq:covex-comb-alpha}
\end{equation}

Using (\ref{eq:covex-comb-alpha}), we finally have 
\[
\begin{array}{c}
\sum_{k\in\mathcal{K}_{{BS}}}\sum_{i_{k}\in\mathcal{G}_{k}}\alpha_{i_{k}}^{\star}\nabla_{\mathbf{w}^{*}}\left(-u_{i_{k}}(\bar{\mathbf{w}})\right)\hfill\smallskip\\
=-\sum_{i_{k}\in\mathcal{I}(\bar{\mathbf{w}})}\alpha_{i_{k}}^{\star}\nabla_{\mathbf{w}^{*}}u_{i_{k}}(\bar{\mathbf{w}})\in\partial(-U(\bar{\mathbf{w}})),
\end{array}
\]
 where $\partial(-U(\bar{\mathbf{w}}))$ is the set of (conjugate) subgradients
of $-U$ at $\bar{\mathbf{w}}$. For every $\mathbf{w}\in\mathcal{W}$,
\[
\begin{array}{l}
-U^{\prime}(\bar{\mathbf{w}};\mathbf{w}-\bar{\mathbf{w}})=\max\left\{ \left\langle \boldsymbol{\xi},\mathbf{w}-\bar{\mathbf{w}}\right\rangle :\boldsymbol{\xi}\in\partial(-U(\bar{\mathbf{w}}))\right\} \\
\geq\left\langle -\sum_{i_{k}\in\mathcal{I}(\bar{\mathbf{w}})}\alpha_{i_{k}}^{\star}\nabla_{\mathbf{w}^{*}}u_{i_{k}}(\bar{\mathbf{w}}),\mathbf{w}-\bar{\mathbf{w}}\right\rangle \\
=-\dfrac{1}{1+\bar\zeta}\\\cdot {\displaystyle \sum_{k\in\mathcal{K}_{\text{{BS}}}}\sum_{i_{k}\in\mathcal{G}_{k}}}\!\!\left\langle \nabla_{\mathbf{w}^{*}}\left(\frac{\bar{{\lambda}}_{i_{k}}}{\bar{{\beta}}_{i_{k}}}\bar{{\mathbf{w}}}_{k}^{H}\mathbf{H}_{i_{k}k}\bar{{\mathbf{w}}}_{k}\!-\bar{{\eta}}_{i_{k}}\,y_{i_{k}}(\bar{{\mathbf{w}}})\right),\mathbf{w}-\bar{\mathbf{w}}\right\rangle \\
\geq0,
\end{array}
\]
where the last inequality is due to $(a^{'})$. This proves that $\bar{\mathbf{w}}$
is d-stationary for (\ref{eq:MMF}).

\end{document}